\documentclass[11pt,reprint,showpacs,showkeys,amsfonts,amssymb,amsmath,nofootinbib]{article}
\usepackage{bm,amsmath,epsf,epsfig,latexsym,amssymb,cite,lmodern}
\usepackage{graphics,psfrag}
\usepackage{hyperref}\textheight 22.3cm
\newcommand{\bg}{\begin{align}}
\newcommand{\eeg}{\end{align}}
\newcommand{\be}{\begin{equation}}
\newcommand{\ee}{\end{equation}}
\newcommand{\ba}{\begin{eqnarray}}
\newcommand{\ea}{\end{eqnarray}}

\newcommand{\nn}{\nonumber}

\begin{document}

\thispagestyle{empty}

\vspace{2cm}
\begin{center}
{\Large{\bf Nucleon-Nucleon scattering from dispersion relations: next-to-next-to-leading order study}}
\end{center}
\vspace{.5cm}

\begin{center}
{\Large J.~A.~Oller}
\end{center}

\begin{center}
{\it  Departamento de F\'{\i}sica. Universidad de Murcia.\\
 E-30071 Murcia, Spain.\\
oller@um.es}
\end{center}
\vspace{1cm}

\begin{abstract}
We study nucleon-nucleon ($NN$) scattering by applying the $N/D$ method in chiral perturbation theory up to 
next-to-next-to-leading  (NNLO) order in the calculation of the imaginary part of the $NN$ 
partial-wave amplitudes along the left-hand-cut, which is the dynamical input for this approach.
 A quite good reproduction of the Nijmegen partial-wave analysis phase shifts and mixing angles is 
obtained, which implies a steady improvement in the accurateness achieved by increasing the chiral order 
in the calculation of the dynamical input.
 A power counting for the subtraction constants is established, which is appropriate for those subtractions 
attached to both the left- and  right-hand cuts.
We discuss that it is not necessary to modify the  $NN$ chiral potential at NNLO to agree with data, but 
instead one should perform the iteration of two-nucleon 
intermediate states to finally achieve analytic and unitarity $NN$ partial-wave amplitudes  in a well-defined way. 
We also confirm at NNLO the long-range correlations between the $NN$ $S$-wave effective ranges and scattering lengths, 
when employing only once-subtracted dispersion relations, that holds up to around 10\% when compared with experimental values.

\end{abstract}


\newpage
\section{Introduction}\label{sec:intro}

Chiral Perturbation Theory (ChPT) is the effective field theory of QCD at low energies \cite{weinberg75,sielder}. 
Its  paradigmatic application is the purely mesonic sector in $SU(2)$.\footnote{Which even presents one corner of 
concern due to the enhanced role of the right-hand-cut in the isoscalar scalar pion-pion scattering 
\cite{npa620,nd,zeros,gamma,alba12,bernasigma,truong1988}, with an important 
impact as well in the pion-nucleon ($\pi N$) sector \cite{sainiosumrule,aco2013,sigma2012}.}
Its extension to the one-baryon sector presents some complications due to the 
large nucleon mass that does not vanish in the chiral limit \cite{sainio88,manohar91}, which posed interesting problems to 
the theory.\footnote{A faster stabilization of the chiral series in this case has been recently accomplished \cite{sigma2012,aco2013,strangeness14} by combining the covariant formalism of the Extended on Mass Shell Regularization Scheme (EOMS) \cite{eoms} with the explicit inclusion 
of the $\Delta(1232)$ in the $\delta$-counting \cite{delta-c} .}
  For reviews on ChPT on these topics see e.g. \cite{bijnens09,ecker,bernard,pich,ulf}.

The extension of ChPT to systems with a larger  baryonic number was  
considered in  Ref.~\cite{weinn}, where the chiral counting is applied to the calculation 
of the multi-nucleon potential. In these cases one also has to face the problem associated with the 
infrared enhancement associated with the small nucleon kinetic energies, which requires 
to resum the infinite string of diagrams due to the iteration of  intermediate multi-nucleon states. 
The extension of the chiral power counting to finite density system, including the contributions of multi-nucleon 
reducible diagrams, is given in Ref.~\cite{finiteden}. For related reviews see e.g. 
\cite{Epelbaum:2008ga,Machleidt:2011zz,Epelbaum:2005pn,Bedaque:2002mn,vanKolck:1999mw}.

 The application of the set up of Ref.~\cite{weinn} to nucleon-nucleon ($NN$) scattering 
has been phenomenologically successful \cite{ordo94,entem,thesis,epe042}. However, the sensitivity 
of the results on the values of the cutoff taken to solve the associated  Lippmann-Schwinger equation for the 
iteration of two-nucleon intermediate states has given rise to a flurry of publications, whose fair 
 and comprehensive consideration is beyond this introduction. For more detailed accounts on this 
respect the reader is referred to \cite{Machleidt:2011zz,pavon06,nogga,kswa,phillipssw,pavon11,longyang,zeoli,Epelbaum:2008ga}.

We continue here  the application of the $N/D$ method \cite{chew} to $NN$ scattering 
 extending the previous work of Refs.~\cite{paper1,paper2,gor2013}. For this method the dynamical 
input is not the $NN$ potential but the discontinuity of a $NN$ partial-wave amplitude along the 
left-hand-cut (LHC), which is denoted in the following by $2i\Delta(A)$. Here $A$ is   
the center of mass (c.m.) three-momentum squared of a $NN$ state. In other words, 
$\Delta(A)$ is the imaginary part of a $NN$ partial-wave amplitude along the LHC, that 
 extends for real $A$ with $A<-M_\pi^2/4$, being $M_\pi$ the pion ($\pi$) mass.  
The function $\Delta(A)$  is due to the multi-exchange of pions driving the finite-range nuclear 
forces, while in a low-energy effective field theory the short-range nuclear forces are accounted for by 
local interactions of zero range that do not contribute to $\Delta(A)$ for finite $A$.
 The two-nucleon irreducible contributions to $\Delta(A)$ are amenable to a straightforward ChPT expansion, 
in much the same way as discussed  in Ref.~\cite{weinn} for the calculation of the chiral $NN$ potential.
However, $\Delta(A)$ has also contributions from two-nucleon reducible diagrams but, 
as explained in Ref.~\cite{gor2013},  these contributions require to cut all 
the pion lines simultaneously when iterating one-pion exchange (OPE). 
In this way, when including an extra $NN$ intermediate state in the iteration of the unitarity two-nucleon 
diagrams their contribution to $\Delta(A)$ starts further away in the LHC. It then results that the $n$th iteration of 
 two-nucleon intermediate states, which at least requires $n+1$ OPE ladders, 
 gives contribution to $\Delta(A)$ only for $A<-(n+1)^2 M_\pi^2/4$. 
This makes that its relevance 
for physical values of $A$ ($A\geq 0$) in the low-energy region clearly dismisses 
with increasing $n$. As a result, because of the chiral expansion together with this other effect that
 numerically suppresses the proliferation of two-nucleon reducible diagrams in the calculation of $\Delta(A)$,  
one can determine this function reliably in ChPT.\footnote{Notice that the suppression 
of the iteration of two-nucleon reducible diagrams only occurs for $\Delta(A)$, and it does not occur to any 
other ``component'' of a $NN$ partial wave amplitude.}

 In Refs.~\cite{paper1,paper2} the $N/D$ method was 
solved with $\Delta(A)$  calculated at leading order (LO) from OPE, while in Ref.~\cite{gor2013}
the NLO contributions to  $\Delta(A)$ were also included. 
These contributions comprise two-nucleon irreducible two-pion exchange and once-iterated OPE, whose sum gives 
the  leading two-pion exchange (TPE). 
Reference~\cite{gor2013} obtained a clear improvement in the reproduction of the phase shifts and mixing angles  
 given by the Nijmegen partial-wave analysis (PWA) \cite{Stoks:1994wp} as compared with the LO 
study, so that a  global and rather good agreement  is achieved at NLO. 
 We want to give one step forward and consider here the next-to-next-to-leading (NNLO) contributions to 
$\Delta(A)$, which are given by the imaginary part along the LHC of the 
two-nucleon irreducible TPE diagrams with a NLO $\pi N$ vertex in Heavy-Baryon ChPT (HBChPT) \cite{peripheral}. 
We see that the chiral expansion within our approach is well behaved, so that there is a steady improvement 
in the reproduction of the Nijmegen PWA results when passing from LO to NLO and then to NNLO, where a quite good
 reproduction of the Nijmegen PWA is  finally obtained. 
This is accomplished in a progressive and smooth way, without violent variations in the results obtained at 
every order.\footnote{This was not the case in previous studies, e.g. in the model calculation of $NN$ scattering by Ref.~\cite{lutz} 
that uses a modified version of the $N/D$ method by truncating the integrals along the LHC with a sharp cutoff.}
 In addition, we deal with convergent integrals by taking  enough number of subtractions so that the above 
referred regulator dependence that arises when solving the Lippmann-Schwinger equation with a chiral $NN$ potential 
is avoided by construction in our approach. An interesting outcome from our study is that we
 corroborate the long-range correlations between the effective range and scattering length for each of the $NN$ $S$ waves,  
$^1S_0$ and $^3S_1$, when only the corresponding scattering length is taken as experimental input. 
These correlations, first noticed in Ref.~\cite{pavon06}, were also obtained in the NLO $N/D$ study of Ref.~\cite{gor2013}, and 
within our approach they are deduced solely from basic principles of $NN$ partial-wave amplitudes, namely, 
 chiral symmetry, unitarity and analyticity.
They  are typically fulfilled at the level of around a $10\%$ when comparing with 
the experimental values for the effective ranges.  
We should say that we can proceed further and include more subtractions, so that we can implement within our formalism 
the exact values of the effective ranges, something not possible in the tight scheme of Ref.~\cite{pavon06}.
 
Regarding the subtraction constants we elaborate below a chiral power counting for them, by 
taking into account the change in their values due to variations in the subtraction point.
 We show that at NLO and NNLO in the calculation of $\Delta(A)$ one properly takes twice-subtracted dispersion relations (DRs). 
Nevertheless, on top of this criterion we impose that one should obtain the proper threshold behavior 
for higher partial waves, as well as having meaningful solutions of the integral equations (IEs) 
that result from the corresponding DRs.\footnote{By a meaningful solution we mean here a mathematical solution to the IE 
that does not depend on the the number of points employed and 
in the arbitrary large extension of the LHC on which they lie  
 when performing the numerical discretization to solve the IE.} 
These two requirements often imply the necessity of taking  more than two subtractions 
in the corresponding DRs relations. 
Regarding the number of subtractions used to guarantee the threshold behavior 
for higher partial waves we use here the formalism developed in Ref.~\cite{gor2013}, so that partial waves with 
orbital angular momentum $\ell\geq 1$ and mixing partial waves with total angular momentum $J\geq 1$ vanish  
at threshold as $ A^\ell$ and $A^J$, respectively. 
This requires to take at least $\ell$ or $J$ subtractions, in order, 
with $\ell-1$ or $J-1$ free parameters, respectively. 
But at the end, as emphasized in Ref.~\cite{gor2013}, 
none or only one  of the resulting subtraction constants for a given partial wave with $\ell>1$ (or $J>1$ for a mixing wave) is necessary to reproduce data. 
This interesting point, which allows to treat easily higher partial waves,  is called in Ref.~\cite{gor2013} the 
principle of maximal smoothness.

  In our study we have also paid special attention to the issue concerning the impact on the results of the rather large size 
of the NLO $\pi N$ counterterms, typically denoted by $c_i$ \cite{ulf},  which first appear 
 in the calculation of $\Delta(A)$ at NNLO. 
It is discussed in Ref.~\cite{epe04} that the $\pi N$ monomials, proportional to the $c_i$ counterterms, 
 produce a too large contribution to the  $NN$ potential at medium and short distances   when it is 
calculated at NNLO in dimensional regularization, which worsens the properties of the chiral expansion. 
Because of this Ref.~\cite{epe04} argued to better use  a cutoff regularization to calculate the NNLO potential, or equivalently, 
to cut  the energy spectral representation of the NNLO $NN$ potential at around the chiral symmetry breaking scale. 
This last point would be equivalent to truncate the full extent of the LHC in our dispersive integrals. 
 However, it is interesting to remark that we do not need to do that in order to obtain a good reproduction of the Nijmegen 
PWA when employing $\Delta(A)$ determined up to NNLO. 
In fact,  we observe that the definitive improvement of our results compared 
with the Born approximation does not arise by 
modifying  the two-nucleon 
irreducible diagrams at NNLO,  but by performing the iteration of two-nucleon unitarity diagrams as required
 by analyticity and  unitarity  in a well-defined way.

After this introduction we review the $N/D$ method for coupled and uncoupled partial waves in 
Sec.~\ref{unformalism}.
 The function $\Delta(A)$, calculated in ChPT up to NNLO, is discussed in Sec.~\ref{delta}, 
where we also elaborate the chiral power counting for the subtraction constants.
 Sections \ref{1s0} to 
\ref{gi5w} are devoted to discuss the application of the $N/D$ method to 
 the different $NN$ partial waves up to $J=5$.
 There it is shown that a quite good reproduction of the Nijmegen PWA phase shifts and mixing angles results.
 In these sections we also compare 
with the Born approximation for higher partial waves and discuss on the relative importance of the different contributions 
to $\Delta(A)$.
 Our concluding remarks are given in Sec.~\ref{conc}. 
Finally, we discuss in Appendix \ref{appen:vs} a method to calculate higher order shape parameters 
 of the $NN$ $S$ waves.

\section{The $N/D$ method}
\label{unformalism}

A detailed presentation of the formalism for the $N/D$ method \cite{chew}
 can be found in Ref.~\cite{gor2013}.
 Here we only reproduce the main facets of the approach. 

\subsection{Uncoupled partial waves}
\label{upw}
An uncoupled $NN$ partial wave is written as the quotient of two functions, where the
 numerator is the function $N(A)$ and the denominator is $D(A)$. 
Then, one writes
\begin{align}
T(A)&=\frac{N(A)}{D(A)}~,
\label{eq.ta}
\end{align}
with $T(A)$ the corresponding $NN$ partial wave in the c.m. frame. In the following we use the spectroscopic notation and denote by 
$^{2S+1}L_J$ the different $NN$ partial waves with $S$ the total spin, $L$ the orbital angular momentum and $J$ the 
total angular momentum. The point for the splitting of $T(A)$ in two functions is because $N(A)$ has only LHC while $D(A)$ 
has only right-hand cut (RHC), also called unitarity cut.
 The following expressions for the discontinuities of the functions $N(A)$ and $D(A)$ 
along their respective cuts then arise,
\begin{align}
\mathrm{Im} D(A)=-\rho(A) N(A)~,~A>0~,\nn \\
\mathrm{Im} N(A)=\Delta(A) D(A)~,~A<L~.
\label{disconts}
\end{align}
Here $L=-M_\pi^2/4$  and it represents the onset of the LHC for $A<L$ due to OPE, 
and $\rho(A)$ is the phase space factor
\begin{align}
\rho(A)=\frac{m \sqrt{A}}{4\pi}~~,
\label{rhodef}
\end{align}
where $m$ is the nucleon mass. 
  The first of the relations in Eq.~\eqref{disconts} is a consequence of elastic unitarity for a $NN$ partial wave, which reads 
\begin{align}
\mathrm{Im}T(A)=\rho |T(A)|^2~,~A>0~.
\end{align}
 In terms of $1/T(A)$ this can be recast simply as 
\begin{align}
\mathrm{Im}\,\frac{1}{T(A)}=-\rho(A)~,~ A>0~.
\label{invTun}
\end{align} 
With this normalization the relation between the $T$ and $S$ matrices is $S(A)=1+2i\rho(A) T(A)$.  
The  discontinuity of a $NN$ partial wave 
$T(A)$ along the LHC is given by $2i \Delta(A)$, which directly implies the second expression in Eq.~\eqref{disconts}.

 Standard DRs for the functions $D(A)$ and $N(A)$ are derived in Ref.~\cite{gor2013} under the assumption that 
the function $D(A)$  does not diverge faster than a polynomial of degree $n_0$  for $A\to \infty$. Then for $n>n_0$ one 
can write \cite{gor2013}
\begin{align}
D(A)&=\sum_{i=1}^n \delta_i (A-C)^{i-1}-\frac{(A-C)^n}{\pi}\int_0^\infty dq^2\frac{\rho(q^2)N(q^2)}{(q^2-A)(q^2-C)^n}~,\nn\\
N(A)&=\sum_{i=1}^n \nu_i (A-C)^{i-1}+\frac{(A-C)^n}{\pi}\int_{-\infty}^L dk^2\frac{\Delta(k^2)D(k^2)}{(k^2-A)(k^2-C)^n}~,
\label{standardr}
\end{align}
where $C$ is the subtraction point. Notice that the same number of subtractions is taken both in $D(A)$ and $N(A)$. 
The argument given in Ref.~\cite{gor2013} makes use of the fact that $N(A)=T(A)D(A)$ and $T(A)$,
 because of unitarity, vanish at least as $A^{-1/2}$ for $A\to +\infty$. 
 As a result if $D(A)$ diverges at most as $A^{n_0}$ then $N(A)$ does not diverge faster than $A^{n_0-1/2}$. 
Here we take into account the
Sugawara and Kanazawa theorem \cite{barton,suga}, as a consequence of which any function 
like $D(A)$ or $N(A)$ with only one cut  of infinite extent along the
 real axis has the same limit for $A\to \infty$ in any direction of the $A$-complex plane. 
 In addition, it is clear from Eq.~\eqref{standardr} and the standard theory of DRs \cite{spearman}, 
that we can take different values for the corresponding subtraction points for each function 
separately. Indeed, for many partial waves  we will take the subtractions for the function $D(A)$   in two 
different subtraction points, one at $C=0$ and the other at $C=-M_\pi^2$. This is motivated by the fact that 
we impose the normalization 
\begin{align}
D(0)=1~,
\label{nor.da}
\end{align}
which can always be done by dividing simultaneously $D(A)$ and $N(A)$ by a constant without altering their ratio 
corresponding to $T(A)$, Eq.~\eqref{eq.ta}. 
 In this way,  one subtraction for $D(A)$ is always taken at $C=0$ in order 
to guarantee straightforwardly the normalization Eq.~\eqref{nor.da}.

To solve $D(A)$ in terms of the input $\Delta(A)$ and the subtraction constants  we substitute   in Eq.~\eqref{standardr} the expression 
for $N(A)$ into the DR of $D(A)$, so that we end with the following IE for $D(A)$ with $A<L$,
\begin{align}
D(A)&=\sum_{i=1}^n \delta_i (A-C)^{n-i}-\sum_{i=1}^n \nu_i\frac{(A-C)^n}{\pi}\int_0^\infty dq^2\frac{\rho(q^2)}{(q^2-A)(q^2-C)^{n-i+1}}\nn\\
&+\frac{(A-C)^n}{\pi^2}\int_{-\infty}^L dk^2\frac{\Delta(k^2)D(k^2)}{(k^2-C)^n}\int_0^\infty dq^2\frac{\rho(q^2)}{(q^2-A)(q^2-k^2)}~.
\label{inteq1}
\end{align}
The key point of the method is to solve this IE numerically which provides the knowledge of $D(A)$ for $A<L$. 
Once $D(A)$ is known  along the LHC  we can calculate all the 
functions $D(A)$, $N(A)$ and $T(A)$ in the whole $A$-complex plane. To obtain $D(A)$ one can use 
Eq.~\eqref{inteq1} and for $N(A)$ one has the second of the DRs in Eq.~\eqref{standardr}. 
Once $N(A)$ and $D(A)$ are known one can calculate $T(A)$ by applying Eq.~\eqref{eq.ta}.

 Notice also that 
the integrations along the RHC in Eq.~\eqref{inteq1} can be done algebraically 
in terms of the function
\begin{align}
g(A,k^2)&\equiv \frac{1}{\pi}\int_0^\infty dq^2\frac{\rho(q^2)}{(q^2-A)(q^2-k^2)}=\frac{i m / 4\pi}{\sqrt{A+i0^+}+\sqrt{k^2+i0^+}}~.
\label{gdef}
\end{align}
 The term $+i0$  is necessary  for negative $A$ or $k^2$, with the prescription 
 $\sqrt{-1\pm i0}=\pm i\pi$. 
We can calculate the other RHC integrals 
 of Eq.~\eqref{inteq1} with higher powers of the factor $(q^2-C)$ in the denominator by simple differentiation 
with respect to $C$ of the function $g(A,k^2)$, 
\begin{align}
\frac{\partial^{p-1}\,g(A,C)}{\partial C^{p-1}}=\frac{(p-1)!}{\pi}\int_0^\infty dq^2\frac{\rho(q^2)}{(q^2-A)(q^2-C)^p}~.
\label{derg}
\end{align}

\subsection{Coupled partial waves}
\label{cpw}

For the case of the triplet partial waves with total angular momentum $J$ we have the mixing between the partial waves 
with $\ell=J-1$ and $\ell'=J+1$, except for the $^3P_0$. In this case we denote the different coupled partial waves 
by $t_{ij}(A)$ with $i,~j=1,~2$, where 1 labels the lower angular momentum $\ell\equiv \ell_1$ and 2 the higher one 
$\ell' \equiv \ell_2$. All of 
them are gathered together in the $2\times 2$ matrix $T(A)$, in terms of which the $S$-matrix reads
\begin{align}
S(A)&=I+2i\rho(A)T(A)\nn\\
& = \left(
 \begin{array}{cc}
 \cos 2\epsilon_J\ e^{2i\delta_1}            & i\sin 2\epsilon_J\ e^{i(\delta_1+\delta_2)} \\ 
i\sin 2\epsilon_J\ e^{i(\delta_1+\delta_2)} &   \cos 2\epsilon_J\ e^{2i\delta_2}
 \end{array} \right)~,
\label{relst}
\end{align}
 where $I$ is the $2\times 2 $ unit matrix, $\epsilon_J$ is the mixing angle, and $\delta_{1}$ and $\delta_2$ are the phase
 shifts for the channels with orbital angular momentum $\ell$ and $\ell'$, in this order. Equation 
\eqref{relst} corresponds to the Stapp parameterization \cite{stapp}.

Now, the $N/D$ method explained for the uncoupled waves in Sec.~\eqref{unformalism} is extended to the coupled channel case 
\cite{paper2,gor2013} by writing down three $N/D$ equations, one for every $t_{ij}(A)$ [notice that because of time reversal 
$t_{12}(A)=t_{21}(A)$]. 
 The main difference with respect to the uncoupled case is that now the discontinuity along the RHC 
of the inverse of $t_{ij}(A)$ does not 
simply correspond to $-\rho(A)$, but it also contains information on the other coupled partial waves. 
 In the following let us employ the notation 
\begin{align}
\mathrm{Im} \frac{1}{t_{ij}(A)} \equiv -\nu_{ij}(A)~,A>0~.
\label{nuij.def}
\end{align}
 From Eq.~\eqref{relst} it is straightforward to obtain  the following expressions for the $\nu_{ij}(A)$ \cite{paper2,gor2013},  
\begin{align}
\nu_{11}(A) & =   \rho(A) \left[ 1- \frac{\frac{1}{2}\sin^2 2\epsilon_J}{1-\cos 2\epsilon_J \cos 2\delta_1} \right]^{-1} ~,\nn\\
\nu_{22}(A) & =   \rho(A) \left[ 1- \frac{\frac{1}{2}\sin^2 2\epsilon_J}{1-\cos 2\epsilon_J \cos 2\delta_2} \right]^{-1}~,\nn \\
\nu_{12}(A) & = 2 \rho(A) \frac{\sin(\delta_1 + \delta_2)}{\sin 2\epsilon_J} \label{nuij}~.
\end{align}
In terms of them we have the analogous  DRs for $D(A)$ and $N(A)$ of Eq.~\eqref{standardr}, but now distinguishing 
between the different $D_{ij}(A)$ and $N_{ij}(A)$ such that $t_{ij}(A)=N_{ij}(A)/D_{ij}(A)$, and employing $ \nu_{ij}(A)$ 
instead of simply $\rho(A)$. The following expressions are obtained \cite{gor2013}:
\begin{align}
D_{ij}(A)&=\sum_{p=1}^n \delta^{(ij)}_p (A-C)^{p-1}-\sum_{p=1}^n \nu^{(ij)}_p\frac{(A-C)^n}{\pi}\int_0^\infty dq^2\frac{\nu_{ij}(q^2)}{(q^2-A)(q^2-C)^{n-p+1}}\nn\\
&+\frac{(A-C)^n}{\pi^2}\int_{-\infty}^L dk^2\frac{\Delta_{ij}(k^2)D_{ij}(k^2)}{(k^2-C)^n}\int_0^\infty dq^2\frac{\nu_{ij}(q^2)}{(q^2-A)(q^2-k^2)}~,\\
N_{ij}(A)&=\sum_{p=1}^n \nu^{(ij)}_p (A-C)^{p-1}+\frac{(A-C)^n}{\pi}\int_{-\infty}^L dk^2\frac{\Delta_{ij}(k^2)D_{ij}(k^2)}{(k^2-A)(k^2-C)^n}~.
\label{standardrcc}
\end{align}
 Here, we also impose the normalization condition at $A=0$,
\begin{align}
D_{ij}(0)=1~.
\end{align}
Of course, the same remark concerning the subtraction point as done in Sec.~\ref{upw} is also in order here. Namely, 
we can use different subtraction points for the functions $D_{ij}(A)$ and $N_{ij}(A)$, as well as to use even different subtraction 
points in the same function, as we will do below for $D_{ij}(A)$. 

\subsection{Higher partial waves}
\label{hpw}

An uncoupled $NN$ partial wave with $\ell\geq 1$ should vanish at threshold as $A^\ell$.
 Similarly for a coupled partial wave we have the analogous results but in terms of $\ell_{ij}\equiv (\ell_i+\ell_j)/2$, with   $i,~j=1,~2$. 
 As discussed in Ref.~\cite{gor2013} this threshold behavior is enforced by taken at least $\ell$ 
or $\ell_{ij}$ subtractions
 at $C=0$ in the DR for $N(A)$ in  Eq.~\eqref{standardr} or Eq.~\eqref{standardrcc}, respectively, 
and setting $\nu_p=0$ ($\nu_p^{(ij)}=0$) for $p=1,\ldots,\ell$ ($\ell_{ij}$). In this way we end with the DRs:
\begin{align}
&\underline{\mathrm{Uncoupled~ case}:}\nn \\
\label{highd}
D(A)&=1+\sum_{p=2}^{\ell}\delta_p A^{p-1}+\frac{A^\ell}{\pi}\int_{-\infty}^L dk^2\frac{\Delta(k^2)D(k^2)}{(k^2)^\ell}g(A,k^2)~,\\
\label{highn}
N(A)&=\frac{A^\ell}{\pi}\int_{-\infty}^\ell dk^2\frac{\Delta(k^2)D(k^2)}{(k^2)^\ell (k^2-A)}~,\\
\label{tayloruc}
\delta_p&=\frac{1}{(p-1)!}D^{(p-1)}(0)~,~p=2,3,\ldots\\
&\underline{\mathrm{Coupled~ case}:}\nn \\
\label{highdcc}
D_{ij}(A)&=1+\sum_{p=2}^{\ell_{ij}}\delta^{(ij)}_p A(A-C)^{p-2}  +\frac{A(A-C)^{\ell_{ij}-1}}{\pi} 
\int_{-\infty}^L\!\! dk^2 \frac{\Delta_{ij}(k^2)D_{ij}(k^2)}{(k^2)^{\ell_{ij}}} g_{ij}(A,k^2,C;\ell_{{ij}-1})~,\\
\label{highncc}
N_{ij}(A)&=\frac{A^{\ell_{ij}}}{\pi}\int_{-\infty}^L \!\!dk^2\frac{\Delta_{ij}(k^2)D_{ij}(k^2)}{(k^{2})^{\ell_{ij}}(k^2-A)}~,\\
\delta_p^{(ij)}&=\frac{(-1)^p}{C^{p-1}}\left[
\sum_{n=0}^{p-2}\frac{(-1)^n}{n!}C^n D^{(n)}_{ij}(C)-1
\right]~,~p=2,3,\ldots
\label{taylor}
\end{align}
where we have denoted the derivative of $D(A)$ of order $n$ by $D^{(n)}(A)$. In addition, we have introduced the function $g_{ij}(A,k^2,C;m)$ defined 
as
\begin{align}
g_{ij}(A,k^2,C;m)&=\frac{1}{\pi}
\int_0^\infty\!\! dq^2
\frac{ \nu_{ij}(q^2) (q^2)^m }{ (q^2-A) (q^2-k^2) (q^2-C)^{m}}~,
\label{gij.a.k2.c} 
\end{align}
which can be expressed algebraically as a combination of $g(A,B)$'s, Eq.~\eqref{gdef}, with 
different arguments.

Although in this way there is  a proliferation of subtraction constants 
(which are not constrained) in the function $D(A)$ as $\ell$ ($\ell_{ij}$)
 grows, most of them play a negligible role. 
This is so because $NN$ partial waves with $\ell$ or $\ell_{ij}$ greater than 2 are 
 quite perturbative \cite{peripheral,gor2013}.  
In practical terms we have found in our NNLO study, as well as in the previous 
 one at NLO \cite{gor2013}, that for higher partial waves
 only $\delta_{\ell}$ (or $\delta^{(ij)}_{\ell_{ij}}$), if  any,  
 is needed to fit data, with the rest of them fixed to zero. Furthermore,
  no significant improvement in the reproduction of data or in the fitted values is observed 
by releasing  $\delta_i$ or $\delta_i^{(ij)}$ with $i<\ell$ or $\ell_{ij}$, respectively, so that the fit is stable. This is called in Ref.~\cite{gor2013} 
the {\it principle of  maximal smoothness} because it implies for the uncoupled case 
that  the derivatives of $D(A)$ at $A=0$ with order $< \ell-1$ are zero, as it follows from Eqs.~\eqref{highd} and 
\eqref{tayloruc}. Similarly, for the coupled case
 it implies that $D_{ij}(C)=1$ and $D_{ij}^{(n)}(C)=0$ for $1\leq n \leq \ell_{ij}-3$, cf. Eqs.~\eqref{highdcc} and \eqref{taylor}. 
 In some cases, it happens that  $\delta_\ell$ or $\delta_{\ell_{ij}}^{(ij)}$ is also zero and then we say that for this partial wave the subtraction constants have the {\it pure perturbative values}. 

We further illustrate in this work the perturbative character of $NN$ partial waves with $\ell\,(\ell_{ij})\geq 3$ by 
comparing the full outcome from the $N/D$ method with the perturbative result corresponding to the leading Born 
approximation, cf. Sec.~\ref{born}. In this case there is no dependence on any of the 
subtraction constants $\delta_p$ or $\delta_p^{(ij)}$ and, indeed, we show below  that the results are typically 
rather similar to the full ones, although the latter reproduce closer the Nijmegen PWA, as one should expect.

\section{The input function $\Delta(A)$}
\label{delta}
The discontinuity along the LHC of a NN partial wave, $2i \Delta(A)$, is taken from the  
calculation of Ref.~\cite{peripheral} in Baryon ChPT (BChPT)  up to ${\cal O}(p^3)$ or NNLO, which includes OPE plus leading and subleading TPE.
 At this order $\Delta(A)$ 
for a given partial wave diverges at most as $\lambda (-A)^{3/2}$ for $A\to-\infty$, with $\lambda$ a constant. 
As discussed in Ref.~\cite{gor2013}, when $\lambda<0$ one can  have solutions for the integral 
equation providing $D(A)$ for $A<L$ in the once-subtracted case even  with a divergent $\Delta(A)$ for  
$A\to-\infty$. 
However, as we will see below $\lambda$ is not always negative and more subtractions are then required.

\subsection{NLO $\pi N$ counterterms}
At NNLO the function $\Delta(A)$ is sensitive to the NLO $\pi N$ ChPT low-energy constants (LECs) 
 $c_1$, $c_3$ and $c_4$. 
 We take their values from different works in the literature, that are summarized in Table~\ref{tab:cis}. 
Within the same reference we distinguish, when appropriate, between those values obtained by 
fitting phase shifts from the  Karlsruhe-Helsinki group (KH) \cite{kh} or the George Washington University group (GW) \cite{gw}. 

\begin{table}
\begin{center}
\begin{tabular}{|r|l|l|l|}
\hline
  Analysis      & $c_1$ [GeV$^{-1}$]   &   $c_3$ [GeV$^{-1}$]   &   $c_4$ [GeV$^{-1}$]  \\  
\hline   
GW-HBChPT \cite{epe12} & $-1.13$  & $-5.51$ & $3.71$\\
KH-HBChPT \cite{epe12} & $-0.75$ & $-4.77$ & $3.34$ \\
\hline
KH \cite{buttiker} & $-0.81\pm 0.12$ & $8\pm 57$ & $3.40\pm 0.04$ \\
\hline
GW-EOMS \cite{aco2013} & $-1.50\pm 0.007$ & $-6.63\pm 0.31$ & $3.68\pm 0.14$ \\
KH-EOMS \cite{aco2013} & $-1.26\pm 0.14$ & $-6.74\pm 0.38$ & $3.74\pm 0.16$ \\
\hline
GW-IR\cite{ir2011}  & $-1.32\pm 14$ & $-6.9\pm 6 $ & $3.66\pm 0.33$ \\
KH-IR \cite{ir2011}  & $-1.08\pm 0.15$ & $-7.0\pm 0.7$ & $3.72\pm 0.32$ \\  
\hline
NN data \cite{rentmeester2003} & $-0.76\pm 0.7$ & $-4.78\pm 0.10$ & $3.96\pm 0.22$ \\
\hline
GW-UChPT \cite{aco2013}& $-1.11\pm 0.02$ & $-4.78\pm 0.04$ & $3.04\pm 0.02$ \\
KH-UChPT \cite{aco2013} & $-1.04 \pm 0.02$ & $-4.48\pm 0.05$ & $3.00\pm 0.02$ \\
\hline
\end{tabular}
\caption{Different sets of values for the ${\cal O}(p^2)$ $ \pi N$ LECs $c_1$, $c_3$ and $c_4$.
\label{tab:cis}}
\end{center}
\end{table}

Reference~\cite{epe12} performs an ${\cal O}(p^4)$ HBChPT study of $\pi N$ scattering data. 
We take its values instead of the ones from the older HBChPT studies at ${\cal O}(p^3)$ and ${\cal O}(p^4)$ \cite{fettes98}. 
 We include too the values from Lorentz covariant BChPT obtained in Ref.~\cite{aco2013} 
by fitting $\pi N$ phase shifts making use of EOMS at ${\cal O}(p^3)$. 
 Furthermore, we show in the table the $c_i$'s obtained in the covariant ${\cal O}(p^3)$ BChPT study of Ref.~\cite{ir2011} 
within Infrared Regularization (IR).
 However, due to the better convergence of the $\pi N$ 
scattering amplitude in EOMS than in IR \cite{sigma2012,aco2013} we give results only for
 the values obtained within EOMS \cite{aco2013}. 
The resulting uncertainty band is already wide enough to 
take into account further uncertainties by considering explicitly the $c_i$'s  from the IR study of 
Ref.~\cite{ir2011}, which indeed are rather close to those obtained in EOMS \cite{aco2013}. 
 We also notice that the values from Ref.~\cite{rentmeester2003}, obtained in a $NN$ scattering study, 
are very similar to those of KH \cite{epe12}, so that in the following we  consider only the latter ones. 
Again the uncertainty estimated  takes into account the variation in the results by 
employing the $c_i$'s from Ref.~\cite{rentmeester2003}.
 The work \cite{buttiker} fixes $c_4$ accurately but its analysis is insensitive to $c_3$, 
precisely the ${\cal O}(p^2)$ $\pi N$ LEC on which our results mostly depend. 
This is  why we do not show results for this set of $c_i$'s either. 
Finally, we also give the resulting values from the fits to $\pi N$ data within Unitarized  EOMS BChPT
 obtained in Ref.~\cite{aco2013}.  
These are the fits  that provide more stable values under the change of the data between KH and GW. 
These values are quite similar to those from the set KH \cite{epe12}. 
In summary, when discussing our results we will take into account the 
 values for the LECs $c_i$  obtained in Refs.~\cite{epe12} and \cite{aco2013}, namely, the rows 2, 3, 5, 6, 10 and 11 
in Table~\ref{tab:cis}.  

\subsection{Number of subtractions in the chiral expansion of $\Delta(A)$}
\label{nschpt}

An interesting point to discuss is   the appropriate number of subtractions for a given chiral 
order in the calculation of $\Delta(A)$. In other terms, we want to establish a chiral power counting 
for the subtraction constants involved in the calculation of the functions $D(A)$ and $N(A)$. 

In the previous works in which we applied the $N/D$ method to study $NN$ interactions \cite{paper1,paper2,gor2013} our main 
criterion for fixing the number of subtractions was to end with a well-defined IE for $D(A)$ with $A<L$. 
 We could also add more subtractions and fit low-energy data with more precision by having more free parameters
 at our disposal, a point actually  used in these works too. 
 However, by having a chiral power counting for the subtraction constants 
one has a connection between the number of subtraction  and the chiral order for the 
calculation of $\Delta(A)$.

A chiral power counting for the subtraction constants can be established by studying their variation when changing the 
subtraction point in the low-energy region.
 Let us consider first the chiral order for the subtraction constants appearing in $N(A)$,  
denoted by $\nu_i$ in Eq.~\eqref{standardr}. 
For definiteness  let us employ a twice-subtracted DR, which reads 
\begin{align}
N(A)&=\nu_1+\nu_2 A+\frac{A^2}{\pi}\int_{-\infty}^L dk^2\frac{\Delta(k^2)D(k^2)}{(k^2)^2(k^2-A)}~.
\label{sc.co1}
\end{align}
Now, let us move the subtraction point from zero to $C={\cal O}(M_\pi^2)$. It is then straightforward to show that the 
previous DR can be rewritten as\footnote{To show this, one can rewrite the factor $A^2/k^2$ in the integral 
of Eq.~\eqref{sc.co1} as  $([A-C]+C)^2/(k^2-C)^2\times (k^2-C)^2/(k^2)^2$ and then isolate the term $(A-C)^2/(k^2-C)^2$. 
The rest of terms can be reabsorbed in the polynomial on the right-hand side (r.h.s.) of Eq.~\eqref{sc.co1}.}
\begin{align}
N(A)&=\nu'_1+\nu'_2 A+\frac{(A-C)^2}{\pi}\int_{-\infty}^L dk^2\frac{\Delta(k^2)D(k^2)}{(k^2-C)^2(k^2-A)}~,\nn\\
\nu'_1&=\nu_1-\frac{C^2}{\pi}\int_{-\infty}^L dk^2 \frac{\Delta(k^2)D(k^2)}{(k^2-C)^2 k^2}~,\nn\\
\nu'_2&=\nu_2+\frac{C}{\pi}\int_{-\infty}^L dk^2 \frac{\Delta(k^2)D(k^2)}{(k^2-C)^2 k^2}\frac{2k^2-C}{k^2}~.
\label{sc.co2}
\end{align}
For  $C={\cal O}(p^2)$, $k^2={\cal O}(p^2)$ because the result of the convergent integral  at low-energies 
is dominated by the low-energy region of the integrand and $D(k^2)=1+\ldots={\cal O}(p^0)$. Furthermore, since at 
leading order $\Delta(k^2)={\cal O}(p^0)$,  it  follows then from Eq.~\eqref{sc.co2} that 
$\nu_1={\cal O}(p^0)$ and $\nu_2= {\cal O}(p^{-2})$. This
 procedure can be easily generalized so that $ \nu_n={\cal O}(p^{-2(n-1)})$. By increasing the chiral order 
in the calculation of $\Delta(A)$ up to ${\cal O}(p^m)$ the $\nu_n$ will receive an extra contribution 
starting at ${\cal O}(p^{-2(n-1)+m})$, as it is also clear from Eq.~\eqref{sc.co2}.
 Now, the point is to demand that for a given $m$ the maximum value of $n$, denoted by $n_0$,
 should not be so large that $-2(n_0-1)+m<0$. 
By this condition we are  requiring that the chiral dimension for a given subtraction  constant
 with $n\leq n_0$  be positive or zero, since 
short-distance physics gives rise to contributions that do not vanish in the chiral limit.\footnote{As a 
result they are counted as ${\cal O}(p^0)$} Then  the raising in the chiral dimension of $\nu_n$ until the nominal one, 
$-2(n-1)+m\geq 0$,  must come from powers of $M_\pi$, $|C|^{\frac{1}{2}}\sim  M_\pi$.\footnote{One could ask about the fact that the chiral dimension for 
the other contributions to $\Delta(A)$ of order $m'<m$ could imply a negative $-2(n-1)+m'$ with $n\leq n_0$.
 This already occurs e.g. in 
the paradigmatic example of ChPT, namely, meson-meson scattering. The point is to realize that these extra long-range physics 
contributions cancel explicitly with other contributions stemming from the rearrangement of the dispersive integral, 
which was done already with less subtractions when including only lower orders in $\Delta(A)$.} This power 
counting coincides with the standard Weinberg chiral power counting \cite{weinn}, that is applied to the calculation
 of the $NN$ potential. It is also  worth noticing 
that $\nu_n$ is multiplied by $(A-C)^{n-1}$, so that the chiral order of the product is always $m$ for any $n$, which 
corresponds to the chiral order of the dispersive integral with the ${\cal O}(p^m)$ contribution of $\Delta(A)$. 

One can proceed analogously also for the function $D(A)$. 
We also exemplify it by writing down a 
twice-subtracted DR for $D(A)$, 
\begin{align}
D(A)&=1+\delta_2 A-\frac{A(A-C)}{\pi}\int_0^\infty dq^2\frac{\rho(q^2)N(q^2)}{q^2(q^2-C)(q^2-A)}~.
\label{sc.co3}
\end{align}
We now move the subtraction point from $C$ to $E$. 
Let us recall that the normalization $D(0)=1$ is fixed 
and this is why we do not change the position of the first subtraction taken at $A=0$.
 As a result of this rewriting we obtain the evolution 
\begin{align}
\delta_2&\rightarrow \delta_2+\frac{C-E}{\pi}\int_0^\infty dq^2\frac{\rho(q^2)N(q^2)}{q^2(q^2-E)(q^2-C)}
\label{sc.co4}
\end{align}
For ascribing the chiral order to $\delta_n$, $n\geq 2$, we have, as before, that
 $C\sim E\sim M_\pi^2$, $q^2={\cal O}(p^2)$. 
Additionally, we count $\rho(q^2)={\cal O}(p^0)$ because it involves the product $m \sqrt{q^2}$ with $m\gg M_\pi$, a large number.  Let us also recall at this point that along the RHC, the extent of the integral in Eq.~\eqref{sc.co3},
 the strong effects due to the infrared enhancement of the $NN$ 
intermediate states \cite{weinn}, which is directly related with 
the large nucleon mass,   should be resummed.
 We then conclude from Eq.~\eqref{sc.co4} that
 $\delta_2={\cal O}(p^{-2})$ for the LO contribution of $N(A)={\cal O}(p^0)$. 
This result can be generalized 
easily to more subtractions so that $\delta_n={\cal O}(p^{-2(n-1)})$. 
However, this chiral order increases when considering higher orders contributions to $N(A)$ stemming in turn 
from  higher orders in the calculation of $\Delta(A)$.
 As just discussed above in connection with Eq.~\eqref{sc.co2}, these ${\cal O}(p^m)$ contributions to
 $\Delta(A)$   give rise to contributions of the same order in $N(A)$. 
Thus, once they are taken into account, one has the corresponding rise in the chiral order of $\delta_n$ so that now 
it counts as $\delta_n={\cal O}(p^{-2(n-1)+m})$.
 In this way,  the chiral orders of $\nu_n$ and $\delta_n$ are the same for the same $n$ and $m$. 
Indeed, this is a necessary result because according with the general 
formalism of Sec.~\ref{unformalism}  the same number of subtractions are taken 
both in $D(A)$ and $N(A)$.  
To satisfy this requirement is also another reason for taking  $\rho={\cal O}(p^0)$ in the 
chiral counting.  
 We also stress that the chiral power counting that we have established for the subtraction 
constants $\delta_n$ corresponds to two-nucleon reducible diagrams,\footnote{As it is apparent 
from the factor $q^2-A$ in the denominator of the RHC integral in Eq.~\eqref{sc.co3}.} 
while the standard Weinberg chiral power counting for  nuclear interactions \cite{weinn} only involves 
two-nucleon irreducible diagrams.

Although we have offered here the arguments for the uncoupled case the same results follow 
for the coupled-channel partial waves because the function $\nu_{ij}(A)$, Eq.~\eqref{nuij.def},
 share the same chiral counting 
as $\rho(A)$, since the $T$-matrix is ${\cal O}(p^0)$.\footnote{With $\rho={\cal O}(p^0)$ it is also true that $\text{Im}t_{ij}={\cal O}(p^0)$ because 
of unitarity  for $A\geq 0$, $\text{Im}t_{ij}=\rho\sum_k t_{ik}t_{jk}^*$~.}  
In summary, for $\Delta(A)$ calculated up to ${\cal O}(p^m)$ we have 
the following power counting for 
the subtractions constants, 
\begin{align}
\nu_n~,~\delta_n&\sim {\cal O}(p^{-2(n-1)+m})~.
\label{summarypwc}
\end{align}

Now, by applying the requirement that $-2(n-1)+m\geq 0$  it results 
that in our present study at NNLO  one should properly take two subtractions ($n=2$) since $m=3$. 
However, on top of this criterion we first require that the resulting IE  has  
well-defined solutions and for this to happen it is necessary to introduce more than two subtractions in 
some $NN$ partial waves, as discussed below. 
In addition, we have to satisfy the right threshold behavior for higher partial waves, which for $\ell\geq 3$ ($J\geq 3$ for 
the mixing partial waves)
requires to take $\ell>2$ ($J>2$ subtractions), cf. Sec.~\ref{hpw}.

\section{Uncoupled $^1 S_0$ wave}
\label{1s0} 

In this section we study the $^1 S_0$ partial wave. We first take the once-subtracted DRs:
\begin{align}
\label{onceD}
D(A)&=1-\nu_1 A  g(A,0)+\frac{A}{\pi}\int_{-\infty}^L dk^2\frac{\Delta(k^2)D(k^2)}{k^2}g(A,k^2)~,\\
\label{onceN}
N(A)&=\nu_1+\frac{A}{\pi}\int_{-\infty}^Ldk^2\frac{\Delta(k^2)D(k^2)}{k^2(k^2-A)}~.
\end{align}
We have one free parameter $ \nu_1$ that can be fixed in terms of the $^1S_0$ scattering length $a_s$ 
\begin{align}
\nu_1=-\frac{4\pi a_s}{m}~,
\label{1s0nu1}
\end{align}
with the experimental value $a_s=-23.76\pm 0.01$~fm \cite{thesis}.  

\begin{figure}
\begin{center}
\includegraphics[width=.7\textwidth]{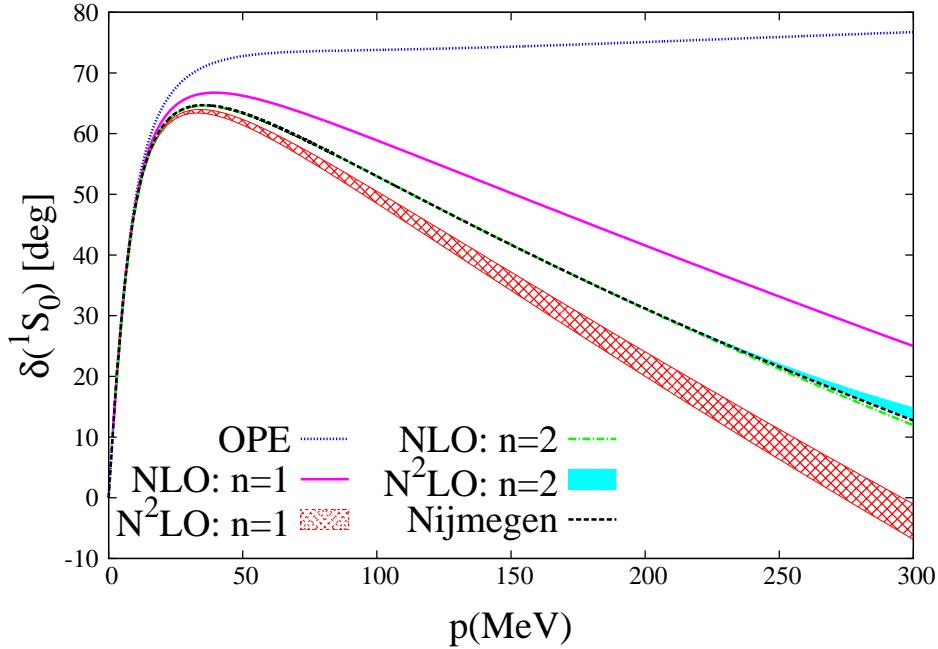} 
\end{center}
\caption[pilf]{\protect {  (Color online.) Phase shifts of the $^1S_0$ $NN$ partial wave where the number 
of subtractions taken is indicated by the value of $n$ given in the legend of each type of line. 
 The once-subtracted DR results are shown by the (red) hatched areas at NNLO, the 
(magenta) solid lines at NLO \cite{gor2013} and the (blue) dotted lines at LO (OPE) \cite{paper1}. 
The twice-subtracted DR results correspond to the (cyan) band at NNLO and 
the (green) dash-dotted  line at NLO  \cite{gor2013}. 
 The Nijmegen PWA phase shifts are shown by the (black) dashed lines.
}
\label{fig:1fp1s0}
}
\end{figure}

The phase shifts obtained by solving the IE of Eq.~\eqref{onceD} are shown in 
 Fig.~\ref{fig:1fp1s0} as a function of the c.m. three-momentum, denoted by $p$ ($p= \sqrt{A}$)  
in the axis of abscissas. 
The (red) hatched area   corresponds to our results from Eqs.~\eqref{onceD}-\eqref{1s0nu1}
 with $\Delta(A)$ calculated up-to-and-including ${\cal O}(p^3)$ contributions and by taking into account the variation in the results from the different values employed for the NLO $\pi N$ ChPT counterterms in Table~\ref{tab:cis}.
Our present results are compared with 
 the neutron-proton ($np$) $^1S_0$ phase shifts of the Nijmegen PWA \cite{Stoks:1994wp}  (black dashed line), 
 the OPE results  of Ref.~\cite{paper1}  (blue dotted line) and the NLO results of Ref.~\cite{gor2013} (magenta solid line).
 As we see,  the Nijmegen PWA phase shifts are better reproduced at lower energies at NNLO than  at smaller orders, 
though one also observes an excess of repulsion at this order.

Next, we work out the effective range expansion (ERE) parameters for the $^1S_0$ . 
By taking into account the relation in our normalization
\begin{align}
\frac{4\pi}{m}\frac{D}{N}=-\frac{1}{a_s}+\frac{1}{2}r_s A+
\sum_{i=1}^{10}v_i A^i - i\sqrt{A}+{\cal O}(A^{11})~,
\label{efr1}
\end{align}
with $r_s$ the $^1S_0$ effective range and the shape parameters $v_i$, $i=2,\ldots,10$. 
 We designate by $I_m$, $m=1,2,\ldots$, the  integral along the LHC,
\begin{align}
 I_{2n}&=\int_{-\infty}^Ldk^2\frac{\Delta(k^2)D(k^2)}{(k^2)^n}~,\nn\\
 I_{2n+1}&=\int_{-\infty}^Ldk^2\frac{\Delta(k^2)D(k^2)}{(k^2)^n\sqrt{-k^2}}~.\nn\\
\end{align}
From Eqs.~\eqref{onceD}, \eqref{onceN} and \eqref{efr1} we derive the following expressions for 
$r_s$ and the shape parameters in the ERE up to $i=4$
\begin{align}
r_s&=-\frac{m (a_s I_3+I_4)}{2 \pi ^2 a_s^2} ~,\nn\\
v_2&=-\frac{m \left(I_4 m (a_s I_3+I_4)+4 \pi ^2 a_s (a_s
   I_5+I_6)\right)}{16 \pi ^4 a_s^3}~,\nn\\
v_3&=-\frac{m \left[16 \pi ^4 a_s^2 (a_s I_7+I_8)+I_4^2 m^2 (a_s I_3+I_4)+4
   \pi ^2 a_s m (a_s I_3 I_6+a_s I_4 I_5+2 I_4 I_6)\right]}{64 \pi
   ^6 a_s^4}~\,\\
v_4&=-\frac{m}{256 \pi ^8 a_s^5}\left[64 \pi ^6 a_s^3 (a_s I_9+I_{10})+16 \pi ^4 a_s^2 m \left(a_s (I_3
   I_8+I_4 I_7+I_5 I_6)+2 I_4 I_8+I_6^2\right)+I_4^3 m^3
   (a_s I_3+I_4)\right.\nn\\
&\left.+4 \pi ^2 a_s I_4 m^2 (2 a_s I_3 I_6+a_s I_4
   I_5+3 I_4 I_6)\right]~.
\label{ere.1s0}
\end{align}
For higher order shape parameters is more efficient to use the numerical method developed in Appendix \ref{appen:vs}, 
to which we refer.

The resulting values for $r_s$ and 
 the shape parameters $v_i$, $i=1,\ldots,6$, are given in 
Table~\ref{table:vs1s0a} and for $v_i$, $i=7,\ldots, 10$ are shown 
in Table~\ref{table:vs1s0b} in the second and third rows for NLO and 
NNLO, respectively. The latter are indicated by NNLO-I.
 These results are compared 
with the results from the calculation based on the NNLO $NN$ potential of Refs.~\cite{epe04} and \cite{thesis},
 and with the Nijmegen PWA values.
 Our results for $v_3$ and $v_4$ are very similar to those obtained in Ref.~\cite{thesis}. The difference between 
\cite{thesis} and \cite{epe04} stems from the fact that in the latter reference a different method to regularize pion exchanges was introduced, the so-called spectral function regularization, instead of the dimensional regularization used in Ref.~\cite{thesis}. We also observe a clear improvement in the reproduction of the ERE parameters from NLO to NNLO. 
At NLO the errors in Tables~\ref{table:vs1s0a} and \ref{table:vs1s0b} reflect the numerical uncertainty in the 
calculation of higher order derivatives. At NNLO in addition they take into account the spread in the results 
from the different sets of $c_i$'s used. 
\begin{table}
\begin{center}
\begin{tabular}{|l|l|l|l|l|l|l|}
\hline
    &$r_s$   & $v_2$  & $v_3$   & $v_4$ & $v_5$ & $v_6$    \\
\hline
NLO & $2.32$ & $-1.08$ & 6.3  & $-36.2$  & 225 & $-1463$  \\
\hline
NNLO-I & $2.92(6)$ & $-0.32(8)$ & $4.9(1)$ & $-27.7(8)$ & $177(4)$  & $-1167(30)$  \\
\hline
NNLO-II & $2.699(4)$ & $-0.657(3)$ & $5.20(2)$  & $-30.39(9)$  & $191.9(6)$ & $-1263(3)$  \\
\hline
Ref.~\cite{thesis} & $2.68$ & $-0.61$ & $5.1$ & $-30.0$ & & \\
\hline
Ref.~\cite{epe04} & $2.62\sim 2.67$ & $-0.52\sim -0.48$ & $4.0\sim 4.2$ & $-20.5\sim -19.9$ & & \\
\hline
Ref.~\cite{Stoks:1994wp} & $2.68$ & $-0.48$ & $4.0$ & $-20.0$ & & \\
\hline
\end{tabular}
\caption{Values for effective range $r_s$ [fm] and the shape parameters 
$v_i$, $i=2,\ldots,6$ in units of fm$^{2i-1}$
 for our present results at NNLO with once-subtracted DRs [Eq.~\eqref{onceD}] (NNLO-I in the third row) 
and with twice-subtracted DRs [Eq.~\eqref{twiceD}] (NNLO-II in the fourth row). 
The second row shows the results at  NLO with once-subtracted DRs [Eq.~\eqref{onceD}].
 We also give the values obtained by using the NNLO $NN$ potential in 
Refs.\cite{thesis} and \cite{epe04} (fifth and
 sixth  rows, respectively). The values corresponding to the Nijmegen PWA \cite{Stoks:1994wp}, 
 as obtained in Refs.~\cite{epe04,thesis}, are given in the last row.  
\label{table:vs1s0a}}
\end{center}
\end{table}

\begin{table}
\begin{center}
\begin{tabular}{|l|l|l|l|l|}
\hline
    &  $v_7\times 10^{-1}$ & $v_8\times 10^{-2}$  & $v_{9}\times 10^{-3}$  & $v_{10}\times 10^{-4}$   \\
\hline
NLO & $985$ & $-681$ & 480  & $-344(3)$  \\
\hline
NNLO-I & $795(18)$ & $-554(12)$ & $393(8)$ & $-284(6)$  \\
\hline
NNLO-II & $857.1(1.9)$ & $-595.7(1.3)$ & $421.7(9)$ & $-304(3)$ \\
\hline
\end{tabular}
\caption{Values for the shape parameter $v_i$, $i=7,\ldots,10$ in units of fm$^{2i-1}$. 
For the meanings of the rows see Table~\ref{table:vs1s0a}.
\label{table:vs1s0b}}
\end{center}
\end{table}

From Eq.~\eqref{ere.1s0}  we can also derive a power series expansion of the ERE parameters as a function of $a_s$, as it was done previously for $r_s$ in Ref.~\cite{gor2013} at NLO. We refer to that reference for further details. The important point is that $D(A)$ satisfies the linear IE of  Eq.~\eqref{onceD} with an inhomogeneous term that is a polynomial of first degree in $a_s$. As a result,  $D(A)=D_0(A)+a_s D_1(A)$, with $D_0(A)$ and $D_1(A)$ independent of $a_s$. This also implies that the different $I_n$ can be expressed as $I_n^{(0)}+a_s I_n^{(1)}$ with $I_n^{(0)}$ and $I_n^{(1)}$ independent of $a_s$. 
In this way, the ERE parameters satisfies the following expansions 
\begin{align}
r_s&=\alpha_0+\frac{\alpha_{-1}}{a_s}+\frac{\alpha_{-2}}{a_s^2}~,\nn\\
v_n&=\sum_{m=-n-1}^0 \frac{v_n^{(m)}}{a_s^m}~,
\label{expvs.1s0}
\end{align}
with the coefficients $\alpha_i$ and $v_n^{(i)}$ independent of $a_s$. 
The relation between $r_s$ and $a_s$ was first realized in Ref.~\cite{pavon06} in the context of $NN$ scattering.\footnote{The correlation between the effective range and the scattering length
 in Eq.~\eqref{expvs.1s0} was derived earlier in atomic physics for Van der Waals potentials
 \cite{flambaum}, and throughly confronted    with data \cite{calle2010}.} The explicit expressions 
of $\alpha_i$ ($i=-2,-1,0)$ in terms of $D_0(A)$ and $D_1(A)$ were given in Ref.~\cite{gor2013}.
 Its values at NNLO are
\begin{align}
\alpha_0&= 2.61\sim 2.73~\text{fm}~,\nn\\
\alpha_{-1}&=-5.93\sim -5.65~\text{fm}^2~,\nn\\
\alpha_{-2}&=5.92\sim 6.12~ \text{fm}^3~.
\label{exp.1s0}
\end{align}
 The expressions for the coefficients $v_n^{(m)}$ in Eq.~\eqref{expvs.1s0} can also be worked straightforwardly 
in terms of $D_0(A)$ and $D_1(A)$ by the interested reader. For conciseness we do not reproduce them here. 
The  results in Eq.~\eqref{exp.1s0}  are perfectly compatible with those obtained in the first entry of 
Ref.~\cite{pavon06}, $\alpha_0=2.59\sim 2.67$~fm, $\alpha_{-1}=-5.85\sim -5.64$~fm$^2$ and $\alpha_{-2}=5.95\sim 
6.09$~fm$^3$. This reference employs the chiral $NN$ potential in a Lippmann-Schwinger equation that is 
 renormalized  with boundary conditions and imposing 
the hypothesis of orthogonality of the wave functions determined with
 different energy.\footnote{Since the potentials involved are singular this orthogonality condition is imposed in 
 the formalism of Ref.~\cite{pavon06}.} 
In our case, however, the expansions in Eq.~\eqref{expvs.1s0} are consequences of basic principles of a $NN$ 
partial wave like unitarity, analyticity and chiral symmetry.  
 The resulting phase shifts in Fig.~\ref{fig:1fp1s0} from Eq.~\eqref{onceD}, and shown by the (red) hatched area,
 are also coincident with those obtained by Ref.~\cite{pavon06}.  
They are also rather similar to those obtained when employing only one contact term in the third entry of Ref.~\cite{phillipssw},
 which studies the independence of its results as a function of the cutoff used to solve the Lippmann-Schwinger equation.
 Nevertheless, in this case  the NNLO chiral potential is calculated by truncating its spectral representation \cite{epe04},  
 while Ref.~\cite{pavon06} uses the dimensional regularized result 
(which requires to take to infinity the cutoff(s) used in Ref.~\cite{phillipssw}.)

Next,  we consider the twice-subtracted DRs: 
\begin{align}
\label{twiceD}
D(A)&=1+\delta_2 A-\nu_1\frac{A(A+M_\pi^2)}{\pi}\int_0^\infty dq^2\frac{\rho(q^2)}{(q^2-A)(q^2+M_\pi^2)q^2}
-\nu_2 A(A+M_\pi^2) g(A,-M_\pi^2)\nn\\
&+\frac{A(A+M_\pi^2)}{\pi^2}\int_{-\infty}^L dk^2\frac{\Delta(k^2)D(k^2)}{(k^2)^2} 
\int_0^\infty dq^2\frac{\rho(q^2)q^2}{(q^2-A)(q^2+M_\pi^2)(q^2-k^2)}
~, \\
\label{twiceN}
N(A)&=\nu_1+\nu_2 A+\frac{A^2}{\pi}\int_{-\infty}^L dk^2\frac{\Delta(k^2)D(k^2)}{(k^2-A)(k^2)^2}~,
\end{align}
where the two subtractions in the function $N(A)$ and one  for $D(A)$ are taken at $C=0$, while
 the other subtraction in $D(A)$ is placed at $C=-M_\pi^2$. 
Taking into account Eq.~\eqref{gdef} it is straightforward to rewrite
\begin{align}
\frac{1}{\pi}\int_0^\infty dq^2\frac{\rho(q^2)q^2}{(q^2-A)(q^2-k^2)(q^2-C)}=\frac{C g(A,C) - k^2 g(A,k^2)}{C-k^2}~.
\end{align} 
 The subtraction constant $\nu_1$ is given by Eq.~\eqref{1s0nu1}, while $\nu_2$ and $\delta_2$ 
are directly fitted to 
 the $np$ Nijmegen PWA phase shifts.\footnote{Since Ref.~\cite{Stoks:1994wp} does not provide 
errors we always perform a least square fit,  without weighting.}  
 The best fit occurs for
\begin{align}
\nu_2&=-23(1)~M_\pi^{-4}\nn\\
\delta_2&=-8.0(3)~M_\pi^{-2}~,
\label{nu2delta2}
\end{align}        
where the intervals of values stem from the uncertainty due to the different values of $c_i$'s taken. 
The reproduction of data is very good, as shown by the (cyan) filled area in Fig.~\ref{fig:1fp1s0} which 
lies on top of the Nijmegen PWA $np$ phase shifts.  In the same figure 
  we show by the (green) dash-dotted line 
the twice-subtracted DR result at NLO,  which reproduces  
the  Nijmegen data equally well as obtained at NNLO, with  
 the fitted values $\nu_2=-11.9$~$M_\pi^{-4}$ and $\delta_2=-4.6~M_\pi^{-2}$. 
   The resulting ERE shape parameters for the fit in Eq.~\eqref{nu2delta2} 
are shown in the fourth rows of Tables~\ref{table:vs1s0a} and \ref{table:vs1s0b}, where 
 we observe a remarkable good agreement with Ref.~\cite{thesis}. 
We predict $r_s=2.70$~fm which is compatible with its experimental value 
$r_s=2.75\pm 0.05$~fm \cite{thesis}. A similar good reproduction of the $^1S_0$ phase shifts is also achieved by Ref.~\cite{phillipssw} in 
terms of two contact terms, although in this case there is a strong sensitivity   on the cutoff employed to 
regularize the Lippmann-Schwinger equation near those values that give rise to poles in the domain of validity 
of the effective field theory. 

 The value of $\nu_2$ in 
Eq.~\eqref{nu2delta2} is rather large, of similar size in absolute value to $\nu_1\simeq 31~M_\pi^{-2}$, Eq.~\eqref{1s0nu1}. A linear correlation between $\nu_2$ and $\delta_2$ can be observed in  a $\chi^2$ contour plot,  along which 
 there is an absolute minimum corresponding to the parameters given 
in Eq.~\eqref{nu2delta2}.  
The subtraction constant $\nu_2$ that results from the once-subtracted DR 
 Eq.~\eqref{onceN}, and that we denote by $\nu_2^{\mathrm{pred}}$, is given by the 
expression
\begin{align}
\nu^{\mathrm{pred}}_2&=\frac{1}{\pi}\int_{-\infty}^L dk^2\frac{\Delta(k^2)D(k^2)}{(k^2)^2}~,
\label{nu2predicted}
\end{align}
with the numerical value $\nu_2^{\mathrm{pred}}\simeq -6.0$, $-6.4$ and $-7.5\pm 0.2~M_\pi^{-4}$ when 
 $\Delta(A)$ is calculated up to ${\cal O}(p^0)$, ${\cal O}(p^2)$ and ${\cal O}(p^3)$, respectively.
 The difference between the predicted and fitted values for $\nu_2$   at NLO  
 is denoted by $\delta\nu_2^{(0)}$.
 The superscript takes into account the chiral order for $\nu_2$, ${\cal O}(p^{-2+m})$ according to the new 
contribution to $\Delta(A)$ of ${\cal O}(p^m)$, Eq.~\eqref{summarypwc}. 
  The value obtained is $\delta \nu_2^{(0)}\simeq -5.5~M_\pi^{-4}$. At NNLO in order 
to calculate  $\delta \nu_2^{(1)}$ one has to subtract  $\delta\nu_2^{(0)}$ to the difference 
between the fitted value in Eq.~\eqref{nu2delta2} and the predicted one from Eq.~\eqref{nu2predicted}. 
 Then, one has $\delta\nu_2^{(1)}\simeq -15+5.5=-9.5~M_\pi^{-4}$.
 This implies that in order to overcome the excess of repulsion at NNLO   one needs
 to incorporate a significant contribution from short-distance physics to give account 
of ``missing physics'', beyond the pure long-range physics\footnote{We mean here the physics driven by the 
 multi-pion exchanges giving rise to the LHC and  to $\Delta(A)$.}
 that stems from the once-subtracted DR case and that is not able to provide an accurate 
reproduction of data as shown in Fig.~\ref{fig:1fp1s0} by the (red) hatched areas. 
  The large value for $\delta\nu_2^{(1)}$ is mainly due to the ${\cal O}(p^2)$ $\pi N$ counterterms 
$c_i$'s, which in turn are dominated by the  $\Delta(1232)$ resonance contribution \cite{bernard93,aco2012}. 
  This can be easily seen by performing a fit to data in which we set  $c_i=0$ for all of them.
 A good reproduction 
of the Nijmegen PWA phase shifts results but now $\delta\nu_2^{(1)}\simeq -1.5~M_\pi^{-4}$, which is 
much smaller than $\delta\nu_2^{(0)}$, with a ratio $\delta\nu_2^{(1)}/\delta\nu_2^{(0)}\sim 30\% \sim 
{\cal O}(p)$. This indicates that once the large  contributions 
 that stem from the  $c_i$ coefficients are discounted a quite natural (baryon) chiral expansion emerges. 
 
Regarding the absolute value of $\delta\nu_2^{(0)}$ one should expect on dimensional grounds that
\begin{align}
|\delta\nu_2^{(0)}|\sim  \frac{4\pi\,|a_s|}{m \Lambda^2}~,
\label{abvalue_v2}
\end{align}
with $\Lambda$ the expansion scale. The factor $4\pi/m$ is due to our normalization, cf. Eq.~\eqref{efr1}. 
There should be also another contribution to $\delta\nu_2^{(0)}$ not proportional to $a_s$, but since 
the scattering length is so large the contribution shown in Eq.~\eqref{abvalue_v2}
 is expected to be the most important. For $\Lambda\simeq 350$~MeV, one would have 
$|\delta\nu_2^{(0)}|\sim 5~M_\pi^{-4}$, which is very similar indeed to the reported value above. This 
value of $\Lambda$ is also consistent with the ratio $\delta\nu_2^{(1)}/\delta\nu_2^{(0)}\sim 1/3$ given above 
as $M_\pi/\Lambda \sim 1/3$.

\begin{figure}
\begin{center}
\begin{tabular}{cc}
\includegraphics[width=.4\textwidth]{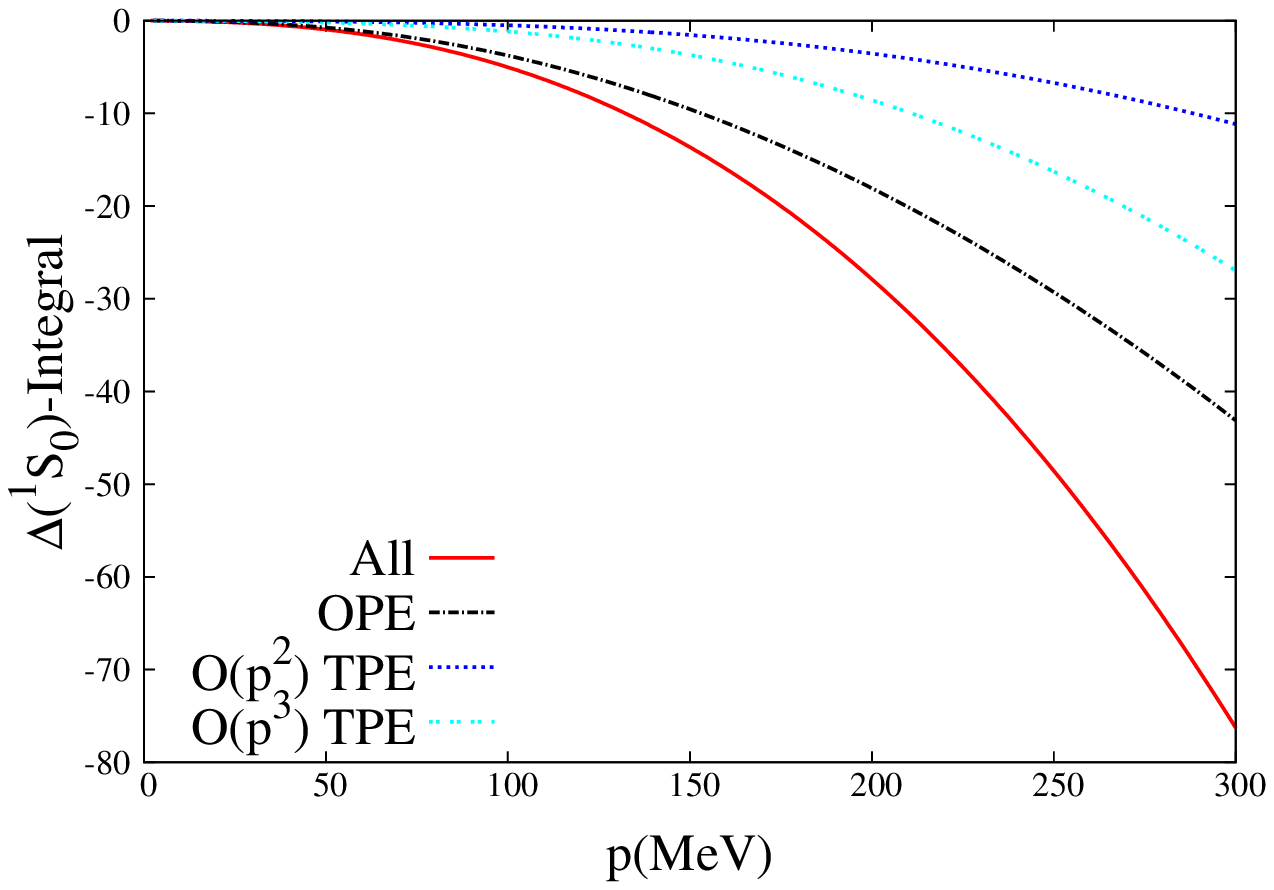} & 
\includegraphics[width=.4\textwidth]{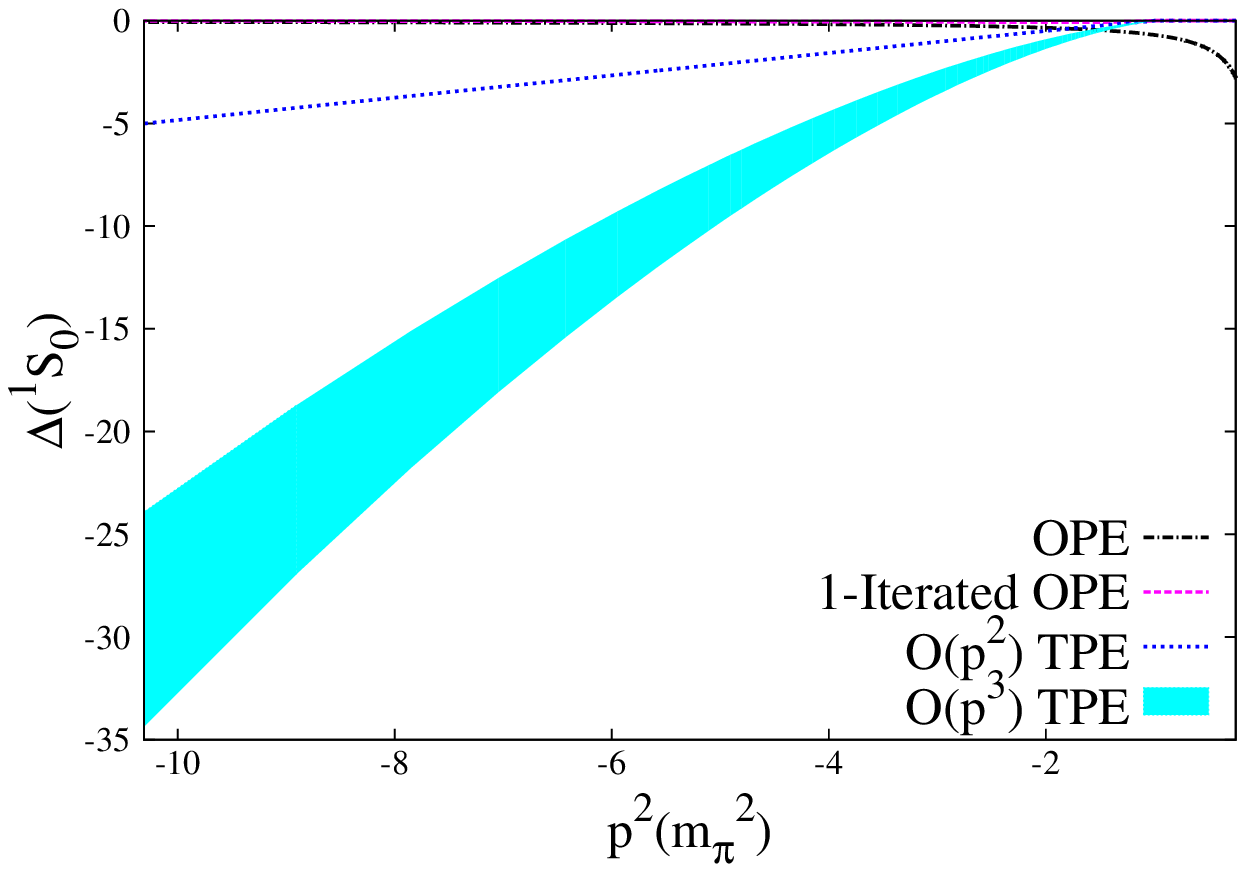}
\end{tabular}
\caption{ { (Color online.) Left panel: different contributions to the integral in Eq.~\eqref{1s0.quanty} for the $^1S_0$.
 Right panel: contributions to $\Delta(A)$. These contributions comprise OPE (black dash-dotted line), 
leading TPE (blue dotted line) and 
the subleading TPE contribution, shown by the (cyan) double-dotted line in the left panel 
and by the (cyan) filled area in the right one. The total result, only shown for the left panel, 
is the (red) solid line.}
\label{fig:1s0quanty}}
\end{center}
\end{figure}

Let us consider now the relevance  of the different contributions to $\Delta(A)$ by evaluating the double
 integral in Eq.~\eqref{twiceD}, namely,
\begin{align}
\frac{A(A+M_\pi^2)}{\pi^2}\int_{-\infty}^L dk^2\frac{\Delta(k^2)D(k^2)}{(k^2)^2}\int_0^\infty dq^2\frac{\rho(q^2)q^2}{(q^2-A)(q^2+M_\pi^2)(q^2-k^2)}~,
\label{1s0.quanty}
\end{align}
 with the full result for $D(A)$ but with $\Delta(A)$ in the integrand of Eq.~\eqref{1s0.quanty} evaluated 
partially with some contributions or all of them. The result of this exercise is given
 in the left panel of Fig.~\ref{fig:1s0quanty} for the $c_i$ coefficients of Ref.~\cite{epe12}, 
 collected in the first row of Table~\ref{tab:cis}. 
 In turn,  we show directly $\Delta(A)$ along the LHC in the right panel of Fig.~\ref{fig:1s0quanty}. 
 The (black) dash-dotted lines correspond to OPE, 
the  (blue) dotted lines take into account the full ${\cal O}(p^2)$ TPE,
 including both two-nucleon reducible and irreducible TPE, and  
 the (cyan) double-dotted lines contain  the ${\cal O}(p^3)$  two-nucleon irreducible TPE. 
In the right panel we show by the (cyan) filled area the variation in the ${\cal O}(p^3)$ irreducible TPE 
contribution by varying between the different sets of $c_i$'s from Refs.~\cite{epe04} and \cite{aco2013}, as discussed above. 
This band indicates a large source of uncertainty  in $\Delta(A)$.
  In the left panel the (red) solid line  results by keeping  all the contributions to $\Delta(A)$, and 
one can quantify from this panel the fact that   the ${\cal O}(p^3)$ irreducible TPE is the largest subleading contribution.
 At $\sqrt{A}=100$~MeV it is around 28\% of the OPE contribution, and it raises with energy
 so that at $\sqrt{A}=200$~MeV it is 44\% and at 300~MeV it becomes 66\%. The increase in energy 
of the relative size of the subleading TPE contribution should be expected because at low energies
 the suppression mechanism due to the earlier onset of the OPE source of $\Delta(A)$ along the LHC at $L$ is more efficient.
 In addition, it is well-known that the $\Delta(1232)$ plays a prominent role in $\pi N$ scattering because
 its proximity to the $\pi N$ threshold and its strong coupling to this channel.
 This manifests in the large size of the LECs $c_3$ and $c_4$ in Table~\ref{tab:cis} due to the $\Delta(1232)$ contribution 
to them, evaluated in Refs.~\cite{bernard93,aco2012}. 
The large impact of the $\Delta(1232)$ is  the well-known reason for the large size of subleading TPE,
 but once its leading effects are taken into account at $ {\cal O}(p^3)$ the chiral expansion
 stabilizes \cite{entem,epe04}, as we have also concluded  in the discussion following Eq.~\eqref{nu2predicted}. 
  In the following, we skip the discussion on the relative importance of the different contributions 
to $\Delta(A)$ for those $NN$ partial waves with a similar situation to the one discussed 
concerning the $^1S_0$. 

\section{Uncoupled $P$ waves}
\label{pw} 
In this section we discuss the application of the method to the  uncoupled $P$ waves. 
 At NNLO one has for these waves that 
\begin{align}
\lambda_P=\lim_{A\to-\infty}\frac{\Delta(A)}{(-A)^{(3/2)}}>0~,
\label{unlambdap}
\end{align}
so that, according to the results of Ref.~\cite{gor2013}, its Proposition 4, a once-subtracted DR for $D(A)$, 
Eq.~\eqref{standardr}, does not converge and more subtractions should be taken. Then, we directly discuss 
the twice- and three-time subtracted DRs. 

The twice-subtracted DRs are given by:
\begin{align}
\label{twiceDNl}
D(A)&=1+\delta_2 A-\nu_2 A^2 g(A,0)
+\frac{A^2}{\pi}\int_{-\infty}^L dk^2\frac{\Delta(k^2)D(k^2)}{(k^2)^2}g(A,k^2)~,\nn \\
N(A)&=\nu_2 A+\frac{A^2}{\pi}\int_{-\infty}^L dk^2\frac{\Delta(k^2)D(k^2)}{(k^2-A)(k^2)^2}~,
\end{align}
with all the subtractions in Eq.~\eqref{standardr} taken at $C=0$. 
 The three-time subtracted DRs are:
\begin{align}
D(A)&=1+\delta_2 A + \delta_3 A(A+M_\pi^2)+(\nu_2-\nu_3 M_\pi^2)  A(A+M_\pi^2)^2  \frac{\partial g(A,-M_\pi^2)}{\partial M_\pi^2}
-\nu_3 \, A(A+M_\pi^2)^2 g(A,-M_\pi^2)\nn\\
&+\frac{A(A+M_\pi^2)^2}{\pi}\int_{-\infty}^Ldk^2 \frac{\Delta(k^2)D(k^2)}{(k^2)^3}g(A,k^2,-M_\pi^2;2)~,\nn\\
N(A)&=\nu_2 A+ \nu_3 A^2+\frac{A^3}{\pi}\int_{-\infty}^L dk^2\frac{\Delta(k^2)D(k^2)}{(k^2-A)(k^2)^3}~.
\label{pw.tdr}
\end{align}
Here all the subtractions in $N(A)$ and one in $D(A)$ are taken at $C=0$, while the other two subtractions 
in $D(A)$ are taken at $C=-M_\pi^2$. This is done in order to avoid handling an infrared diverging 
integral along the RHC multiplying $\nu_2$ that would result if all the subtractions were taken at $C=0$. 
The function $g(A,k^2,C;m)$ appearing in Eq.~\eqref{pw.tdr} is defined as
\begin{align}
g(A,k^2,C;m)=\int_0^\infty\!\! dq^2\frac{\rho(q^2) (q^2)^m}{(q^2-A)(q^2-k^2)(q^2-C)^m}~.
\label{def.el.gm}
\end{align}

In all the cases the subtraction constant $\nu_2$ is fixed in terms of the 
scattering volume, $a_V$, 
\begin{align}
\nu_2=4\pi a_V/m~.
\label{pnu2fix}
\end{align}
For $a_V$ we take the values $0.890$, $-0.543$ and $-0.939~M_\pi^{-3}$ for the partial waves $^3P_0$, $^3P_1$ and $^1P_1$, 
in order, as deduced from Ref.~\cite{Stoks:1994wp}.

\subsection{$^3P_0$ wave}
\label{3p0}

\begin{figure}
\begin{center}
\includegraphics[width=.5\textwidth]{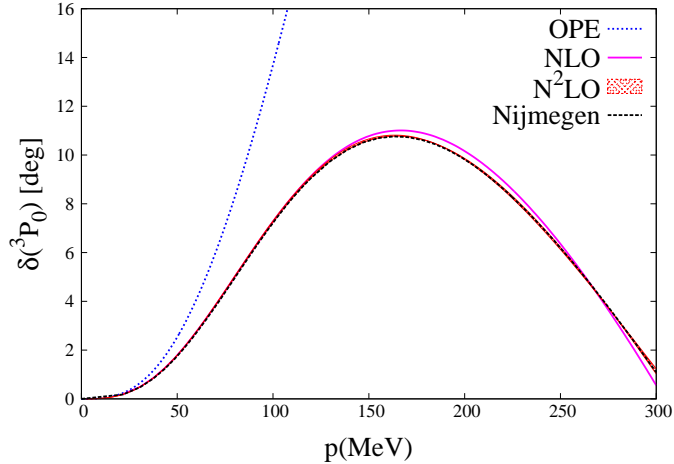} 
\caption{ {\small }
\label{fig:3p0}
 Phase shifts of the $^3P_0$ $NN$ partial wave. 
 The three-time subtracted DR results at NNLO are shown by the (red) hatched area and the twice-subtracted DR results  at NLO \cite{gor2013} are given by the (magenta) solid line. 
The (blue) dotted line corresponds to the OPE results \cite{paper1} and 
the Nijmegen PWA phase shifts are shown by the (black) dashed lines.}
\end{center}
\end{figure}

 For the $^3P_0$ wave the twice-subtracted DRs at NNLO, Eq.~\eqref{twiceDNl}, do not provide 
stable results under the increase in absolute value 
of the lower limit of integration along the LHC. 
However, the three-time subtracted DRs, Eq.~\eqref{pw.tdr}, are convergent and provide meaningful results. 
 Notice that, as stated in Sec.~\ref{nschpt}, 
on top of the number of subtractions required by the chiral counting, two at NNLO,
 we impose the requirement of having well-defined IEs providing stable 
 solutions. 
Regarding the subtractions constants $\nu_3$, $\delta_2$ and $\delta_3$ in Eq.~\eqref{pw.tdr}, 
we can fix $\nu_3=0$ because it plays a negligible role in the fits and, if released,  the fit remains stable. 
The fitted values for $\delta_2$ and $\delta_3$  are
\begin{align}
\delta_2&= 2.82(5)~M_\pi^{-2}\nn\\
\delta_3&=0.18(6) ~M_\pi^{-4}~,
\end{align}
where the intervals of values take into account the dispersion in the results 
that stems from the different sets of $c_i$'s in Table~\ref{tab:cis}.
 The phase shifts calculated, shown by the (red) hatched area 
in Fig.~\ref{fig:3p0}, reproduce exactly the Nijmegen PWA phase shifts \cite{Stoks:1994wp}, 
given by the (black) dashed line. Indeed, the two lines overlap each other. 
The results with different sets of values for the $c_i$ counterterms cannot be distinguished  either between each other. 
 The (magenta) solid line shows the results with twice-subtracted DRs at NLO \cite{gor2013}, 
which are already almost on top of the data, and the OPE results \cite{paper1} are shown by the (blue) dotted line. 
 We have also checked that a tree-time-subtracted DR at LO and NLO provide already a prefect reproduction of data as well. 
Then, the wave $^3P_0$ studied at ${\cal O}(p^3)$ is not a good partial wave to learn above chiral dynamics, because 
 independently of order up to which $\Delta(A)$ is calculated the reproduction of data is excellent when 
three-subtractions are taken.

\subsection{$^3P_1$ wave}
\label{3p1}

\begin{figure}
\begin{center}
\includegraphics[width=.5\textwidth]{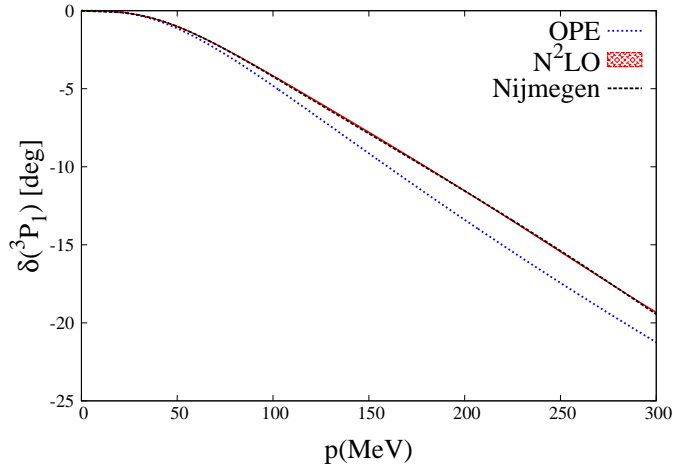} 
\caption{ {\small }
\label{fig:3p1}
 Phase shifts of the $^3P_1$ $NN$ partial wave. 
 The three-time subtracted DR results at NNLO are shown by the (red) hatched area. 
The (blue) dotted line corresponds to the OPE results \cite{paper1} and
 the Nijmegen PWA phase shifts are shown by the (black) dashed lines.}
\end{center}
\end{figure}

For this partial wave the situation is similar to that discussed for  the $^3P_0$. 
The twice-subtracted DRs, Eq.~\eqref{twiceDNl}, do not provide stable results and we have 
to consider then the three-time subtracted DRs, Eq.~\eqref{pw.tdr}. 
 The free parameters are $\delta_2$ and $\delta_3$, with $\nu_3$ fixed to 0 (the fit is stable if this 
subtraction constant is released). The fitted values  are
\begin{align}
\delta_2& = 2.7(1)~M_\pi^{-2},\nn\\
\delta_3& = 0.47(3)~M_\pi^{-4}~.
\end{align}

The resulting phase shifts are  shown in  Fig.~\ref{fig:3p1} by the (red) hatched area and 
 reproduce perfectly the Nijmegen PWA phase shifts (shown by 
the black dashed line),  independently of the set of values
 for the $c_i$'s chosen from Refs.~\cite{epe12,aco2013} in Table~\ref{tab:cis}. 
 At NLO \cite{gor2013} it is also necessary to take three-subtracted DRs in order to obtain stable results and the reproduction 
of data is equally perfect. 
This is why we have not included the NLO results in Fig.~\ref{fig:3p1}. 
 Similarly to the $^3P_0$ case, we cannot discern the impact of chiral dynamics at ${\cal O}(p^3)$ once three-time subtracted DRs 
are considered.

\subsection{$^1P_1$ wave}
\label{1p1}

\begin{figure}
\begin{center}
\includegraphics[width=.5\textwidth]{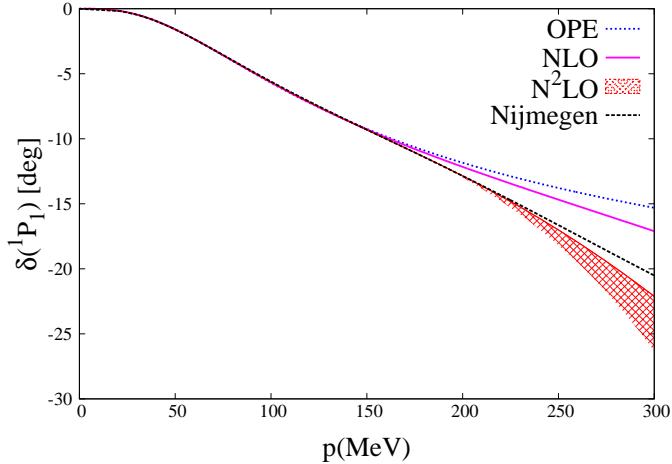} 
\caption{ {\small }
\label{fig:1p1}
 Phase shifts of the $^1P_1$ $NN$ partial wave. 
 The twice subtracted DR results at NNLO are shown by the (red) hatched area, while at NLO \cite{gor2013}
correspond to  the (magenta) solid line. 
The (blue) dotted line corresponds to the OPE results \cite{paper1} and 
the Nijmegen PWA phase shifts are shown by the (black) dashed lines.}
\end{center}
\end{figure}
 
For this partial wave the twice-subtracted DR results from Eq.~\eqref{twiceDNl} are quite stable  at low energies. 
The free parameters are now $\nu_2$ and $\delta_2$. 
The resulting fitted value for $\delta_2$ to  the Nijmegen PWA phase shifts  is  
\begin{align}
\delta_2 =0.4(1)~M_\pi^{-2}~,
\end{align}
with the variation in the value due to the set of $c_i$'s taken [$\nu_2$ is given by Eq.~\eqref{pnu2fix}]. 
We show by the (red) hatched area in Fig.~\ref{fig:1p1} our results by employing the different $c_i$ sets of values. 
For this case the curves obtained with the $c_i$ from \cite{aco2013}, by reproducing the $\pi N$ phase shifts with 
Lorentz covariant EOMS BChPT, are the closest to data and determine the upper
 limit of the hatched area in Fig.~\ref{fig:1p1}. 
 The improvement in the reproduction of data for the $^1P_1$ partial wave by the twice-subtracted DRs at NNLO compared with the 
results obtained at NLO with the same number of subtractions 
  (hatched area versus (magenta) solid line)  is a notorious effect from $\pi N$ physics. One should 
notice that for the $^1P_1$ wave the dispersive integral on the r.h.s. of Eq.~\eqref{twiceDNl} for the function $D(A)$ is clearly 
dominated by the OPE contribution
 This is the reason why for the $^1P_1$ one does not need to take three subtractions but two are enough. 
 Although, as much as for the other partial waves discussed until now, the ${\cal O}(p^3)$ two-nucleon irreducible 
TPE is the dominant contribution between the subleading effects to $\Delta(A)$.

\section{Uncoupled $D$ waves}
\label{dw}

\begin{figure}
\begin{center}
\begin{tabular}{cc}
\includegraphics[width=.4\textwidth]{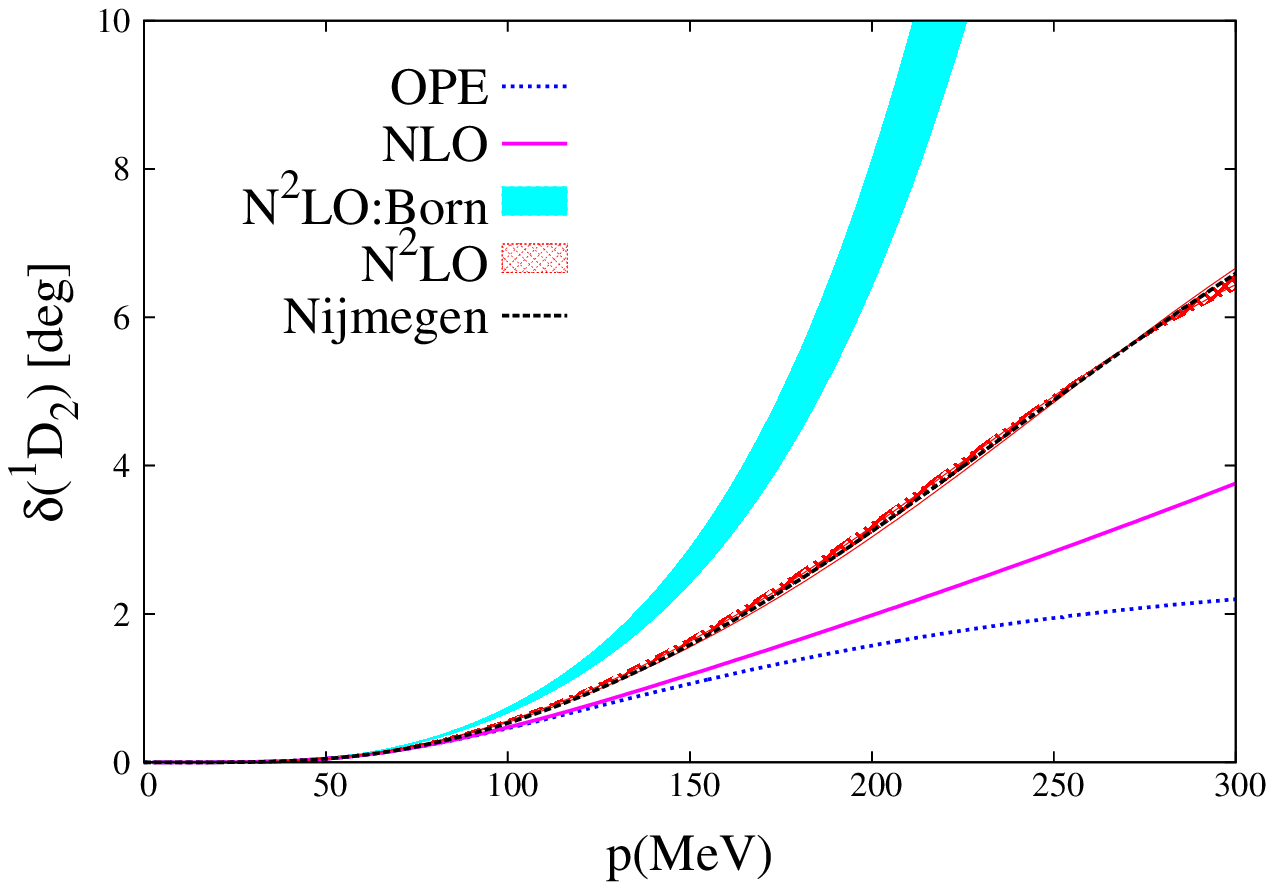} & 
\includegraphics[width=.4\textwidth]{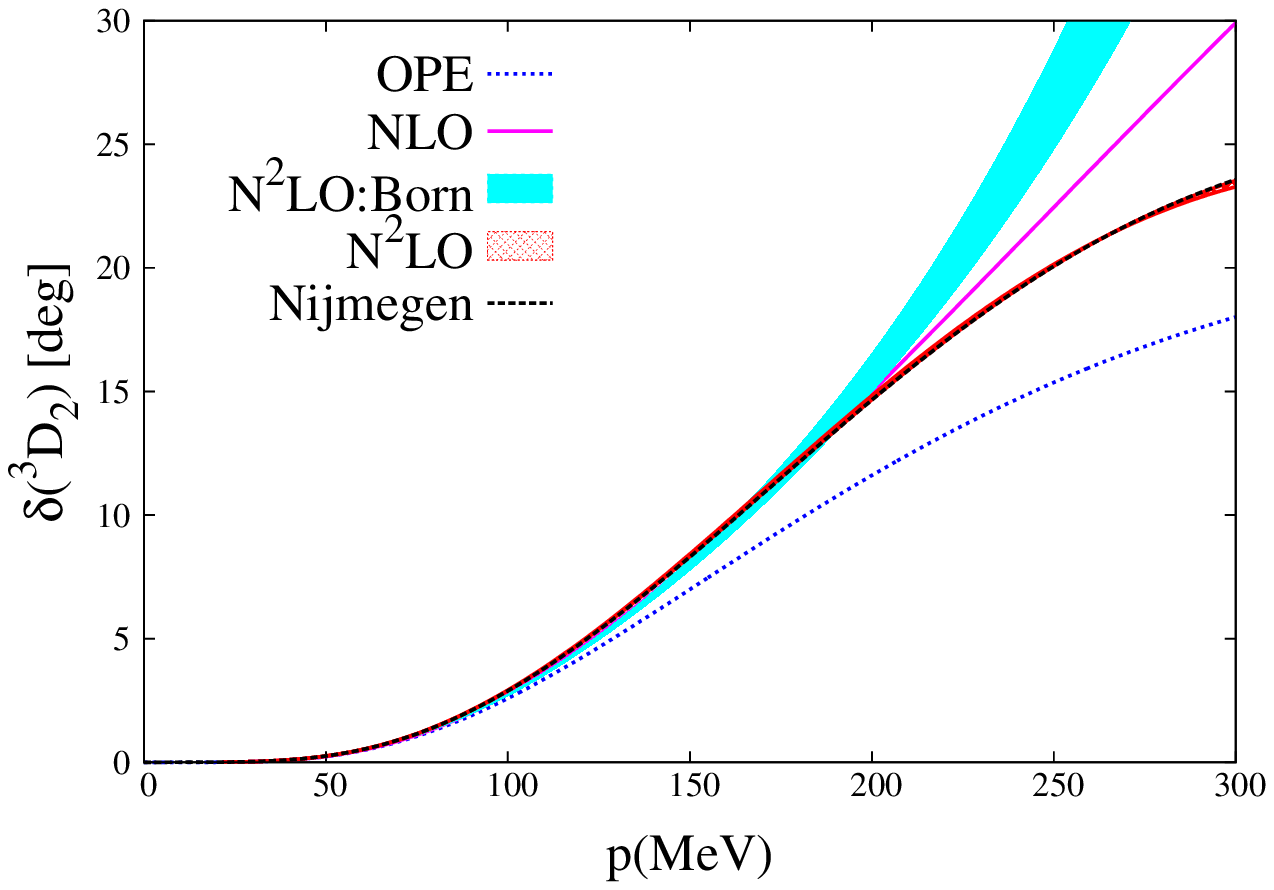}  
\end{tabular}
\caption[pilf]{\protect { (Color online.) Phase shifts for $^1D_2$ (left panel) and $^3D_2$ (right panel). 
 The (red) hatched areas  correspond to the NNLO results while the (magenta) solid lines are 
the NLO outcome \cite{gor2013}. In both cases twice-subtracted DRs are used. The 
phase shifts in the Born approximation   are shown by the (cyan) filled bands, 
the OPE result from Ref.~\cite{paper1} is the (blue) dotted lines and 
 the Nijmegen PWA phase shifts are given by the (black) dashed lines.}
\label{fig:dw} }
\end{center}
\end{figure}

Here, we discuss the $D$ waves. In order to preserve the right threshold behavior we 
employ the twice-subtracted DRs of Eqs.~\eqref{highd} and \eqref{highn} with $\ell=2$. 
For the uncoupled $D$ waves one has that
\begin{align}
\lambda_D&=\lim_{A\to -\infty}\frac{\Delta(A)}{(-A)^{3/2}}<0
\end{align}
and for this sign we do not have numerical problems 
in the solution of the resulting IE even for diverging $\Delta(A)$  \cite{gor2013}.

\begin{align}
\label{twiceDNl_Dw}
D(A)&=1+\delta_2 A
+\frac{A^2}{\pi}\int_{-\infty}^L dk^2\frac{\Delta(k^2)D(k^2)}{(k^2)^2}g(A,k^2)~,\nn \\
N(A)&=\frac{A^2}{\pi}\int_{-\infty}^L dk^2\frac{\Delta(k^2)D(k^2)}{(k^2-A)(k^2)^2}~.
\end{align}

The only free parameter per partial wave is $\delta_2=D^{(1)}(0)$ which is  fitted to the 
Nijmegen PWA phase shifts. Taking into account the 
different sets of values for the $c_i$ counterterms we have the following 
results,
\begin{align}
^1D_2:& ~D^{(1)}(0)= 0.07(1) ~M_\pi^{-2},\nn\\
^3D_2:& ~D^{(1)}(0)= -0.017(3)~M_\pi^{-2}~.
\end{align} 
The reproduction of data is excellent as shown by the (red) 
hatched areas in Fig.~\ref{fig:dw}, where the phase shifts for the $^1D_2$ are given in the left panel 
and those of the $^3D_2$ in the right one. Our results indeed overlap the Nijmegen PWA phase shifts given by 
the (black) dashed lines.  
Reference~\cite{gor2013} obtained the (magenta) solid line making use also of twice-subtracted DRs at NLO. 
 We see a remarkable   improvement from NLO to NNLO due to the inclusion of NLO  $\pi N$ dynamics, 
particularly for the $^1D_2$ partial wave.  

\subsection{Perturbative and Born approximation phase shifts}
\label{born}

The higher is the orbital angular momentum $\ell$ the more perturbative is expected to be the 
corresponding $NN$ partial wave. 
This statement was studied in detail in the perturbative study of  
 Ref.~\cite{peripheral} by making use 
of the one-loop approximation in BChPT.
 Indeed, we can easily obtain  from 
our formalism both the leading perturbative solution to the 
IEs of the $N/D$ method in powers of $\Delta(A)$,  as well as the leading term in the Born series approximation 
for the chiral $NN$ amplitude calculated up to ${\cal O}(p^3)$ in Ref.~\cite{peripheral}. 
The point is that for a 
weak  interaction  (small $\Delta(A)$ at low three-momentum) one can expect that $D(A)\simeq 1$ at low energies.
 It is then reasonable to consider that substituting $D(A)\to 1$ in the integral on the r.h.s.  of 
Eq.~\eqref{highn} would be meaningful to calculate $N(A)$, 
 because we have a rapid converging integral due to the factor $(k^2)^\ell$ in the denominator for a sufficiently 
large value of $\ell$.\footnote{Of course, the precise meaning of this statement could vary from one case to 
the other due to characteristic facets of the considered  partial wave.}  The perturbative 
result for $N(A)$, denoted by $N_p(A)$, is then
\begin{align}
N^{(p)}(A)=\frac{A^\ell}{\pi}\int_{-\infty}^L dk^2\frac{\Delta(k^2)}{(k^2)^\ell (k^2-A)}~.
\label{eq.hw.per}
\end{align}
Had we included only the two-nucleon irreducible contributions to  $\Delta(A)$, which is then 
denoted as $\Delta_B(A)$,   
the previous integral becomes  the DR representation of the $NN$ potential that 
we denominate $N_B(A)$,
\begin{align}
N_B(A)=\frac{A^\ell}{\pi}\int_{-\infty}^L dk^2\frac{\Delta_B(k^2)}{(k^2)^\ell (k^2-A)}~.
\label{eq.nborn}
\end{align}
This is due to the fact that the $NN$ potential projected in a given 
partial wave  is an analytical function that only has LHC and it can be 
written in terms of a  DR  along the latter cut. 
 We have checked numerically that  the DR representation 
Eq.~\eqref{eq.nborn} for the $NN$ potential coincides
 with its explicit partial wave decomposition taking into account 
the expressions given in Ref.~\cite{peripheral}.  
 In our notation the relation between $N_B(A)$ and the phase shifts in the Born approximation, $\delta_B(A)$, reads
\begin{align}
\delta_B(A)=\rho(A) N_B(A)~.
\label{deltab}
\end{align}
An analogous  expression holds for the perturbative phase shifts $\delta^{(p)}(A)$ calculated in terms of $N^{(p)}(A)$. 
 The difference between the perturbative phase shifts and the Born approximation ones 
for $\ell\geq 2$ is typically not very significant and quite small. 
In the following we compare our full results  with $\delta_B(A)$, 
since these phase shifts can be also calculated straightforwardly in potential models.
We proceed in the same way for the coupled channel case as well by evaluating $N_{ij}(A)$ in the Born approximation 
by substituting  $D_{ij}(A)\to 1$ in Eq.~\eqref{highncc}, and keeping only the two-nucleon irreducible contributions 
to $ \Delta_{ij}(A)$.

Turning back to the uncoupled $D$ waves we also show in Fig.~\ref{fig:dw} the leading Born approximation phase shifts 
obtained from the NNLO two-nucleon  irreducible contributions to $\Delta(A)$ by the (cyan) filled areas.
 One observes that these curves are quite different from our full results 
given by the hatched areas. This clearly indicates that the perturbative 
treatment of the $NN$ $D$ waves is not accurate.

\section{Uncoupled $F$ waves}
\label{fw}

\begin{figure}
\begin{center}
\begin{tabular}{cc}
\includegraphics[width=.4\textwidth]{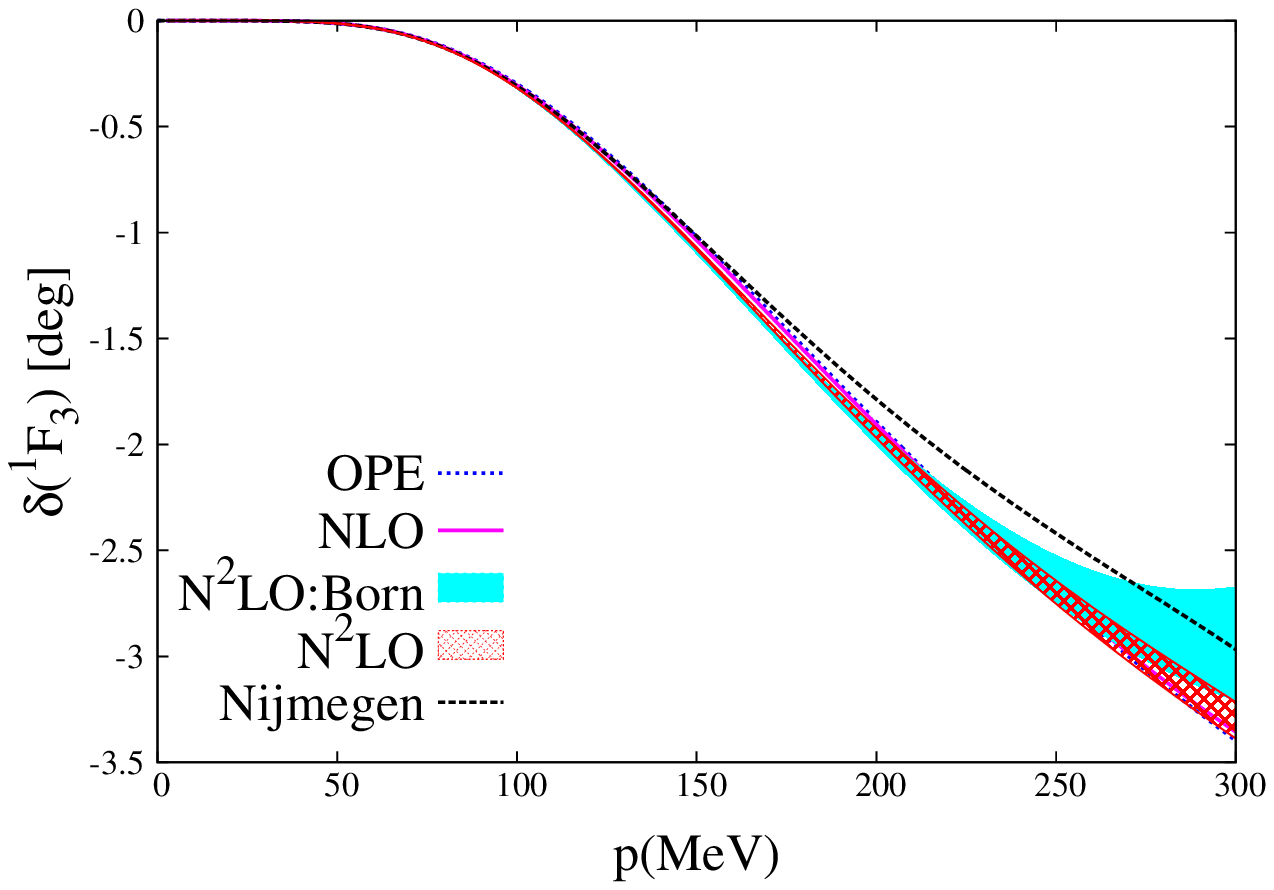} & 
\includegraphics[width=.4\textwidth]{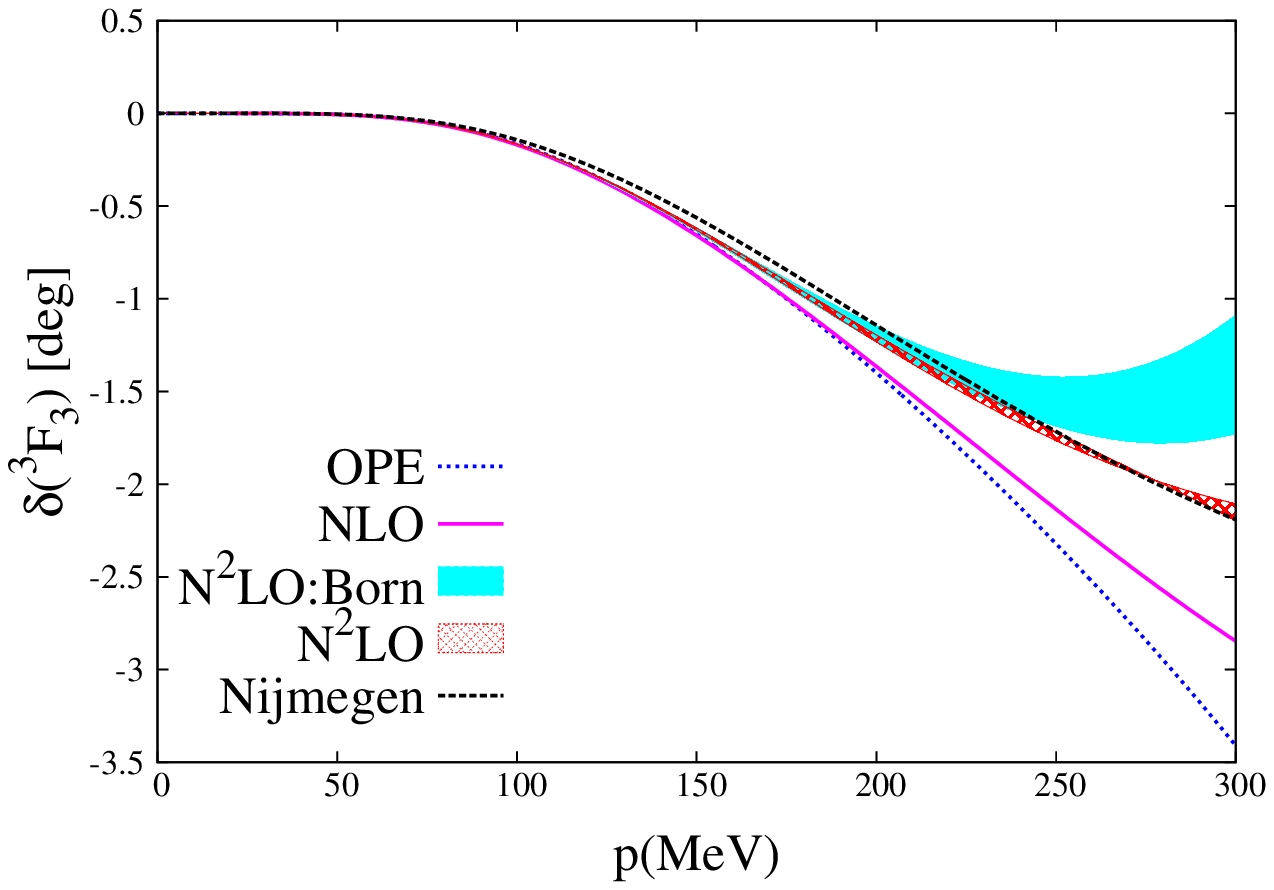}  
\end{tabular}
\caption[pilf]{\protect { (Color online.) Phase shifts for $^1F_3$ (left panel) and $^3F_3$ (right panel). 
 The (red) hatched areas  correspond to the NNLO results while the (magenta) solid lines are 
the NLO outcome. In both cases three-time-subtracted DRs are used.
The (cyan) filled bands give $\delta_B(A)$, 
the OPE result from Ref.~\cite{paper1} is the (blue) dotted lines and 
the Nijmegen PWA phase shifts correspond to the (black) dashed lines.}
\label{fig:fw} }
\end{center}
\end{figure}

For the $F$ waves we have three subtractions with two free parameters $\delta_2$ and $\delta_3$. We fix $\delta_2=0$ in the following (according to the principle of maximal smoothness) and fit $\delta_3$ to data. 
 At NNLO the fitted values for $D^{(2)}(0)=2 \delta_3$, Eq.~\eqref{tayloruc}, are: 
\begin{align}
^1F_3:~& D^{(2)}(0)= 0.057(3)~M_\pi^{-4}~,\nn\\
^3F_3:~& D^{(2)}(0)= 0.035(5)~M_\pi^{-4}~,
\end{align}
where the variation in the values is due to the different sets of $c_i$ counterterms employed. 
The NNLO results are shown by the (red) hatched areas in Fig.~\ref{fig:fw} which reproduce the 
Nijmegen PWA phase shifts (black dashed line) better than the NLO results (magenta lines) and 
the perturbative phase shifts (cyan filled areas).  This improvement 
is particularly noticeable for the $^3F_3$ partial wave.

We also observe that for the $F$ waves the phase shifts in the leading Born approximation, Eq.~\eqref{deltab}, 
run much closer to our full results than for the $D$ waves, which clearly indicates that 
$F$ waves are more perturbative. 
 Nevertheless, the relative deviation of the perturbation results compared 
with the full solution is still around a 50\% at the end of the interval shown in Fig.~\ref{fig:fw}. 
 A similar conclusion on the more perturbative nature of the $F$ waves 
was also reached in the pure perturbative study of Ref.~\cite{peripheral} by comparing with experimental data.  
 However, here we can also compare with the full unambiguous solution of the corresponding IE. 
 For example, we can learn from Fig.~\ref{fig:fw}
 that the widths of the (cyan) filled bands for the Born approximation results 
reflect a much larger dependence on the $c_i$ 
coefficients than the one corresponding to the full nonperturbative results given by the (red) hatched
 areas.
 Thus, within our approach the failure reported in Refs.~\cite{epe04,thesis} to reproduce simultaneously 
the $D$ and $F$ waves by using the NNLO chiral potential 
calculated in dimensional regularization in Ref.~\cite{peripheral}  
 because the large values of the $c_i$  counterterms does not happen. 
Namely, we are able to 
describe properly both the  uncoupled $D$ and $F$ waves, Figs.~\ref{fig:dw} and \ref{fig:fw}, respectively, 
and the dependence on the precise set of $c_i$'s taken is quite mild for the full results. 
Indeed our calculation at NNLO describe the Nijmegen PWA phase shifts better than  the NLO ones \cite{gor2013},
 which is not the case for all of these waves in  Ref.~\cite{thesis} 
based on the (modified) Weinberg approach when comparing their NLO and NNLO results.

\begin{figure}
\begin{center}
\begin{tabular}{cc}
\includegraphics[width=.4\textwidth]{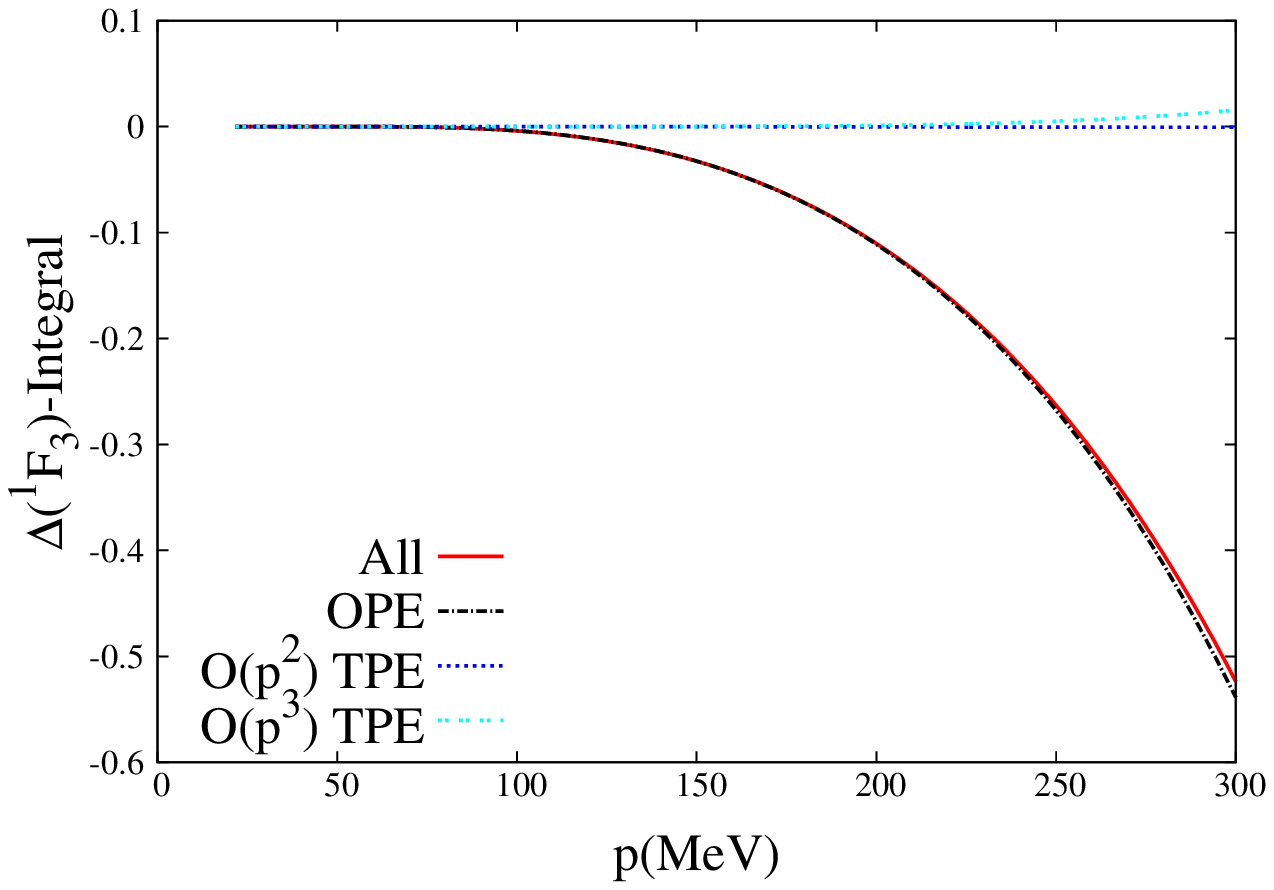} & 
\includegraphics[width=.4\textwidth]{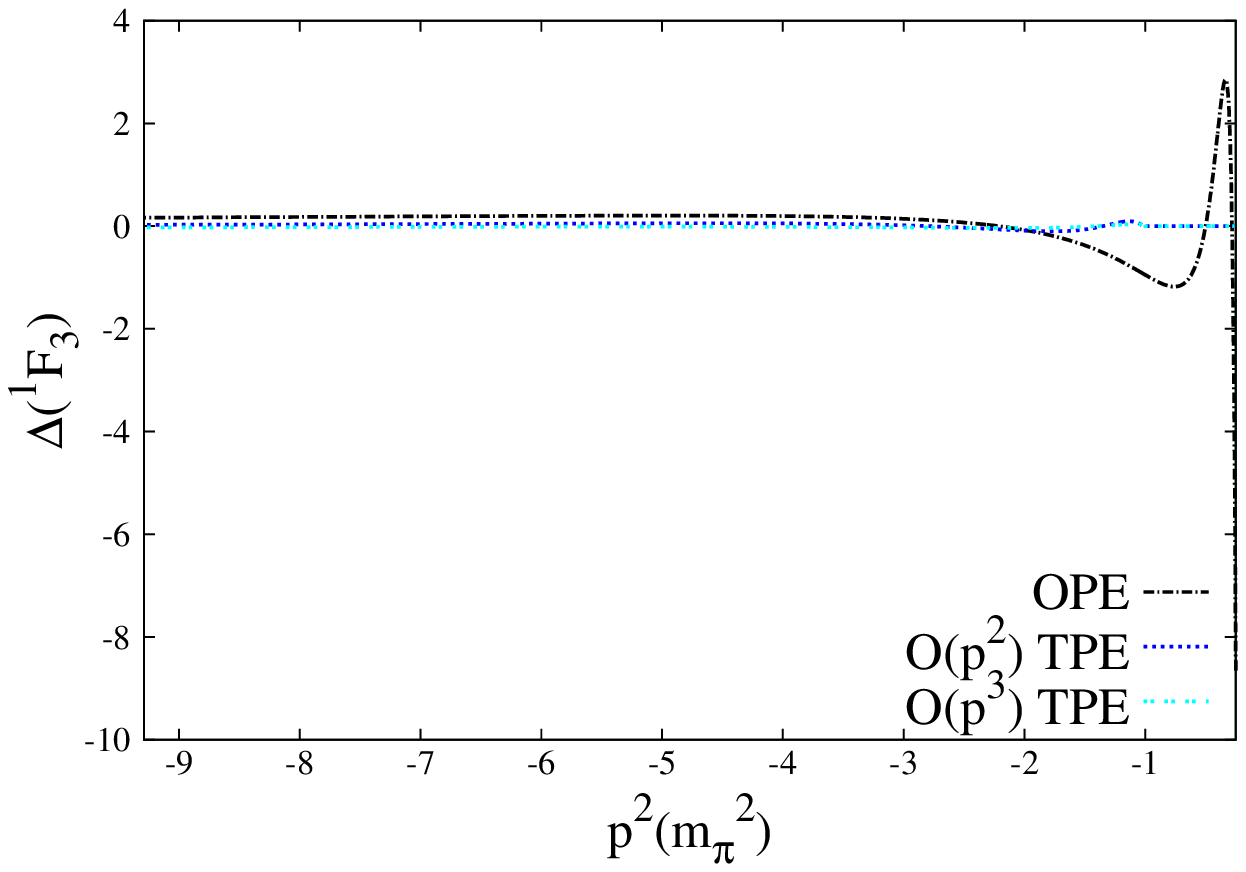} \\
\includegraphics[width=.4\textwidth]{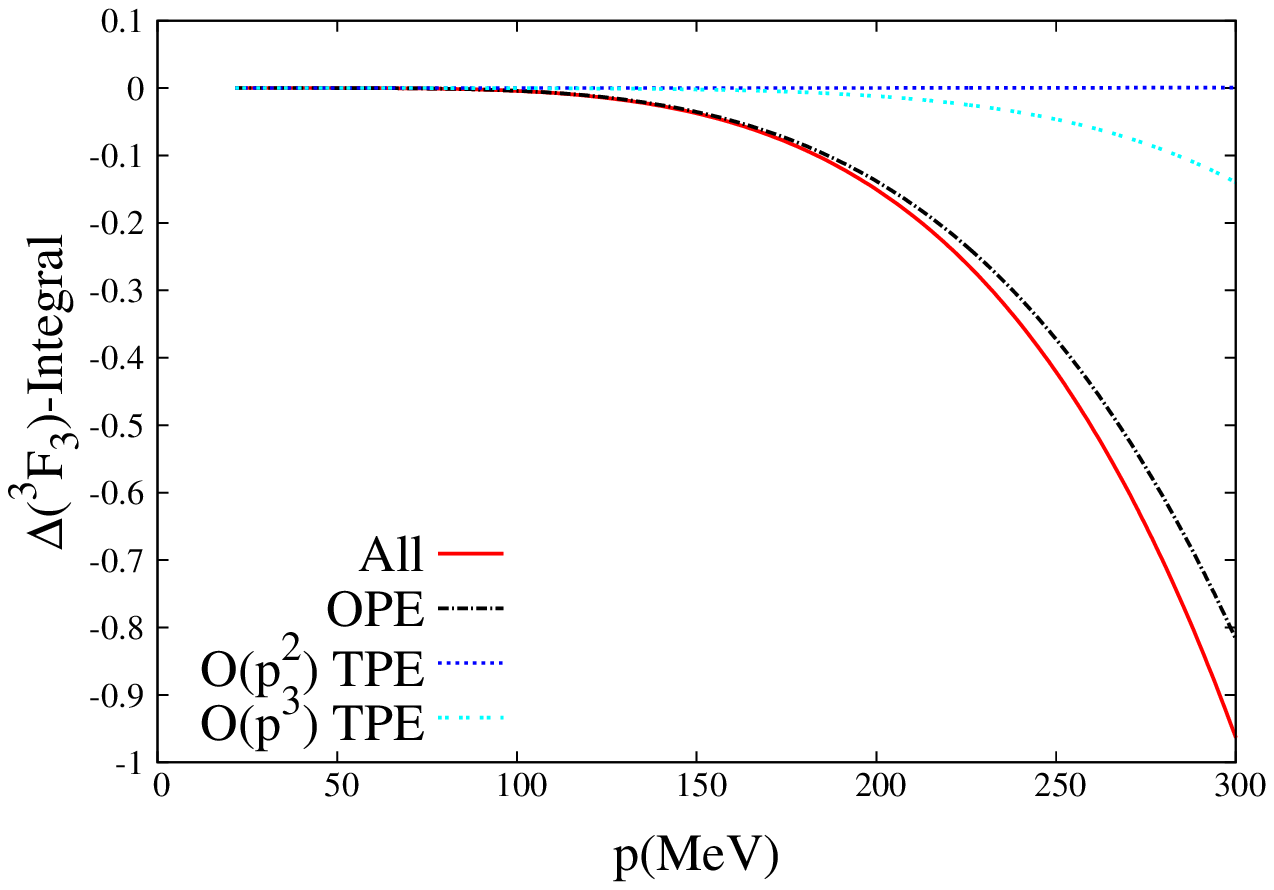} & 
\includegraphics[width=.4\textwidth]{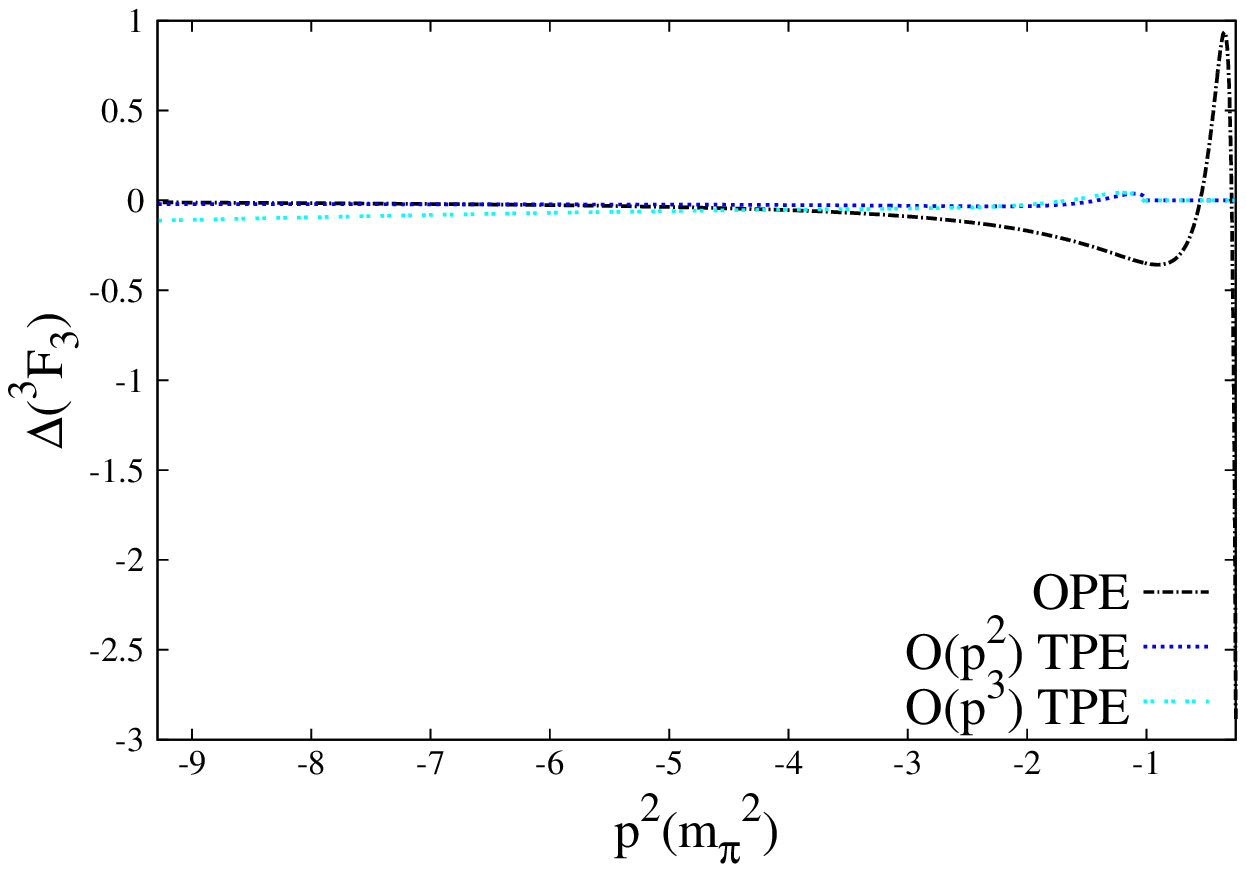} \\
\end{tabular}
\caption{ { (Color online.) Left panel: different contributions to the integral on the 
r.h.s. of Eq.~\eqref{highd} for $\ell=3$.
 The meanings of the lines are the same as in Fig.~\ref{fig:1s0quanty}. 
For definiteness we consider the $c_i$'s given in the last row of Table~\ref{tab:cis}.  }
\label{fig:unFquanty}}
\end{center}
\end{figure}

The increase in the perturbative character of the $F$ waves can also be seen by considering the 
relevance of the different contributions of $\Delta(A)$ to the integral on the r.h.s. of
 Eq.~\eqref{highd}, proceeding in a similar way to that already performed for the $^1S_0$ partial wave 
in Sec.~\ref{1s0}.
 The result is shown in the left panels of Fig.~\ref{fig:unFquanty}, where
 the first row corresponds to $^1F_3$ and the second to $^3F_3$.
 In the right panels we show directly the different contributions to $\Delta(A)$. 
The meanings of the lines 
in Fig.~\ref{fig:unFquanty} are the same as in Fig.~\ref{fig:1s0quanty}, though here 
 the $c_i$'s are taken from Ref.~\cite{aco2013}, given in the last row of Table~\ref{tab:cis}, 
which is enough for the present purposes. 
Notice, that now a qualitative different situation is found with respect to what is shown in 
Fig.~\ref{fig:1s0quanty}, that also holds for the $P$ and $D$ waves discussed in Secs.~\ref{pw} and \ref{dw}. 
 For the $F$ and higher waves the subleading two-nucleon irreducible TPE contribution is much less important and 
OPE is by far the dominant contribution, as it should correspond to a perturbative high-$\ell$ wave.

\section{Uncoupled $G$ waves}
\label{gw}

\begin{figure}
\begin{center}
\begin{tabular}{cc}
\includegraphics[width=.4\textwidth]{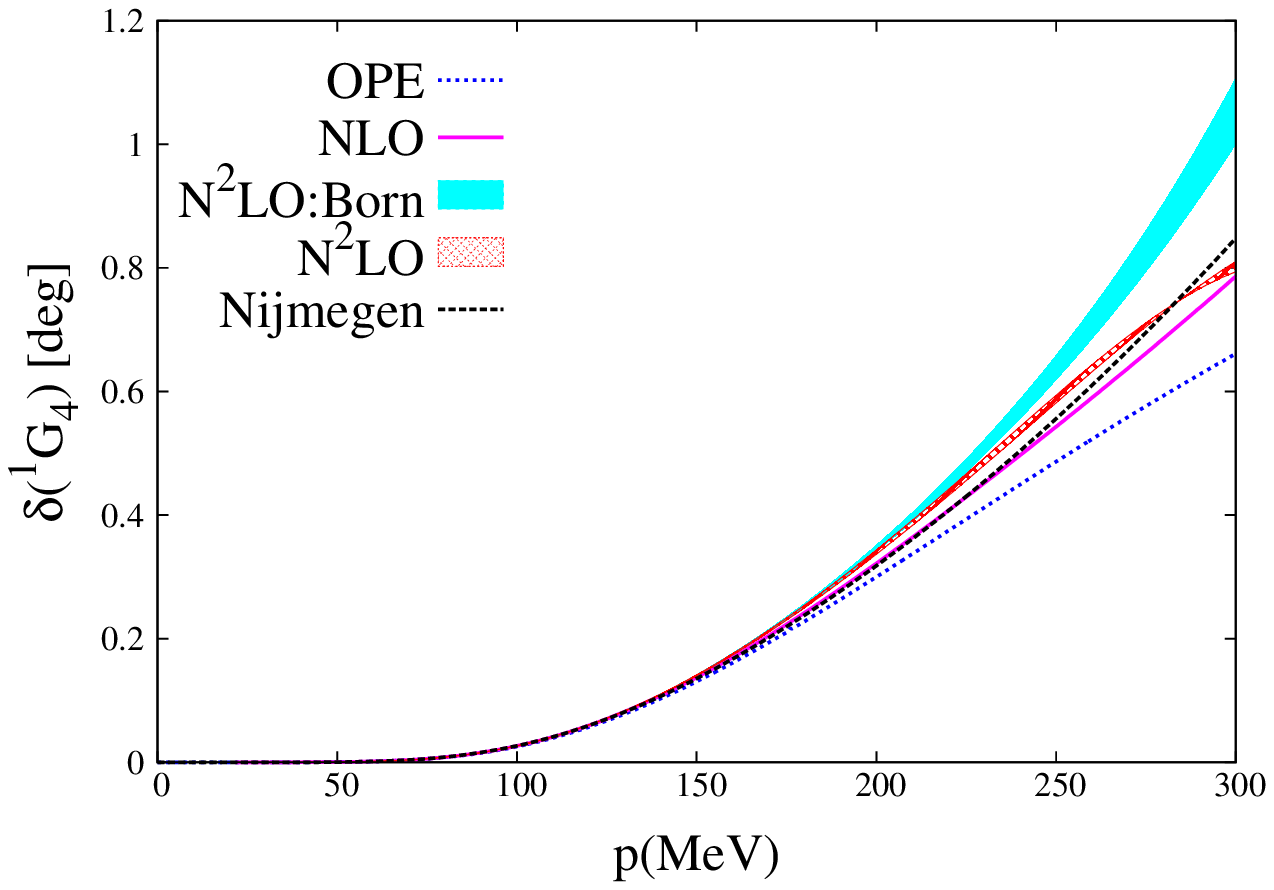} & 
\includegraphics[width=.4\textwidth]{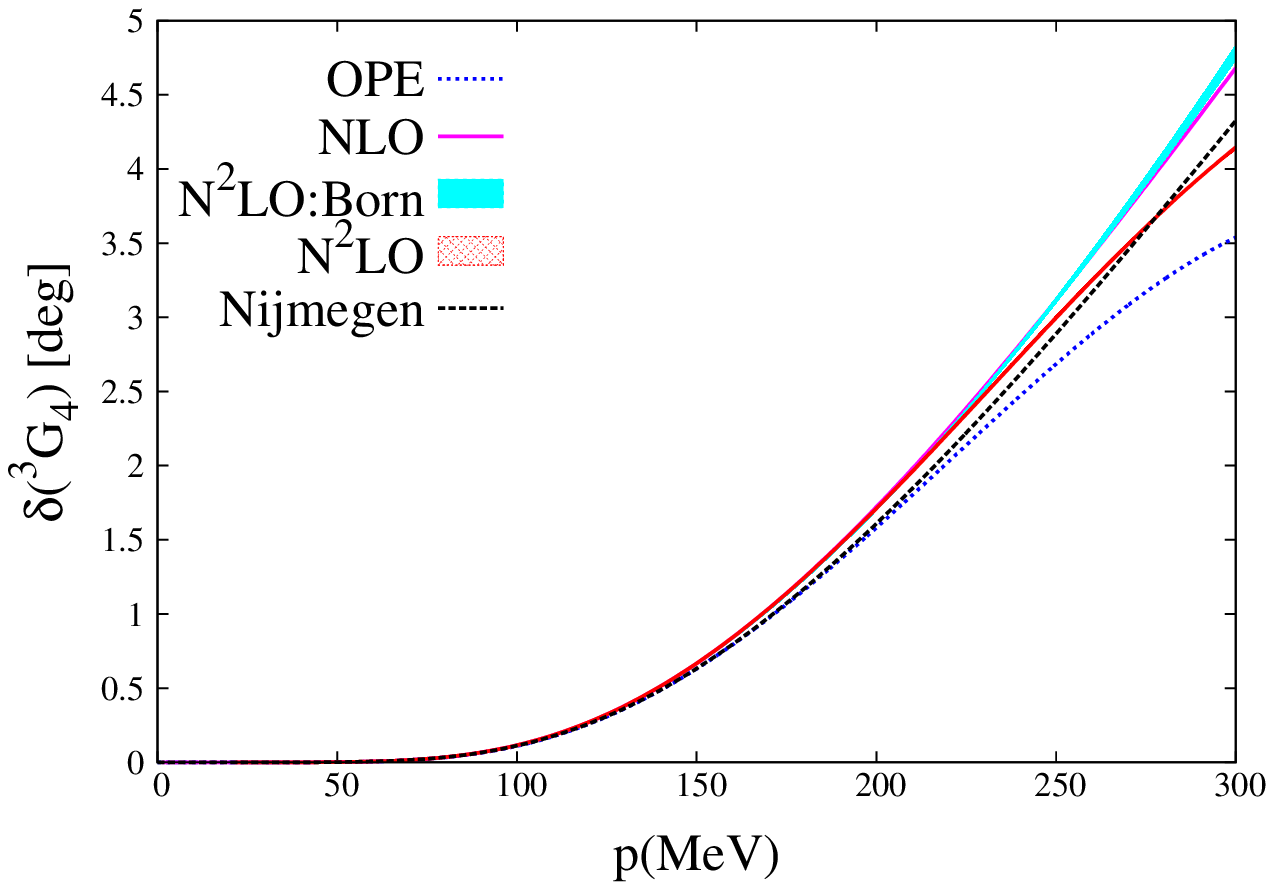}  
\end{tabular}
\caption[pilf]{\protect { (Color online.) Phase shifts for $^1G_4$ (left panel) and $^3G_4$ (right panel). 
 The (red) hatched areas  correspond to the NNLO results while the (magenta) solid lines are 
the NLO outcome. In both cases four-time--subtracted DRs are used. 
 The  (cyan) filled areas represent
 the outcome from the leading Born approximation,  
 the OPE result from Ref.~\cite{paper1} is the (blue) dotted lines 
and the Nijmegen PWA analysis is the (black) dashed lines.}
\label{fig:gw} }
\end{center}
\end{figure}

For the $G$ waves we have four subtractions of which $\delta_i$ $(i=2,3,4)$ are free but, according to the 
principle of maximal smoothness, all of them are fixed to 0 except 
$\delta_4=D^{(3)}(0)/3!$ that is fitted to data. 
At  NNLO the fitted values for $D^{(3)}(0)$ are: 
\begin{align}
^1G_4:~& D^{(3)}(0)=-0.014(2)~M_\pi^{-6}~,\nn\\
^3G_4:~& D^{(3)}(0)=-0.055(5)~M_\pi^{-6}~,
\end{align}
where the variation in the values is due to the different sets of $c_i$ counterterms employed. 
The corresponding results are shown by the (red) hatched areas in Fig.~\ref{fig:gw}. 
For both partial waves the actual dependence on the $c_i$ coefficients for the resulting phase shifts 
is almost negligible and the hatched  areas degenerate to lines. 
  The low-energy results are very similar at NLO and NNLO and reproduce the Nijmegen 
PWA phase shifts quite well.
 These results are better than the perturbative ones in the Born approximation,  Eq.~\eqref{deltab}, which  
are shown by the (cyan) filled areas. 
As indicated for the uncoupled $F$ waves here OPE overwhelmingly dominates 
 the different contribution to the dispersive integral on the r.h.s. of Eq.~\eqref{highd}. 
This indicates that these waves are rather perturbative, though still we observe differences 
around 30\% for $p\lesssim 300$~MeV in Fig.~\ref{fig:gw} between the full and perturbative results.

\section{Uncoupled $H$ waves}
\label{hw}

\begin{figure}
\begin{center}
\begin{tabular}{cc}
\includegraphics[width=.4\textwidth]{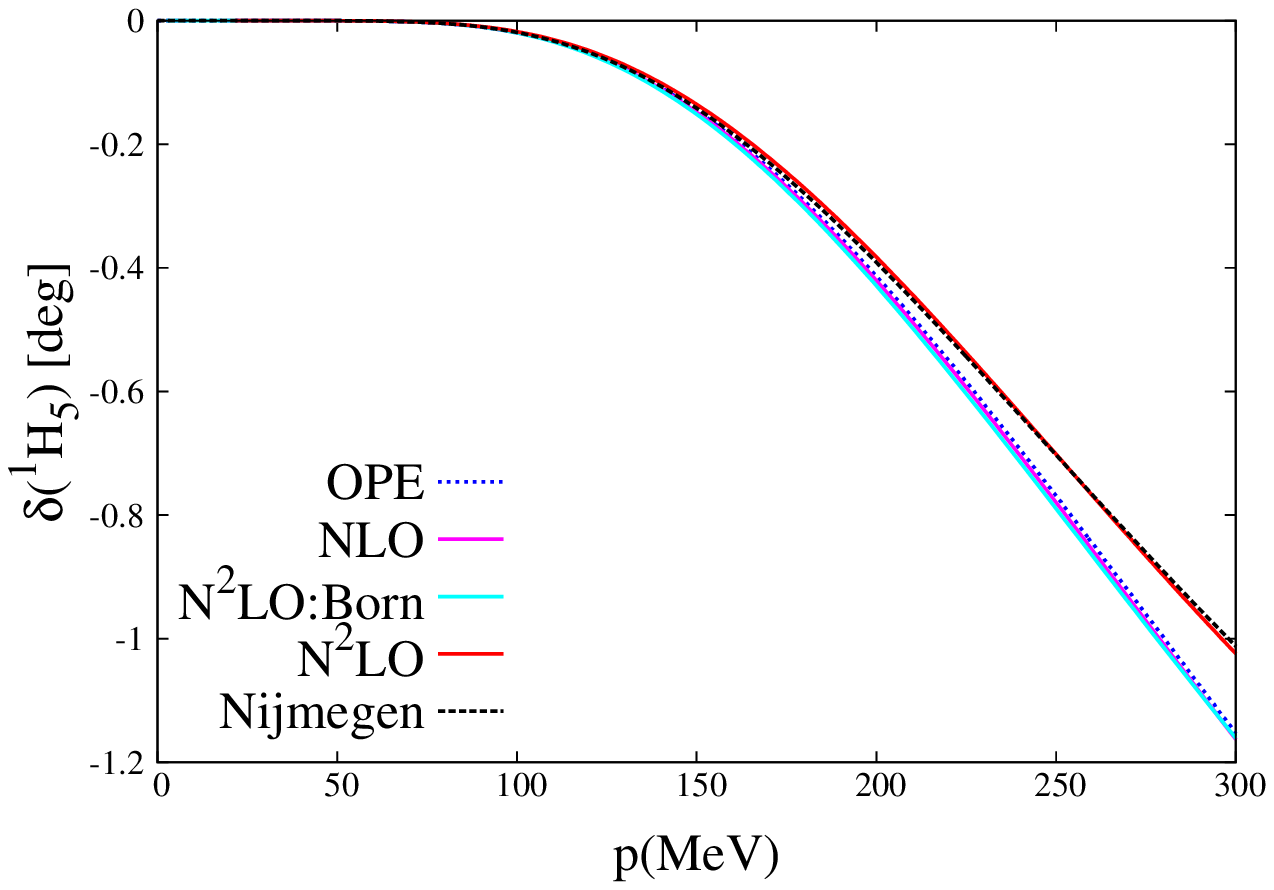} & 
\includegraphics[width=.4\textwidth]{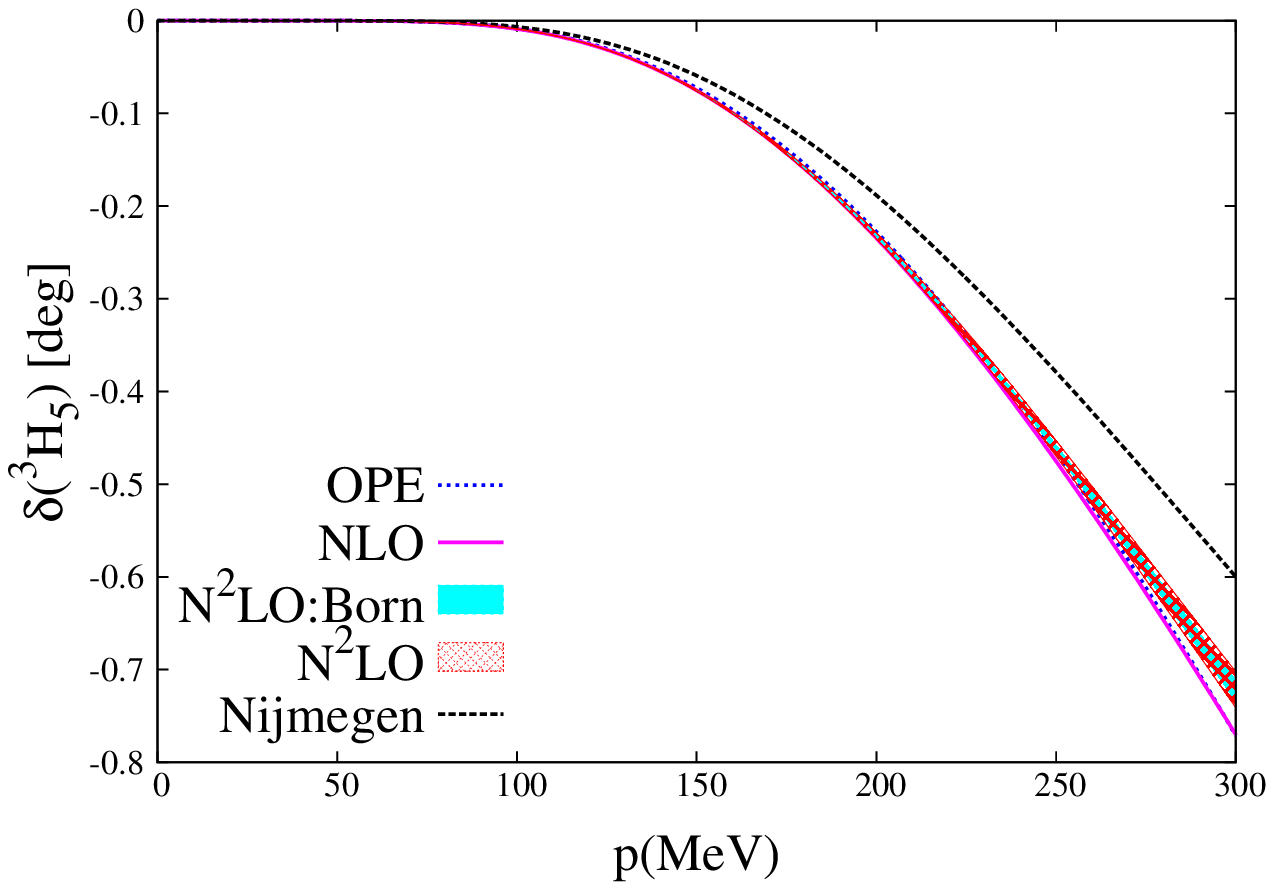}  
\end{tabular}
\caption[pilf]{\protect { (Color online.) Phase shifts for $^1H_5$ (left panel) and $^3H_5$ (right panel). 
 The (red) hatched areas  correspond to the NNLO results while the (magenta) solid lines are 
the NLO outcome. 
 The (cyan) filled bands correspond to $\delta_B(A)$, 
the OPE result from Ref.~\cite{paper1} is the (blue) dotted lines and  
the Nijmegen PWA is the (black) dashed lines.}
\label{fig:hw} }
\end{center}
\end{figure}

For the case of the uncoupled $H$ waves, $^1H_5$ and $^3H_5$, we apply the five-time subtracted DRs of  
Eqs.~\eqref{highd} and \eqref{highn} with $\ell=5$. 
We fit $\delta_5=D^{(4)}(0)/4 !$ to the Nijmegen PWA phase shifts, which for $\ell\geq 5$ 
correspond to  those obtained from the $NN$ potential model of Ref.~\cite{obe}, while $\delta_{2,3,4}$ are fixed to 0 (principle of maximal smoothness). 
We obtain the fitted values: 
\begin{align}
^1H_5:~D^{(4)}(0)&=0.156~ M_\pi^{-8}~,\nn\\
^3H_5:~D^{(4)}(0)&=0.066~M_\pi^{-8}.
\end{align} 
The resulting fit is stable if  we release $\delta_i$ $(i=2,3,4)$. 
The  phase shifts obtained are shown by the (red) hatched areas in  Fig.~\ref{fig:hw}  by taking into account the spread of the results 
depending of the set of $c_i$'s chosen. 
In this figure the left panel corresponds to  $^1H_5$ and  the right one to $^3H_5$.  
 For the former the resulting curve indeed  overlaps the Nijmegen PWA phase shifts \cite{Stoks:1994wp}. 
We also show by the (cyan) filled bands the phase shifts in the leading Born approximation
 which run rather close to 
the full results, indeed for the $^3H_5$ case the (cyan) filled band is overlapped by the (red) hatched one.   
This clearly indicates the perturbative nature for the $H$ waves. 
For them it is also true that OPE overwhelmingly dominates 
the dispersive integral on the r.h.s. of Eq.~\eqref{highd}, which is also the expected behavior 
for a perturbative partial wave.
Notice that for the $^1H_5$ wave the dependence on the actual values of the 
$c_i$ coefficients is so small that at the end the hatched and filled areas collapse to lines.
 For the $^3H_5$ case there is a visible, albeit small, dependence on the set of $c_i$'s employed. 
In both cases the NNLO results reproduce the Nijmegen PWA phase shifts closer than the NLO and OPE results.

\begin{figure}
\begin{center}
\begin{tabular}{cc}
\includegraphics[width=.4\textwidth]{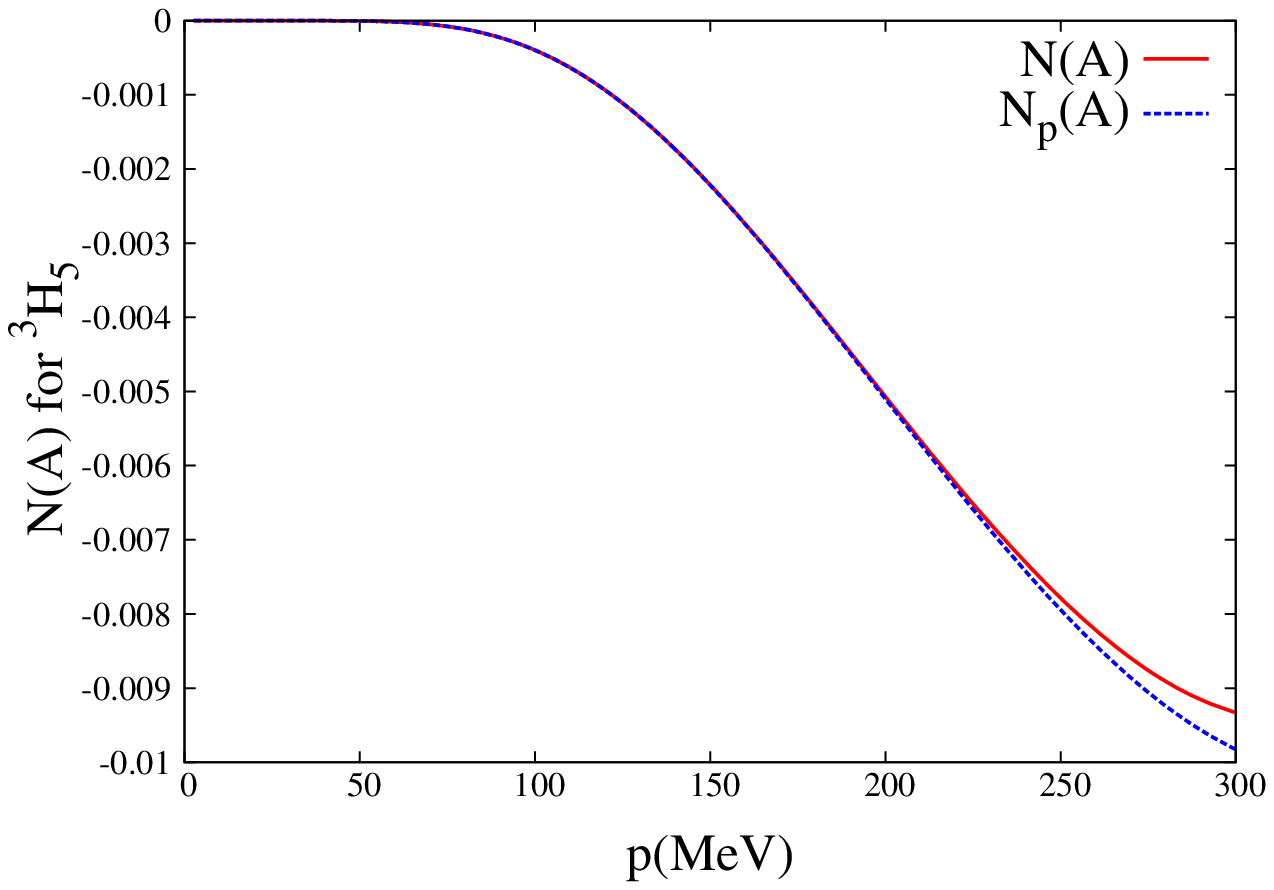} & 
\includegraphics[width=.4\textwidth]{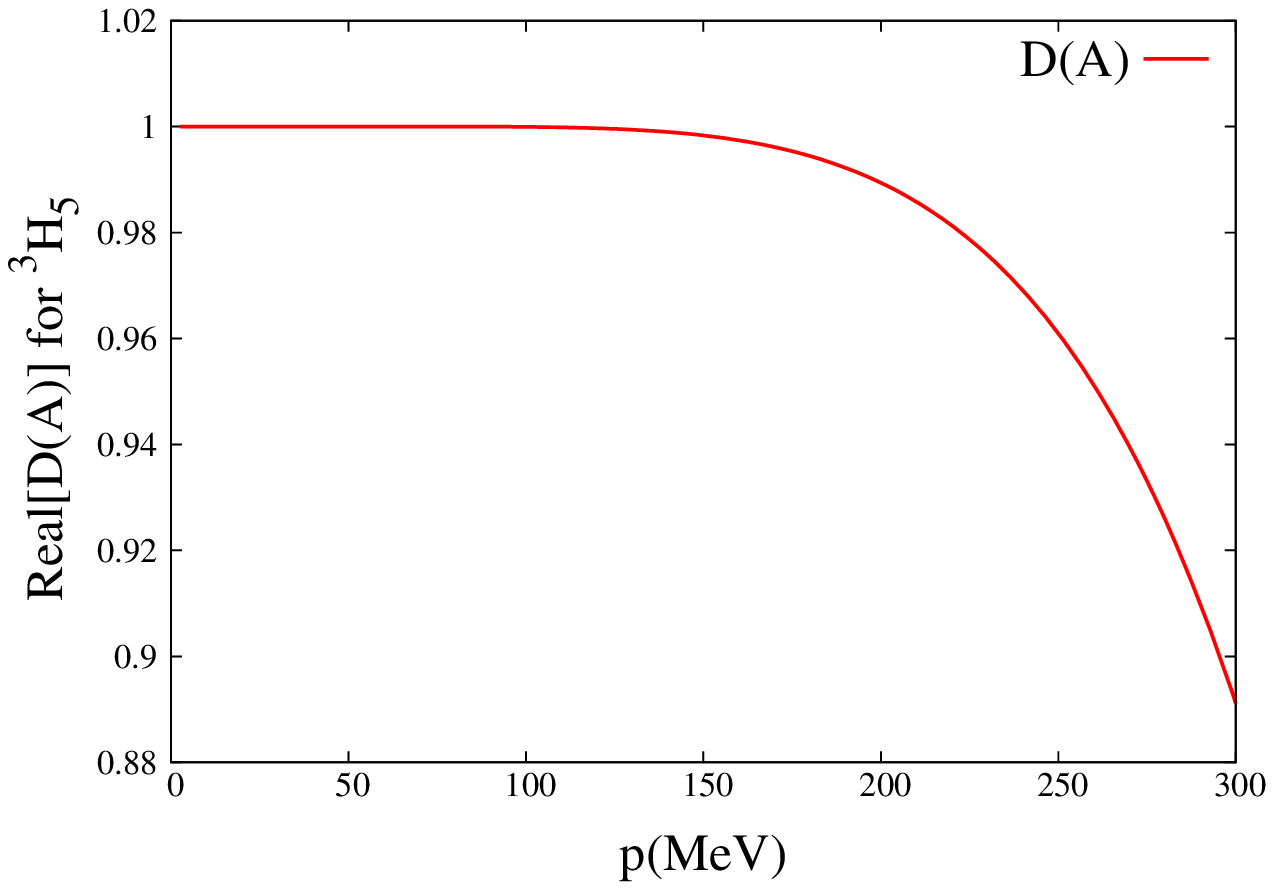}  
\end{tabular}
\caption[pilf]{\protect { (Color online.) The functions $N(A)$ and $N_p(A)$ are shown by the (red) solid and (blue) 
dash-dotted lines in the left panel, respectively. The real part of the function $D(A)$ is plotted in the right panel. }
\label{fig:hw.nd} }
\end{center}
\end{figure}

It is interesting to discuss in this case the behavior of the function $N(A)$ compared with $N_p(A)$, given in Eq.~\eqref{eq.hw.per}. 
The main point is that here both $N(A)$ and $D(A)$ have a zero at around 450~MeV.
 We consider only the $^3H_5$ wave because a similar discussion would follow for $^1H_5$ as well, 
that we skip for brevity.
 In the left panel of Fig.~\ref{fig:hw.nd} we show by the (red) solid line the full $N(A)$ 
 and by the (blue) dashed line the perturbative result $N_p(A)$. 
We see that they  are very similar, as expected for a partial wave with an $\ell$ as high as 5.
 In addition, we display  in the right panel of the same figure the real part of $D(A)$ from Eq.~\eqref{highd}, 
which is very close to 1, as expected for a situation with a weak interaction as well. 
All these curves are obtained by employing the  $c_i$'s from Ref.\cite{aco2013}.
 A bit higher in energy both  $N_p(A)$ and  $N(A)$ have a zero 
 at around $\sqrt{A}=450$~MeV.
 Since $T(A)=N(A)/D(A)$ this would imply that $T(A)=0$ at that energy, which is at odds 
with the values of the phase shifts given by the Nijmegen PWA \cite{Stoks:1994wp} that do not vanish at this point. 
The only remedy is that  $D(A)$ is also zero at the same point so that 
 one had a limit 0/0 that is finally finite. 
This is indeed the case and it is is the reason why 
 $D(A)$ starts to decrease for $\sqrt{A}>200$~MeV in Fig.~\ref{fig:hw.nd}.

Another question of interest to think about is  what have we  gained by solving exactly Eq.~\eqref{highd} instead of 
 using only the perturbative solution, Eq.~\eqref{eq.hw.per}, or the Born approximation, Eq.~\eqref{eq.nborn}, with 
the related $ \delta_B(A)$, Eq.~\eqref{deltab}?.  
The main point that one should consider 
 in connection with this question is that by solving the full and nonperturbative Eq.~\eqref{highd} 
(furthermore, in good agreement with data) one can then state that Eq.~\eqref{eq.hw.per} is a  perturbation of a 
 well-defined and existing nonperturbative solution. 
By solving exactly Eq.~\eqref{highd} we have needed to consider explicitly $\delta_5$ as a free parameter  for the uncoupled 
$H$ waves  and fit it to the Nijmegen PWA. 
Indeed, $\delta_5$ is not only necessary for a good fit, 
 but it is also required in order to keep $D(A)\simeq 1$ at low 
three-momentum. 
Otherwise, the contribution from the dispersive integral to $D(A)$ on the r.h.s. of Eq.~\eqref{highn} would 
be too large and negative and would render a too strong function $N(A)$  in plain disagreement with $N_p(A)$. 
Notice as well that in the case of the partial wave $^1H_5$ a better reproduction of data is achieved than with $\delta_B(A)$. 
 It is also worth recalling the previous finding in Sec.~\ref{fw} for the $F$ waves, 
where the full results show a much smaller dependence on 
the set of $c_i$ coefficients used than the perturbative or Born approximation phase shifts, cf. Fig.~\ref{fig:fw}.

\section{Coupled $^3S_1-{^3D_1}$ waves}
\label{sd12w}

We start our study of the $^3S_1-{^3D_1}$ coupled-partial-wave system in terms of just one free parameter, 
that we choose as the pole position of the deuteron in the $A$-complex plane, $k^2_d=-m E_d$, with $E_d= 2.225$~MeV 
the deuteron binding energy. 
 Thus we implement once-subtracted DRs for the $^3S_1$ and twice-subtracted ones for the $^3D_1$. 
In the case of the mixing partial wave we have a mixed situation with a once-subtracted DR 
for $N_{12}(A)$ and a twice-subtracted one for $D_{12}(A)$.
 In this way we guarantee both the right threshold behavior as well as the experimental 
 deuteron-pole position in all the partial waves. 
We write now explicitly  the DRs considered. For the $^3S_1$ one has,
\begin{align}
\label{3s1_a}
D_{11}(A)&=1-\frac{A}{k_d^2}\frac{g_{11}(A,0)}{g_{11}(k_d^2,0)}
+\frac{A}{\pi}\int_{-\infty}^L dk^2 \frac{\Delta_{11}(k^2)D_{11}(k^2)}{k^2}
\Bigg[g_{11}(A,k^2)-g_{11}(A,0)\frac{g_{11}(k_d^2,k^2)}{g_{11}(k_d^2,0)}\Bigg]~,\nn\\
N_{11}(A)&=\nu_1^{(11)}+\frac{A}{\pi}\int_{-\infty}^L dk^2\frac{\Delta_{11}(k^2)D_{11}(k^2)}{k^2(k^2-A)}~,
\end{align}
with all the subtractions taken at $A=0$ and the new function $g_{ij}(A)$ is defined as
 \begin{align}
g_{ij}(A,k^2)&=\frac{1}{\pi}\int_0^\infty dq^2\frac{\nu_{ij}(q^2)}{(q^2-A)(q^2-k^2)}~,
\label{gij}
\end{align}
 The subtraction constant $\nu_1$ in $N_{11}(A)$ is fixed by imposing that $D_{11}(k_d^2)=0$,
\begin{align}
\nu_1^{(11)}&=\frac{1}{k_d^2 \,g_{11}(k_d^2,0)}\Bigg[
1+\frac{k_d^2}{\pi}\int_{-\infty}^L dk^2 \frac{\Delta_{11}(k^2) D_{11}(k^2)}{k^2}g_{11}(k^2,k_d^2)
\Bigg]~,
\label{once_nu0}
\end{align}
a result that is  already implemented in Eq.~\eqref{3s1_a} for $D_{11}(A)$.
 
The corresponding DRs for the $^3D_1$ and the mixing wave can be grouped together in the same form,
\begin{align}
\label{3sd1_a}
D_{ij}(A)&=1-\frac{A}{k_d^2}+\frac{A(A-k_d^2)}{\pi}\int_{-\infty}^L dk^2\frac{\Delta_{ij}(k^2)D_{ij}(k^2)}{(k^2)^{\ell_{ij}}}
g_{ij}^{(d)}(A,k^2;\ell_{ij})~,\nn\\
N_{ij}(A)&=\frac{A^{\ell_{ij}}}{\pi}\int_{-\infty}^L dk^2\frac{\Delta_{ij}(k^2)D_{ij}(k^2)}{(k^2)^{\ell_{ij}}(k^2-A)}~.
\end{align}
where $\ell_{12}=1$ and $\ell_{22}=2$ and all the subtractions for the $N_{ij}(A)$ are taken at $A=0$,
 while in the function $D(A)$ one is taken at $A=0$ and 
the other at $A=k_d^2$. The  function $g^{(d)}_{ij}(A,k^2;m)$ is defined as
\begin{align}
g^{(d)}_{ij}(A,k^2;m)&=\frac{1}{\pi}\int_0^\infty dq^2\frac{\nu_{ij}(q^2)(q^2)^{m-1}}{(q^2-A)(q^2-k^2)(q^2-k_d^2)}~.
\label{gijd}
\end{align}

\begin{figure}[h]
\begin{center}
\begin{tabular}{cc}
\includegraphics[width=.4\textwidth]{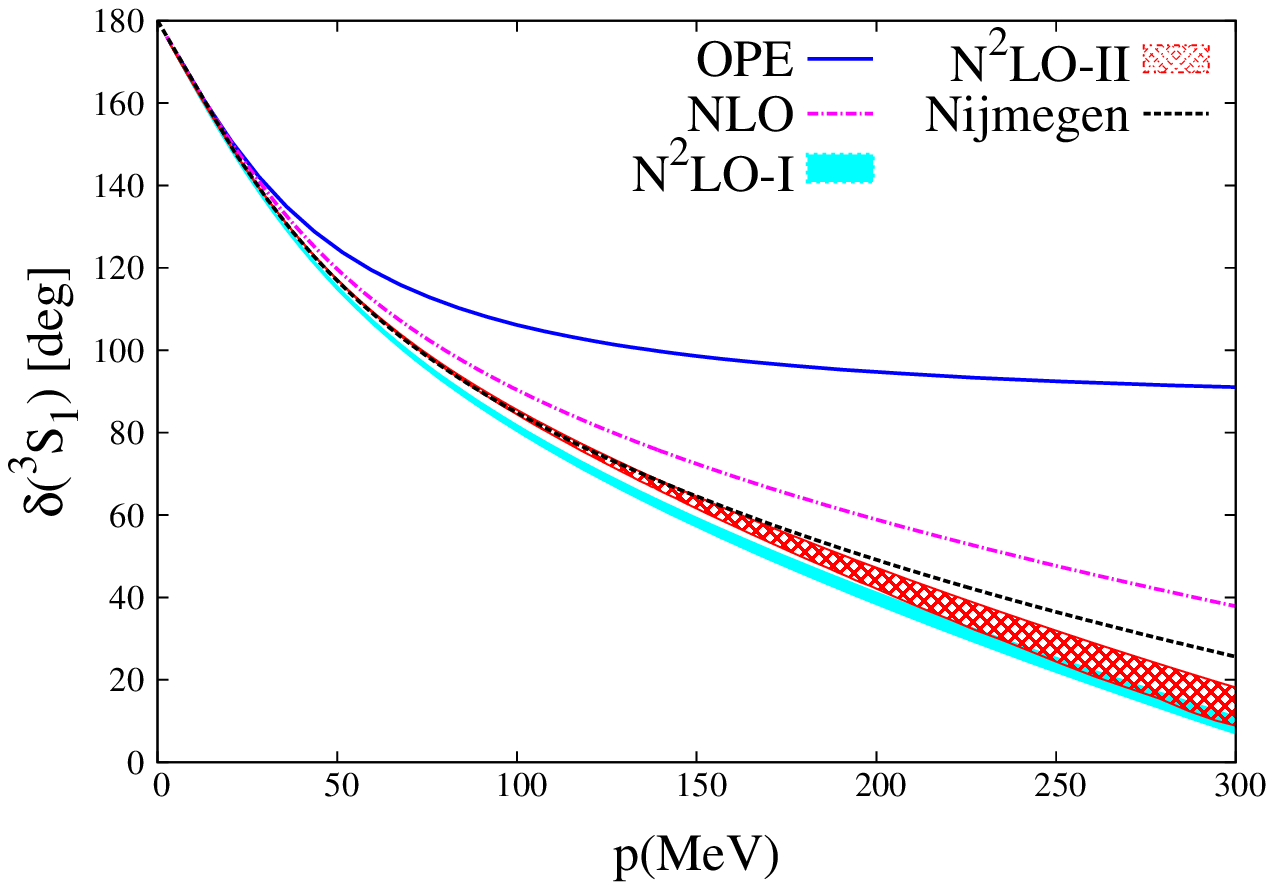} & 
\includegraphics[width=.4\textwidth]{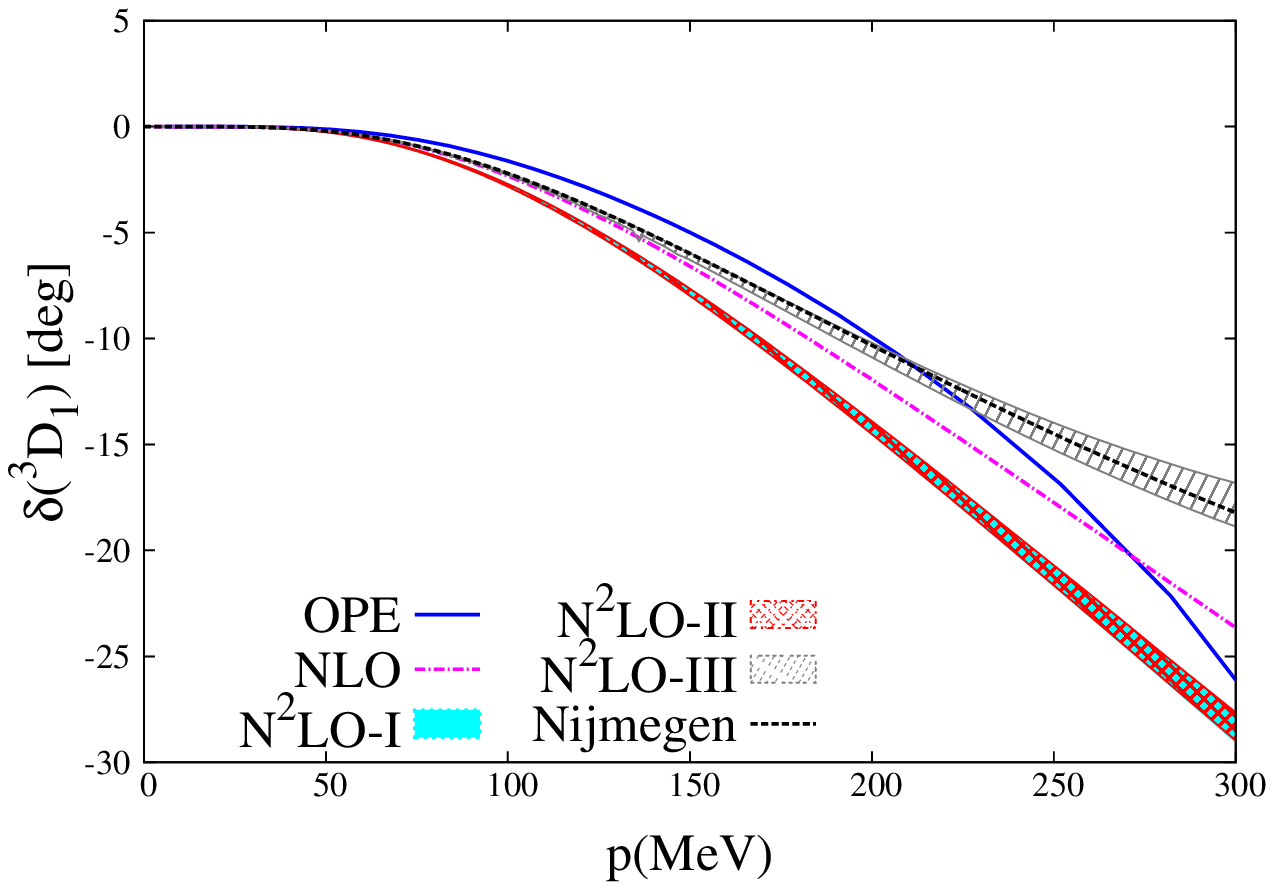}\\  
\includegraphics[width=.4\textwidth]{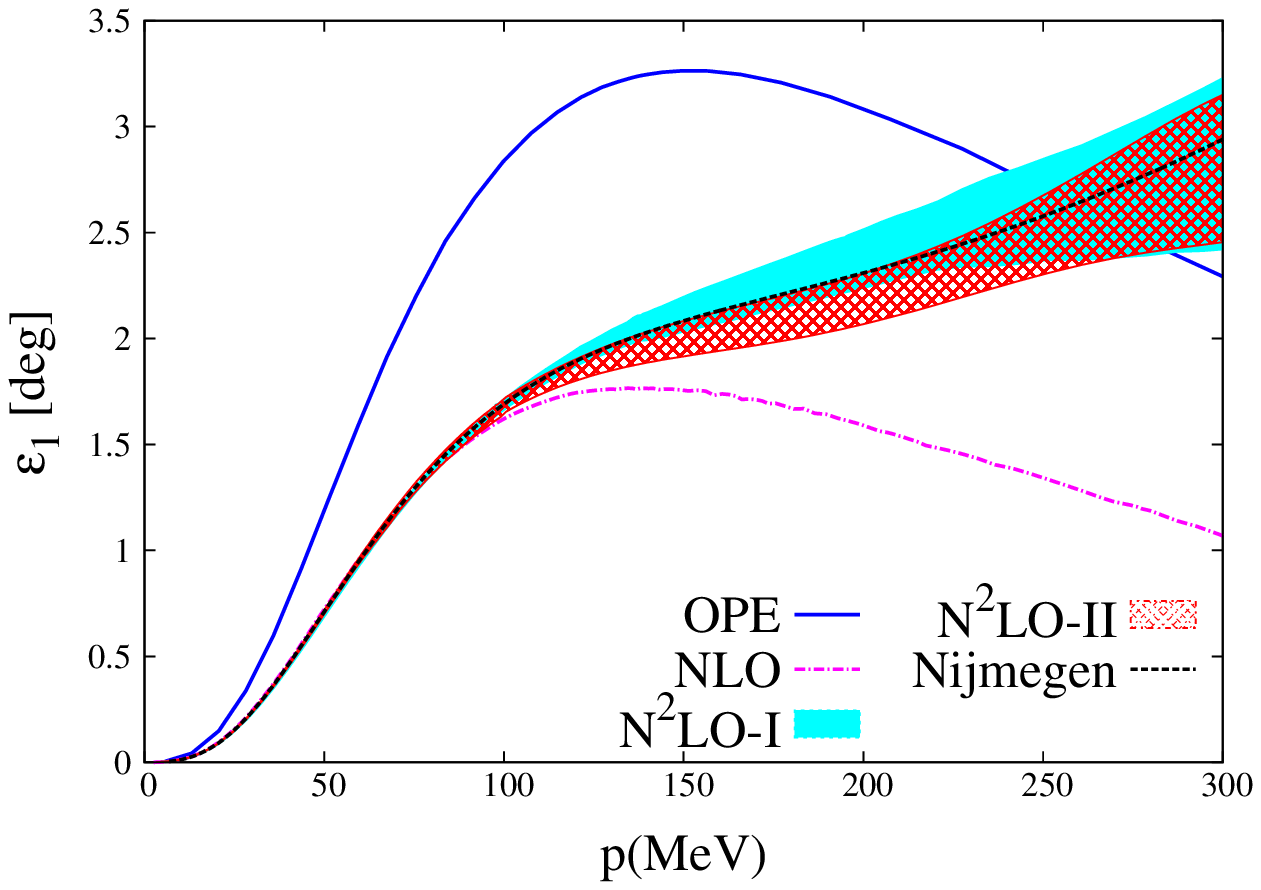}  
\end{tabular}
\caption[pilf]{\protect { (Color online.) From left to right and top to bottom:
 Phase shifts for $^3S_1$, $^3D_1$ and the mixing angle  $\epsilon_1$, respectively.
The (cyan) filled areas  correspond to the NNLO-I outcome obtained by solving 
 Eqs.~\eqref{3s1_a} and \eqref{3sd1_a}. The  hatched areas with (red) crossed lines are the NNLO-II results 
 that stem from Eqs.~\eqref{2subs.3s1}, \eqref{2subs.mix} and \eqref{3sd1_a}.  
In addition, for the $^3D_1$ we show by the hatched areas with (gray) parallel lines the results obtained 
by employing three-time subtracted DRs for $^3D_1$, Eq.~\eqref{3d1.extra}. 
As usual, the (magenta) dot-dashed lines are the NLO phase shifts and mixing angle, 
 the  LO ones are given by the  (blue) dotted lines and the 
Nijmegen PWA results correspond to the (black) dashed lines.}
\label{fig:3sd1_a} }
\end{center}
\end{figure}

The results obtained by solving the IEs for the functions $D_{ij}(A)$ along the LHC
 from Eqs.~\eqref{3s1_a} and \eqref{3sd1_a} are 
shown in Fig.~\ref{fig:3sd1_a} by the (cyan) filled areas.
 These results are indicated as NNLO-I and 
 all the subtraction constants are fixed in terms of $k_d^2$, without any other freedom. 
 The spread in the results originates by taking  different  sets of $c_i$'s from Refs.~\cite{epe12,aco2013} 
and varying the input in the iterative procedure.
 The present NNLO calculation from Eqs.~\eqref{3s1_a} and \eqref{3sd1_a}
 reproduces the  Nijmegen PWA  mixing angle $\epsilon_1$ much better than the NLO 
result from the same set of equations,  which is shown by the (magenta) dot-dashed lines.
 This improvement in the description of $\epsilon_1$ 
when passing from NLO to NNLO is also seen in Ref.~\cite{thesis} by 
employing the Weinberg scheme. 
The $^3S_1$ phase shifts are also reproduced better at NNLO than at NLO, while 
the $^3D_1$ phase shifts are somewhat worse described by the former. 
Our results for the  $^3S_1$ and $^3D_1$ phase shifts are quite similar to those obtained 
in Ref.~\cite{pavon06}, but not for $\epsilon_1$ where our outcome is closer to the Nijmegen PWA. 
The comparison is not so straightforward with the results of Ref.~\cite{phillipssw},  which depend
 very much on the type of chiral $NN$ potential used.
For the $^3S_1-{^3D_1}$ coupled partial waves we do not show the Born approximation results  in Fig.~\ref{fig:3sd1_a} because 
they are specially poor, see e.g. Refs.\cite{epe04,peripheral} for the $^3D_1$ phase shifts.

We can also predict from Eqs.\eqref{3s1_a} and \eqref{3sd1_a} the $^3S_1$ scattering length ($a_t$) and effective range ($r_t$). The former is given in terms of $\nu_1^{(11)}$, Eq.~\eqref{once_nu0}, as 
\begin{align}
a_t&=-\frac{m \nu_1^{(11)}}{4\pi}~.
\end{align} 
Regarding $r_t$ we can proceed similarly as discussed in detail in Ref.~\cite{gor2013} 
where the following expression  is derived, 
\begin{align}
r_t=-\frac{m}{2\pi^2 a_t}\int_{-\infty}^L dk^2\frac{\Delta_{11}(k^2)D_{11}(k^2)}{(k^2)^2}\left\{\frac{1}{a_t}+\frac{4\pi k^2}{m}g_{11}(0,k^2)  \right\}-\frac{8}{m}\int_0^\infty dq^2\frac{\nu_{11}(q^2)-\rho(q^2)}{(q^2)^2}~,
\label{3s1.rt}
\end{align}
This equation also exhibits a correlation between $r_t$ and $a_t$, although in a more complicated manner than 
for the $^1S_0$ partial wave, as shown in Eq.~\eqref{expvs.1s0}, because $\nu_{11}(A)$ depends nonlinearly on $D_{11}(A)$. 
 
Another observable that we also consider is the slope at threshold of $\epsilon_1$, 
indicated as $a_\varepsilon$, and defined by  
\begin{align}
a_\varepsilon=\lim_{A\to 0^+}\frac{\sin 2\epsilon_1}{A^\frac{3}{2}}=1.128~M_\pi^{-3}~,
\label{avarepsilon}
\end{align}
where the numerical value is deduced from the Nijmegen PWA phase shifts. From the DRs in Eq.~\eqref{3sd1_a} we obtain 
the following expression for $a_\varepsilon$,
\begin{align}
a_\varepsilon&=\frac{m}{4\pi^2}\int_{-\infty}^L dk^2\frac{\Delta_{ij}(k^2)D_{ij}(k^2)}{(k^2)^2}~.
\label{aep.predicted}
\end{align}

It is also interesting to diagonalize the $^3S_1-{^3D_1}$ $S$-matrix around the deuteron pole position. This can 
be done by means of a real orthogonal matrix \cite{swart},
\begin{align}
{\cal O}&=\left(
\begin{array}{ll}
 \cos \varepsilon_1 & -\sin\varepsilon_1 \\
\sin\varepsilon_1 & \cos \varepsilon_1
\end{array}
\right)~.
\label{mat.ort}
\end{align}
Such that
\begin{align}
S&={\cal O}\left(
\begin{array}{ll}
S_0 & 0 \\
0 & S_2
\end{array}
\right) {\cal O}^T~,
\label{eigen3s1}
\end{align}
with $S_0$ and $S_2$ the $S$-matrix eigenvalues.  The asymptotic $D/S$ ratio of the deuteron, $\eta$, 
can be expressed in terms of $\varepsilon_1$ as
\begin{align}
\eta=-\tan \varepsilon_1~.
\label{eta.ep}
\end{align}
The residue of $S_0$ at the deuteron pole position is denoted by $N_p^2$,
\begin{align}
N_p^2&=\lim_{A\rightarrow k_d^2} \left(\sqrt{-k_d^2}+i\sqrt{A}\right) S_0~.
\end{align} 

As discussed in  Ref.~\cite{cohen}  the shape parameters are a good testing ground for the range of 
applicability of the underlying EFT. 
We then study our results for the shape parameters of the lowest 
eigenphase $\delta_0$ (also called $^3S_1$ eigenphase), Eq.~\eqref{eigen3s1}, with the diagonalization of the 
$S$-matrix performed in the physical region $A\geq 0$,\footnote{This can also be done in terms of an 
orthogonal matrix Eq.~\eqref{mat.ort} because of two-body unitarity.}
 \begin{align}
\sqrt{A}\,\text{cot}\delta_0=-\frac{1}{a_t}+\frac{1}{2}r_t A +\sum_{i=2}^{10}v_i A^i+{\cal O}(A^{11})~.
\label{3s1.ere}
\end{align}

\begin{table}
\begin{center}
\begin{tabular}{|l|l|l|l|l|l|}
\hline
    & $a_t$ [fm]  & $r_t$ [fm]  & $\eta$ & $ N_p^2$ [fm$^{-1}$] & $a_\varepsilon$ [$M_\pi^{-3}$] \\
\hline
NLO     & 5.22 & 1.47 & 0.0295 & 0.714 & 1.372 \\
\hline
NNLO-I & $5.52(3)$ & $1.89(3)$ & $0.0242(3)$ & $0.818(10)$& $1.270(9)$\\
\hline
NNLO-II & $5.5424^\star$ & $ 1.759^\star$ & $0.02535(13)$ & $0.78173(2)$ & $1.293(8)$\\
\hline
Ref.~\cite{swart}& $5.4194(20)$ & $1.7536(25)$ & $0.0253(2)$ &  $0.7830(15)$ & \\
\hline
Ref.~\cite{thesis} & $5.424$ & $1.753$ & 0.0245 & & \\ 
\hline
\end{tabular}
\caption{Values for $a_t$, $r_t$, $\eta$, $N_p^2$ and $a_\varepsilon$. The results 
 predicted from Eqs.~\eqref{3s1_a} 
 and \eqref{3sd1_a} are given in the second (NLO) and third row (NNLO-I). 
The values given in the fourth row (NNLO-II) are obtained once $a_t$ and $r_t$ are fixed to the experimental 
 figures, which is indicated by a star on top of the values. 
We also show the results from Refs.~\cite{swart} and \cite{thesis} in the fifth and sixth rows, respectively. 
\label{table:eta}}
\end{center}
\end{table}

The scattering length and effective range in the previous equation are the same as given above because 
coupled-wave effects with the $^3D_1$ only affects the shape parameters $v_i$, $i\geq 2$. The values obtained at NLO and NNLO from Eqs.~\eqref{3s1_a} and \eqref{3sd1_a} for $a_t$, $r_t$, $\eta$, 
$N_p^2$ and $a_\varepsilon$ are shown in Table~\ref{table:eta} in the 
second and third rows, respectively. 
We observe that the numbers at NNLO (indicated by NNLO-I) are already 
 rather close to those of Ref.~\cite{swart}, obtained from the Nijmegen PWA of $n p$ data,
 and Ref.~\cite{thesis}. 
It is interesting to remark that our value for $r_t$ is a prediction  in terms of only one  
subtraction constant (fixed by the deuteron pole position) and 
$NN$ forces stemming from $\pi N$ physics. This value deviates from experiment 
 $r_t=1.759\pm 0.005$~fm \cite{thesis} around a $10\%$ at NNLO ($\sim 20\%$  at NLO), while the relative 
 experimental error is  around $3\%$. 
 Other determinations for the parameter $\eta$,  not shown in Table~\ref{table:eta}, are $\eta=0.0256(4)$ \cite{rodning}, 
$\eta=0.0271(4)$ \cite{ericson:82}, 
$\eta=0.0263(13)$ \cite{conzett:79} and 
$\eta=0.0268(7)$ \cite{martorell}.

\begin{table}
\begin{center}
\begin{tabular}{|l|l|l|l|l|l|}
\hline
    & $v_2$  & $v_3$   & $v_4$ & $v_5$ & $v_6$    \\
\hline
NLO & -0.10572(12)& 0.8818(11) & $-5.427(11)$  & 36.73(11) & $-259.9(1.1)$ \\
\hline
NNLO-I & 0.157(22)& 0.645(9) & $-3.41(13)$  & 23.2(8) & $-161(6)$ \\
\hline
NNLO-II & $0.0848(4)$ & $0.762(7)$ & $-4.33(2)$ & $29.0(2)$ & $-198(2)$\\
\hline
Ref.~\cite{swart} & $0.040(7)$ & $0.673(2)$ & $-3.95(5)$ & $27.0(3)$ & \\
\hline
Ref.~\cite{thesis} & $0.046$ & $0.67$ & $-3.9$ & & \\
\hline
\end{tabular}
\caption{Values for the shape parameters $v_i$, $i=2,\ldots,6$ in units of fm$^{2i-1}$.
 The results 
 predicted from Eqs.~\eqref{3s1_a}
 and \eqref{3sd1_a} are given in the second  (NLO) and third row (NNLO-I). 
The errors for the NLO results correspond  entirely to the numerical accuracy in the calculation. 
Those values corresponding to NNLO-II are given in the fourth row. 
The values from Refs.~\cite{swart} and \cite{thesis} appear in the fifth and sixth rows, in order.
\label{table:vs3s1a}}
\end{center}
\end{table}
\begin{table}
\begin{center}
\begin{tabular}{|l|l|l|l|l|}
\hline
    &  $v_7$ & $v_8\times 10^{-1}$  & $v_{9}\times 10^{-2}$  & $v_{10}\times 10^{-3}$   \\
\hline
NLO &  1867(11) & $-1375(11)$ & $1008(11)$ & $-760(12)$ \\
\hline
NNLO-I & 1161(41) & $-840(30)$ & $625(22)$ & $-463(17)$ \\ 
\hline
NNLO-II& $1426(13)$ & $-1015(15)$ & $764(17)$ & $-545(20)$\\
\hline 
\end{tabular}
\caption{Values for the shape parameters $v_i$, $i=7,\ldots,10$ in units of fm$^{2i-1}$. 
For the meanings of the rows see Table~\ref{table:vs3s1a}.
\label{table:vs3s1b}}
\end{center}
\end{table}

The values for the shape parameters $v_i$, $i=2,\ldots,6$, are given in Table~\ref{table:vs3s1a} and 
for $i=7,\ldots,10$ in Table~\ref{table:vs3s1b}. Up to 
our knowledge the values of the shape parameters with $i>5$ were not given before. We detailed in 
Appendix \ref{appen:vs} the numerical method that allows us to perform the appropriate 
derivatives up to so high order.\footnote{For example in Ref.~\cite{swart} it is stated that their numerical 
set up is not precise enough to calculate  $v_6$ and that it already casts doubts about the numerical 
 accuracy for $v_5$.}  
 We could  have also given shape parameters of even higher orders
 within a numerical precision of a few per cent, 
but this is skipped because its apparent little relevance in practice.
 One can appreciate the numerical precision in the calculation of the shape parameters by considering the errors in 
 Tables~\ref{table:vs3s1a} and \ref{table:vs3s1b} for the NLO results, which entirely  correspond to  the numerical accuracy.
 Notice that for the highest shape parameter shown,  $v_{10}$, its relative error is 1.5$\%$, 
just slightly worse than for $v_9$ with a relative error of 1.1\%. 
We then see that by increasing the order of the shape parameter 
the numerical accuracy only worsens little by little. 
Morever,  the errors at NNLO take into account additionally 
the variation in the results from the different sets of $c_i$'s employed and the dependence in the 
input for starting the iterative process.
 For the shape parameters with large order, $i\geq 5$,  their absolute values increase 
typically as ${\cal O}(1/M_\pi)^{2i-1}$, which is the expected behavior for long-range interactions
 mediated by OPE. It is clear from Table~\ref{table:vs3s1a} that the shape parameters $v_i$, $i=2,\ldots,5$ 
predicted by the NNLO-I calculation (third row) are typically
 closer to the values of Refs.~\cite{swart,thesis} than those at NLO (second row). 
This is a positive feature indicating a well-behaved expansion of the results obtained by 
applying the $N/D$ method with the discontinuity $\Delta(A)$ expanded in BChPT.

According to  the power counting for the subtraction constants, Eq.~\eqref{summarypwc}, at NNLO it is appropriate
 to consider twice-subtracted DRs. 
For the $^3S_1-{^3D_1}$ system this implies to take into account two more free parameters for the $^3S_1$ 
wave and one more for the mixing partial wave.
 The three parameters for the $^3S_1$ wave are fixed in terms 
of the experimental values of $k_d^2$, $r_t$ and $a_t$.
The DR for the $^3D_1$ wave is the same as in Eq.~\eqref{3sd1_a}. 
 The twice-subtracted DRs taken now regarding the $^3S_1$ partial wave are
\begin{align}
D_{11}(A)&=1-\frac{A}{k_d^2}-\nu_1^{(11)} \,A(A-k_d^2) g_{11}^{(d)}(A,0;1)
-\nu_2^{(11)} \,A(A-k_d^2) g_{11}(A,k_d^2)\nn\\
&+\frac{A(A-k_d^2)}{\pi}\int_{-\infty}^L dk^2\frac{\Delta_{11}(k^2)D_{11}(k^2)}{(k^2)^2} 
g_{11}^{(d)}(A,k^2;2)~,\nn\\
N_{11}(A)&=\nu_1^{(11)}+\nu_2^{(11)}\,A+\frac{A^2}{\pi}\int_{-\infty}^L dk^2\frac{\Delta_{11}(k^2)D_{11}(k^2)}{(k^2)^2(k^2-A)}~,\nn\\
\nu_1^{(11)}&=-\frac{4\pi a_t}{m}~,\nn\\
\nu_2^{(11)}&=\frac{\nu_1^{(11)}}{\nu_1^{(11)} \,k_d^2\, g_{11}(0,k_d^2)-1}\left\{
\frac{1}{k_d^2}+a_t\Bigg(
\frac{4 k_d^2}{m}\int_0^\infty dq^2\frac{\nu_{11}(q^2)-\rho(q^2)}{(q^2)^2(q^2-k_d^2)}+\frac{1}{\sqrt{-k_d^2}}
-\frac{r_t}{2}\Bigg)\right. \nn\\
&\left. +\frac{k_d^2}{\pi}\int_{-\infty}^L dk^2 \frac{\Delta_{11}(k^2)D_{11}(k^2)}{(k^2)^2}g_{11}(k_d^2,k^2)
\right\}~.
\label{2subs.3s1}
\end{align}

For the mixing partial wave the DRs are 
\begin{align}
D_{12}(A)&=1-\frac{A}{k_d^2}-\nu_2^{(12)} A(A-k_d^2)g_{12}(A,k_d^2)
+\frac{A(A-k_d^2)}{\pi}\int_{-\infty}^L dk^2\frac{\Delta_{12}(k^2)D_{12}(k^2)}{(k^2)^2}g_{12}^{(d)}(A,k^2;2)~,\nn\\
N_{12}(A)&=\nu_2^{(12)} A+\frac{A^2}{\pi}\int_{-\infty}^L dk^2\frac{\Delta_{12}(k^2)D_{12}(k^2)}{(k^2)^2(k^2-A)}~,
\label{2subs.mix}
\end{align}
 The results obtained by solving the IEs of Eqs.~\eqref{2subs.3s1}, \eqref{2subs.mix} and Eq.~\eqref{3sd1_a} 
with $\ell_{22}=2$ are denoted in the following by NNLO-II and correspond to the (red) hatched areas with crossed lines 
in Fig.~\ref{fig:3sd1_a}. 
It turns out that we cannot obtain a solution of the resulting IE for $D_{12}(A)$ by 
implementing any arbitrary value for $\nu_2^{(12)}$.
 We have further checked this statement by employing the following expression for $\nu_2^{(12)}$,
\begin{align}
\nu_{2}^{(12)}&=\frac{\Theta}{2\pi}\int_{-\infty}^L dk^2\frac{\Delta_{ij}(k^2)D_{ij}(k^2)}{(k^2)^2}~.
\end{align}
Here, the integral is the same as in Eq.~\eqref{aep.predicted}, so that if we take $ \Theta=1$ we would simply 
rewrite the IE of Eq.~\eqref{3sd1_a} in terms of twice-subtracted DRs. Then, we  vary $\Theta$ and whenever we find 
a meaningful solution the obtained value for $a_\varepsilon=m\nu_2^{(12)}/2\pi$ is always basically the same,  
$a_\varepsilon\simeq 1.30~M_\pi^{-3}$.
 In our opinion this difficulty in our approach to reproduce the value for $a_\varepsilon$ 
 that follows from the Nijmegen PWA, Eq.~\eqref{avarepsilon}, casts doubts on this number. 
  Notice that  the calculated values for $\epsilon_1$ at low momentum, 
e.g. for $\sqrt{A}\lesssim 100$~MeV, lie on top of the curve for the Nijmegen PWA results
 as shown in the third panel of Fig.~\ref{fig:3sd1_a} by the coincident hatched and filled areas that overlap the Nijmegen PWA line. 
  The phase shifts and $\epsilon_1$  are quite similar to the NNLO-I results in terms of just one free parameter.
Nevertheless,  the $^3S_1$ phase shifts for  NNLO-II are closer to the Nijmegen PWA ones  
at lower energies, but the change for this S-wave by going from once- to twice-subtracted DRs
 is much less notorious than in the case of the partial wave $^1S_0$, discussed in Sec.~\ref{1s0}. 
We can also see in the fourth row of
 Table~\ref{table:eta} that the NNLO-II values for $\eta$ and $N_p^2$ are compatible 
with those of  Ref.~\cite{swart}, which is quite remarkable.
 The value for $a_\varepsilon$ mentioned above is shown in 
the last column of the same table.
 The shape parameters are shown in the forth rows of Tables~\ref{table:vs3s1a} 
and \ref{table:vs3s1b}, where we observe 
a better agreement with the numbers given in Ref.~\cite{swart} for 
 $v_4$ and $v_5$ than for $v_2$ and $v_3$. The variation of the values
 between NNLO-I and NNLO-II for the higher order shape parameters allows us to guess 
in a  conservative way the systematic uncertainty affecting their calculation. 

On the other hand, we would like  to elaborate further on the fact that at NNLO the results for the $^3D_1$ phase 
shifts do not still offer a good reproduction of the Nijmegen PWA ones, being even worse than those obtained 
at NLO.
 In Ref.~\cite{epe04} one can find a discussion on the difficulties arisen in their calculation 
 because of the large values of the  NLO $\pi N$ 
counterterms, namely $c_3$ and $c_4$, in order to reproduce simultaneously the $D$ and $F$ waves 
within the Weinberg scheme using the NNLO  chiral potential  
calculated in dimensional regularization. 
 Considering this observation 
 we obtain that when all the $c_i=0$ our NNLO result 
for $\delta_2$ is then essentially the same as the NLO one in Fig.~\ref{fig:3sd1_a}, 
corresponding to the (magenta) dot-dashed line. 
 In view of this, we study now the influence in the results by including one more subtraction in the DRs for $^3D_1$ 
with the aim of determining whether this worsening 
is an effect that can be counterbalanced in a natural way at ${\cal O}(p^4)$. 
 In this way we use the same twice-subtracted DRs for $^3S_1$ and the mixing partial wave 
given in Eqs.~\eqref{2subs.3s1} and \eqref{2subs.mix}, respectively, while  
 the following three-time subtracted DRs are used for the $^3D_1$
\begin{align}
D_{22}(A)&=1-\frac{A}{k_d^2}+\delta_3^{(22)} A(A-k_d^2)
-\nu_3^{(22)} A(A-k_d^2)^2\frac{\partial g_{22}^{(d)}(A,0;2)}{\partial k_d^2}\nn\\
&+\frac{A(A-k_d^2)^2}{\pi}\int_{-\infty}^L dk^2 \frac{\Delta_{22}(k^2)D_{22}(k^2)}{(k^2)^3} 
\frac{\partial g_{22}^{(d)}(A,k^2;3)}{\partial k_d^2}~,\nn\\
N_{22}(A)&=\nu_3^{(22)} A^2+\frac{A^3}{\pi}\int_{-\infty}^L dk^2 \frac{\Delta_{22}(k^2)D_{22}(k^2)}{(k^2)^3(k^2-A)}~,
\label{3d1.extra}
\end{align}
with two additional subtraction constants  $\delta_3^{(22)}$ and $\nu_3^{(22)}$. 
Considering the results  obtained from the twice-subtracted DRs for all the waves in the system
 $^3S_1-{^3D_1}$, and denoting by  $\hat{D}_{22}(A)$ the function $D_{22}(A)$ obtained then, we have 
the following predictions for the subtraction constants $\delta_3^{(22)}$ and $\nu_3^{(22)}$, 
\begin{align}
\nu_3^{\mathrm{pred}}&=\frac{1}{\pi}\int_{-\infty}^L dk^2 \frac{\Delta_{22}(k^2)\hat{D}_{22}(k^2)}{(k^2)^3}~,\nn\\
\delta_3^{\mathrm{pred}}&=\frac{1}{\pi}\int_{-\infty}^L dk^2 \frac{\Delta_{22}(k^2) \hat{D}_{22}(k^2)}{(k^2)^2}g_{22}^{(d)}(k^2,k_d^2;2)~.
\label{nudelta2pre}
\end{align}
The numerical values that stem from the previous expressions are 
$\delta_3^{\mathrm{pred}}\simeq 1~m_\pi^{-4}$ and $\nu_3^{\mathrm{pred}}\simeq -2.5~m_\pi^{-6}$~. A fit 
to the $^3D_1$ phase shifts only requires to vary $\nu_3^{(22)}$ around that value with the final 
result $\nu^{(22)}_3= -2.05(5)$~$m_\pi^{-6}$, while 
$\delta^{(22)}_3$ stays put.
 Then, it is only necessary a relatively small change of around 20\% in 
 $\nu_3^{(22)}$ 
 from the one predicted by the twice-subtracted DRs in Eq.~\eqref{nudelta2pre} in order to end with a much better
 reproduction of the 
$^3D_1$ phase shifts that is compatible with the Nijmegen PWA, as shown 
 by the hatched areas with (gray) parallel lines in Fig.~\ref{fig:3sd1_a} (denoted as 
NNLO-III results).
 Since the reproduction of 
the $^3S_1$ phase shifts and mixing angle $\epsilon_1$ is the same as the one obtained already in terms of the 
twice-subtracted DRs, the so-called NNLO-II results, we do not show them 
nor the values for the other parameters given  in Tables~\ref{table:eta}, \ref{table:vs3s1a} and \ref{table:vs3s1b}, 
that would be also basically coincident with the NNLO-II ones in these tables. 

 We now elaborate on the difference between the value of $\nu_3^{(22)}$ fitted and the one predicted, $\nu_3^{\mathrm{pred}}$. 
According to the power counting of Sec.~\ref{nschpt}, cf. Eq.~\eqref{summarypwc}, $\nu_3^{(22)}={\cal O}(p^{-1})$
 in our present NNLO calculation. If we consider that this difference is an effect that stems from the ${\cal O}(p^4)$ 
contributions to $\Delta(A)$, which are not considered here yet, one would have that  
$\delta\nu_3 \equiv \nu_3^{(22)}-\nu_3^{\mathrm{pred}}\simeq 0.6~M_\pi^{-6} 
={\cal O}(p^0)$. It also follows then that nominally
 $\delta \nu_3/\nu_3^{\mathrm{pred}}={\cal O}(p)$ and taking into account the 
 numerical values
\begin{align}
\frac{\delta \nu_3}{\nu_3^{\mathrm{pred}}}=0.23\sim \frac{M_\pi}{\Lambda}~,
\end{align}
we can estimate that $\Lambda\sim 4 M_\pi$, which is 
 similar to the estimate of $\Lambda$ obtained in Sec.~\ref{1s0} 
for  the  $^1S_0$ partial wave.
 As a result, $\delta \nu_3$ is consistent with a naturally sized ${\cal O}(p^4)$ effect.

The fact that the matrix of limiting values
\begin{align}
M_{ij}=\lim_{A\to -\infty}\frac{\Delta_{ij}(A)}{(-A)^{3/2}}
\label{mij.3s1}
\end{align}
has two negative eigenvalues  is certainly related with the possibility of obtaining 
meaningful DRs with only one free parameter as first obtained in this section. 
We base this statement on the necessity condition of Ref.~\cite{gor2013} 
in order to obtain meaningful once-subtracted DRs for $\lambda<0$,
 a condition also introduced in Sec.~\ref{nschpt}. 
Indeed, since the mixing between different partial waves is very small 
these eigenvalues are given in good approximation by $M_{11}$ and $M_{22}$; this rule applies 
indeed  not only to the $^3S_1-{^3D_1}$ coupled waves but to any other one.

\section{Coupled $^3P_2-{^3F_2}$ waves}
\label{pf2w}

\begin{figure}[h]
\begin{center}
\begin{tabular}{cc}
\includegraphics[width=.4\textwidth]{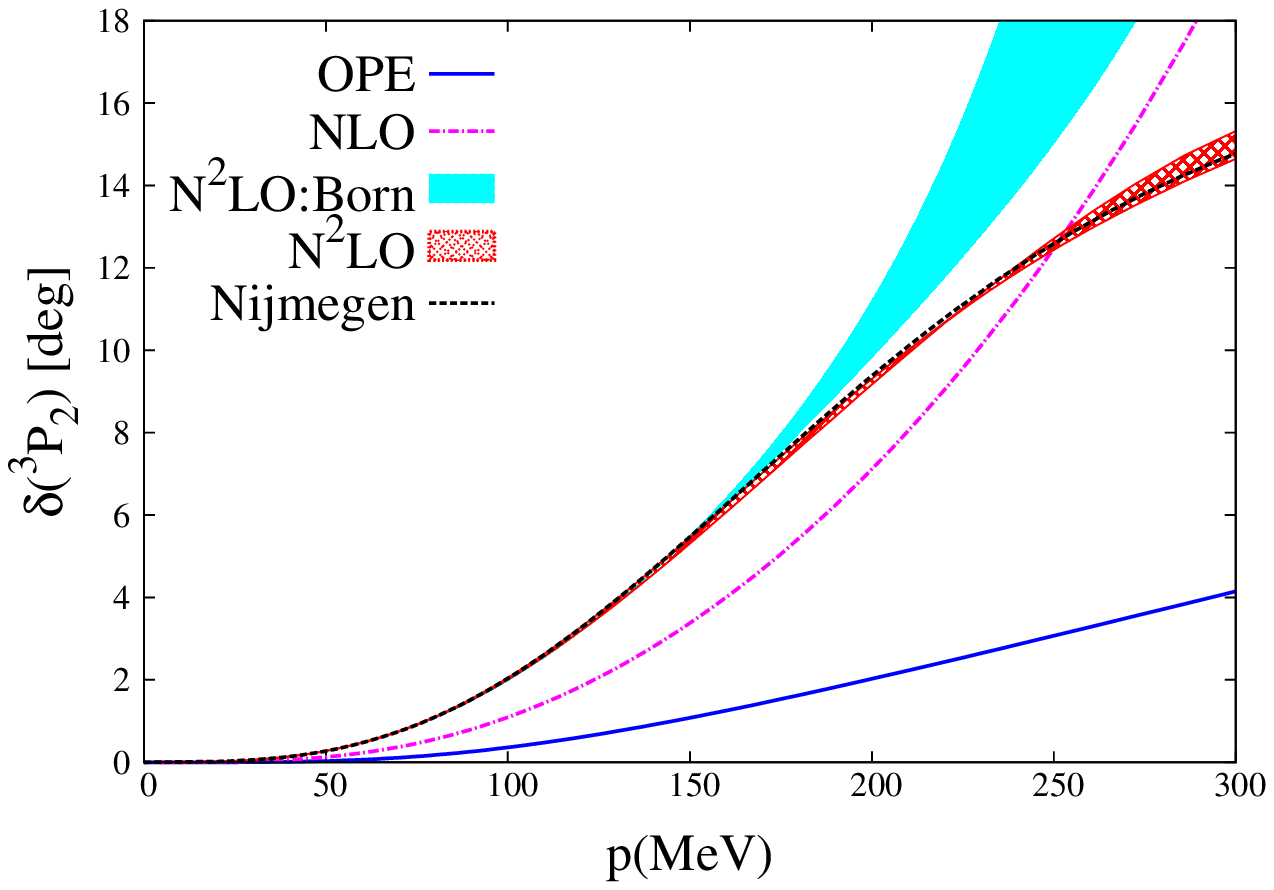} & 
\includegraphics[width=.4\textwidth]{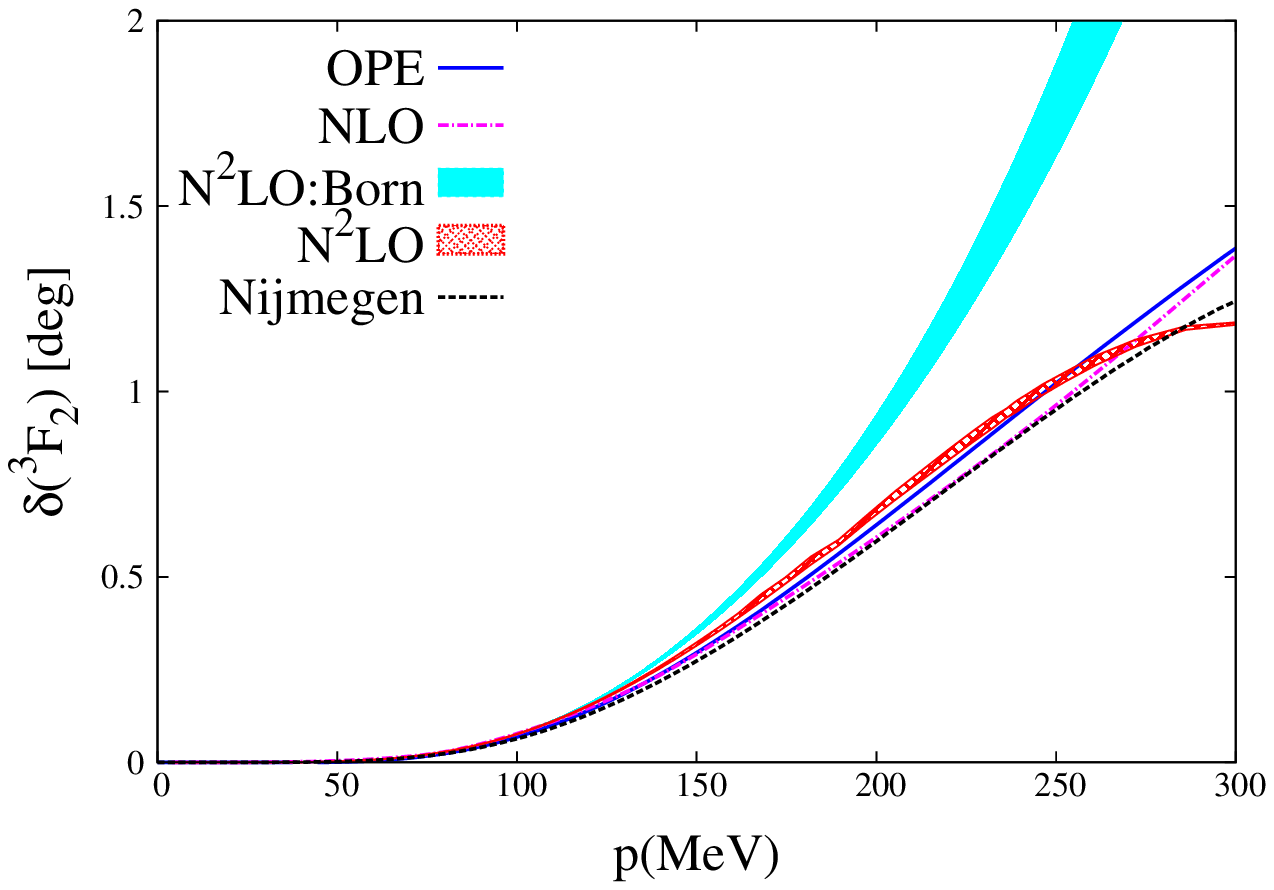}\\  
\includegraphics[width=.4\textwidth]{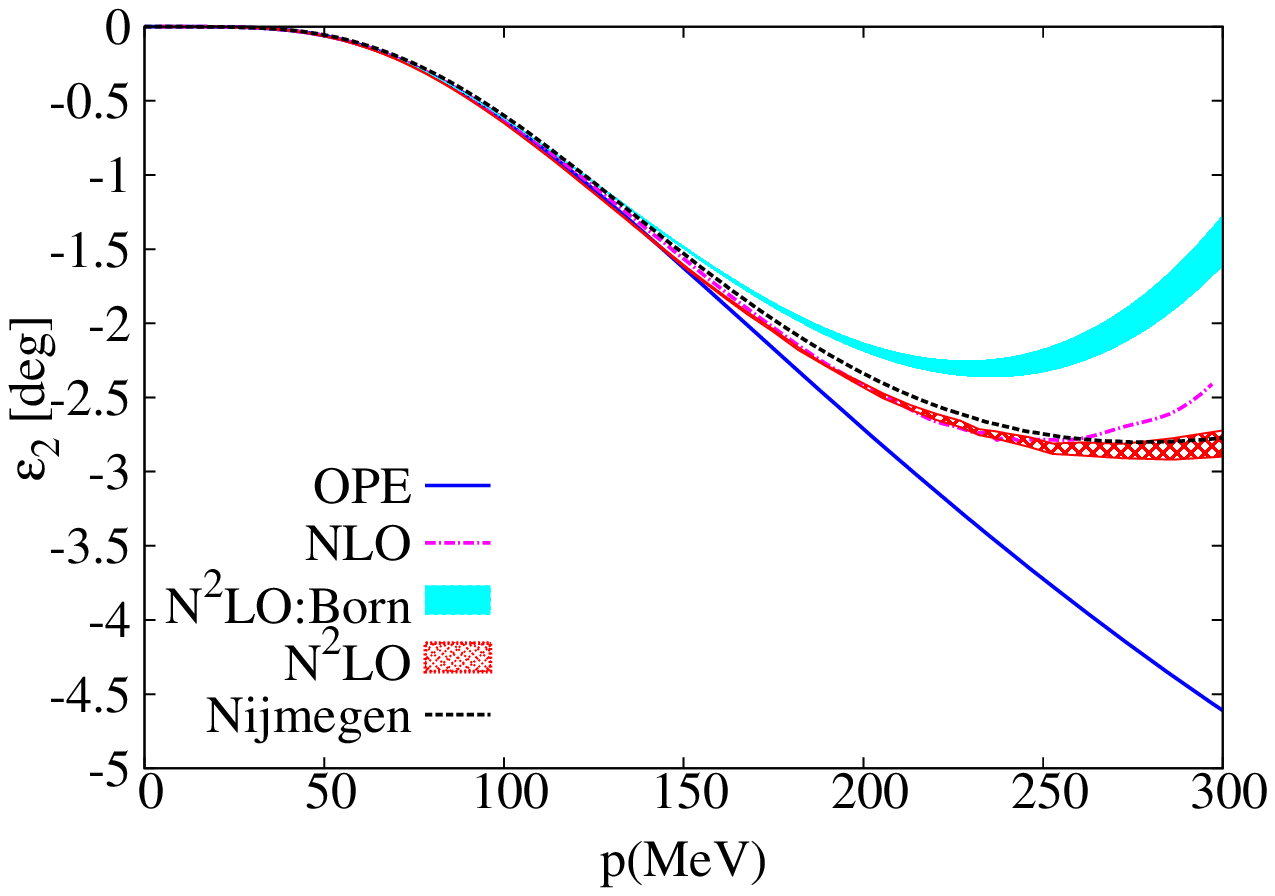}  
\end{tabular}
\caption[pilf]{\protect { (Color online.) From top to bottom and left to right: Phase shifts for $^3P_2$, $^3F_2$ and the mixing angle  $\epsilon_2$, respectively.
The (red) hatched areas  correspond to the NNLO results and the (cyan) filled bands are the leading Born approximation results.
 The NLO phase shifts and mixing angle are shown by the (magenta) dot-dashed lines and the LO ones are given by the 
 (blue) dotted lines.
 The Nijmegen PWA phase shifts correspond to the (black) dashed lines.
} 
\label{fig:3pf2} }
\end{center}
\end{figure}

We dedicate this section to the study of the coupled wave system $^3P_2-{^3F_2}$. By direct computation 
one has in this case that 
\begin{align}
\lambda_{11}=\lim_{A\to -\infty}\frac{\Delta_{11}(A)}{(-A)^{3/2}}>0~,
\label{couplambdap}
\end{align}
which requires one to consider DRs with more than one subtraction for the $^3P_2$ wave \cite{gor2013}.
 Indeed, similarly to the $^3P_0$ and $^3P_1$ partial waves, studied in Secs.~\ref{3p0} and \ref{3p1}, respectively,  
we need to take at least three subtractions in the DRs for the $^3P_2$ wave in order to obtain 
stable and meaningful results.
 Thus,   we have the following  three-time subtracted DRs for the $^3P_2$ wave, 
\begin{align}
\label{d.3p2}
D_{11}(A)&=1+\delta^{(11)}_2 A+\delta^{(11)}_3 A(A-C)-\nu^{(11)}_{2}\frac{A(A-C)^2}{\pi}\int_0^\infty dq^2\frac{\nu_{11}(q^2)}{(q^2-A)(q^2-C)^2}\nn\\
&-\nu^{(11)}_3\frac{A(A-C)^2}{\pi}\int_0^\infty dq^2\frac{\nu_{11}(q^2)q^2}{(q^2-A)(q^2-C)^2}\nn\\
&+\frac{A(A-C)^2}{\pi}\int_{-\infty}^L dk^2
\frac{\Delta_{11}(k^2)D_{11}(k^2)}{(k^2)^3} g_{11}(A,k^2,C;2)~,\\
\label{n.3p2}
N_{11}(A)&=\nu^{(11)}_2 A+\nu^{(11)}_3 A^2+\frac{A^3}{\pi}\int_{-\infty}^L dk^2 \frac{\Delta_{11}(k^2)D_{11}(k^2)}{(k^2)^3(k^2-A)}~.
\end{align}
With respect to the mixing and $^3F_2$  partial waves we use the standard formalism 
 for the coupled waves  given in Eqs.~\eqref{highdcc} and \eqref{highncc} with $\ell_{12}=2$ and $\ell_{22}=3$, respectively. 
As a result  2 and 3 subtractions are taken in order.

 As usual for the $P$ waves, we fix $\nu_2^{(11)}=4\pi a_V/m$  by requiring the exact reproduction  of the $^3P_2$ scattering volume 
extracted from the Nijmegen PWA  \cite{Stoks:1994wp},
\begin{align}
a_V=0.0964~M_\pi^{-3}~,
\end{align}
while $\nu_3^{(11)}$ is fitted to the results of this PWA. 
Regarding the subtraction constants $\delta_i^{(11)}$, $i=1,$~2, we  follow the principle of maximal smoothness in virtue of which we  fix $\delta^{(11)}_{2}=0$
 and fit $D_{11}^{(1)}(-M_\pi^2)$.\footnote{In the following we use 
$D_{ij}^{p-2}(-M_\pi^2)$ as free parameter in terms of which one can calculate $\delta_{p}^{(ij)}$ from Eq.~\eqref{taylor}.}  
 The resulting fitted values are: 
\begin{align}
D_{11}^{(11)}(-M_\pi^2)&=0.025(5)~M_\pi^{-2}~,\nn\\
\nu_3^{(11)}&=0.155(5)~M_\pi^{-6}~,\\
D_{22}^{(11)}(-M_\pi^2)&=0.011(4)~M_\pi^{-2}~,
\end{align}
 with the interval of values reflecting the dependence on the $c_i$'s chosen.
 The free parameter associated with  the mixing wave is fixed to its pure
 perturbative value, cf. Sec.~\ref{hpw}, $D_{12}(-M_\pi^2)=1$.

All in all the resulting phase shifts are shown by the (red) hatched areas in Fig.~\ref{fig:3pf2}. 
There we see a clear improvement at NNLO in the reproduction 
of the $^3P_2$ phase shifts  compared with the results at NLO, given by the (magenta) dot-dashed lines, so that now the (red) hatched area overlaps the Nijmegen PWA phase shifts.
 The $^3F_2$ phase shifts and mixing angle $\epsilon_2$ are reproduced with a similar quality to that already achieved at NLO.
 We also give by the (cyan) filled bands the results obtained by the leading Born approximation, Eq.~\eqref{deltab}, 
with $\Delta(A)$ calculated at NNLO. Due to the fact that the latter diverges as $(-A)^{3/2}$ for $A\to-\infty$ at least two 
subtractions have to be taken in the DR for $N_B(A)$, Eq.~\eqref{eq.nborn}. 
This is immediately accomplished for the $D$ and 
higher partial waves but for a $P$-wave with $\ell=1$ one needs to include one 
extra subtraction.
 In particular, for our present case we use Eq.~\eqref{n.3p2} with $D_{11}(A)\to 1$ and with $\Delta_{11}(k^2)$
 restricted to its two-nucleon irreducible contributions, with the subtraction constants $\nu^{(11)}_{2}$ and $\nu^{(11)}_{3}$ 
taking the same values as discussed before. 
We see that our full results provide a clear improvement in the reproduction of the Nijmegen PWA phase shifts and 
mixing angle  with respect to the Born approximation. 
 One should mention that the Born approximation  phase shifts for 
$^3F_2$ and $^3F_3$ have a striking resemblance to the full NNLO results of Ref.~\cite{thesis} obtained within the Weinberg 
scheme. 
 We have obtained this improvement without dismissing the strength of the TPE at NNLO, as advocated in Ref.~\cite{epe04}. 
 This makes that our full results are not so much sensitive to the particular set of $c_i$'s taken as 
previously thought in the literature from the results of Refs.~\cite{thesis,epe04}.

\section{Coupled $^3D_3-{^3G_3}$ waves}
\label{dg3w}

\begin{figure}[h]
\begin{center}
\begin{tabular}{cc}
\includegraphics[width=.4\textwidth]{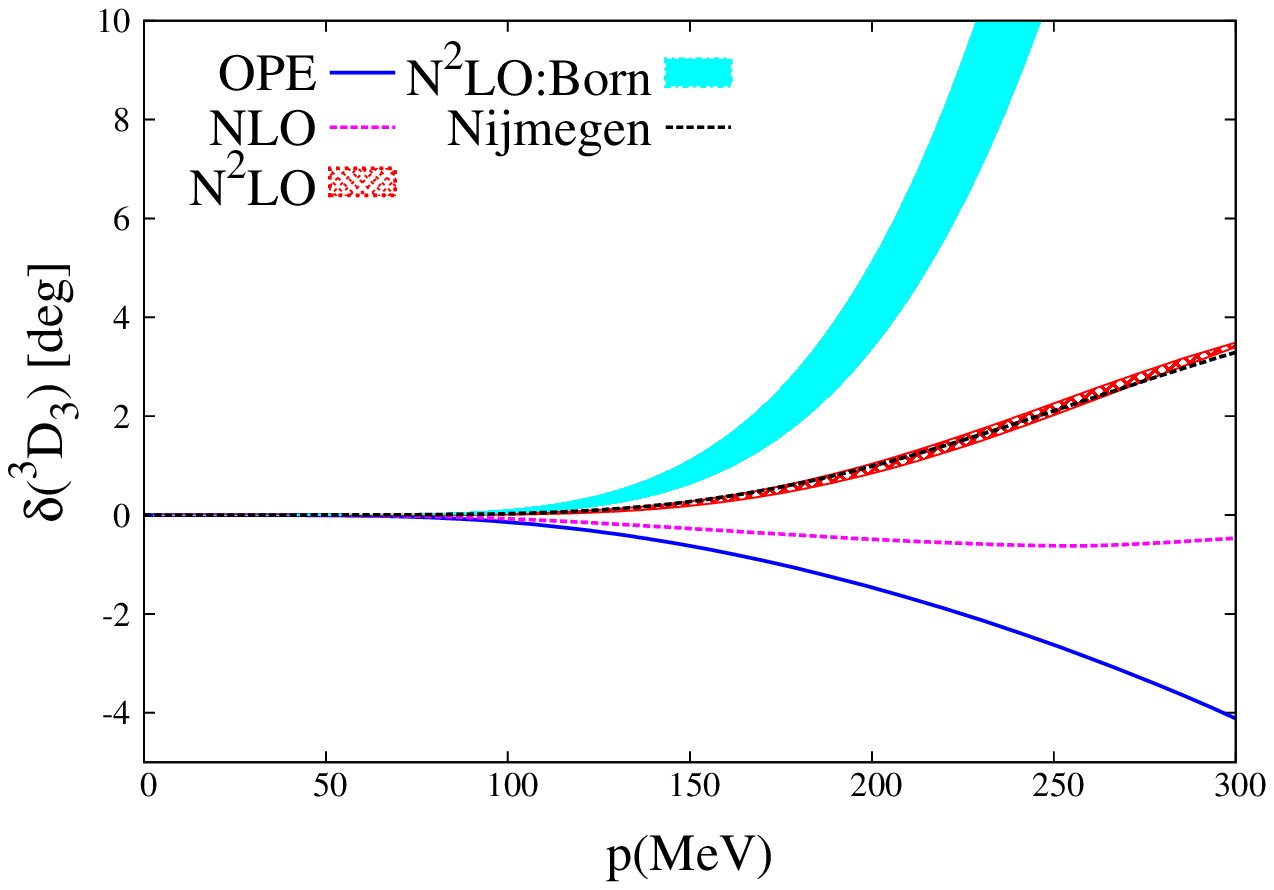} & 
\includegraphics[width=.4\textwidth]{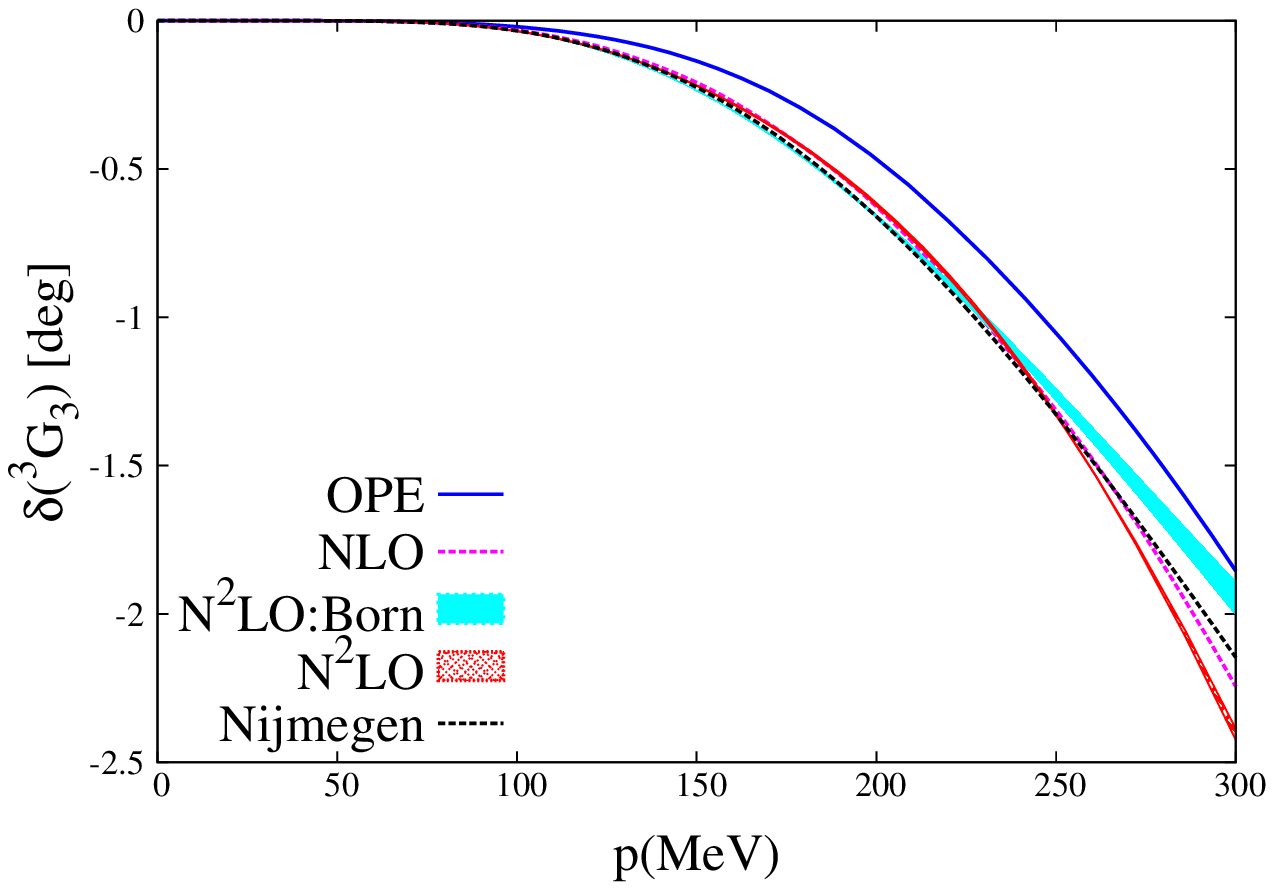}\\  
\includegraphics[width=.4\textwidth]{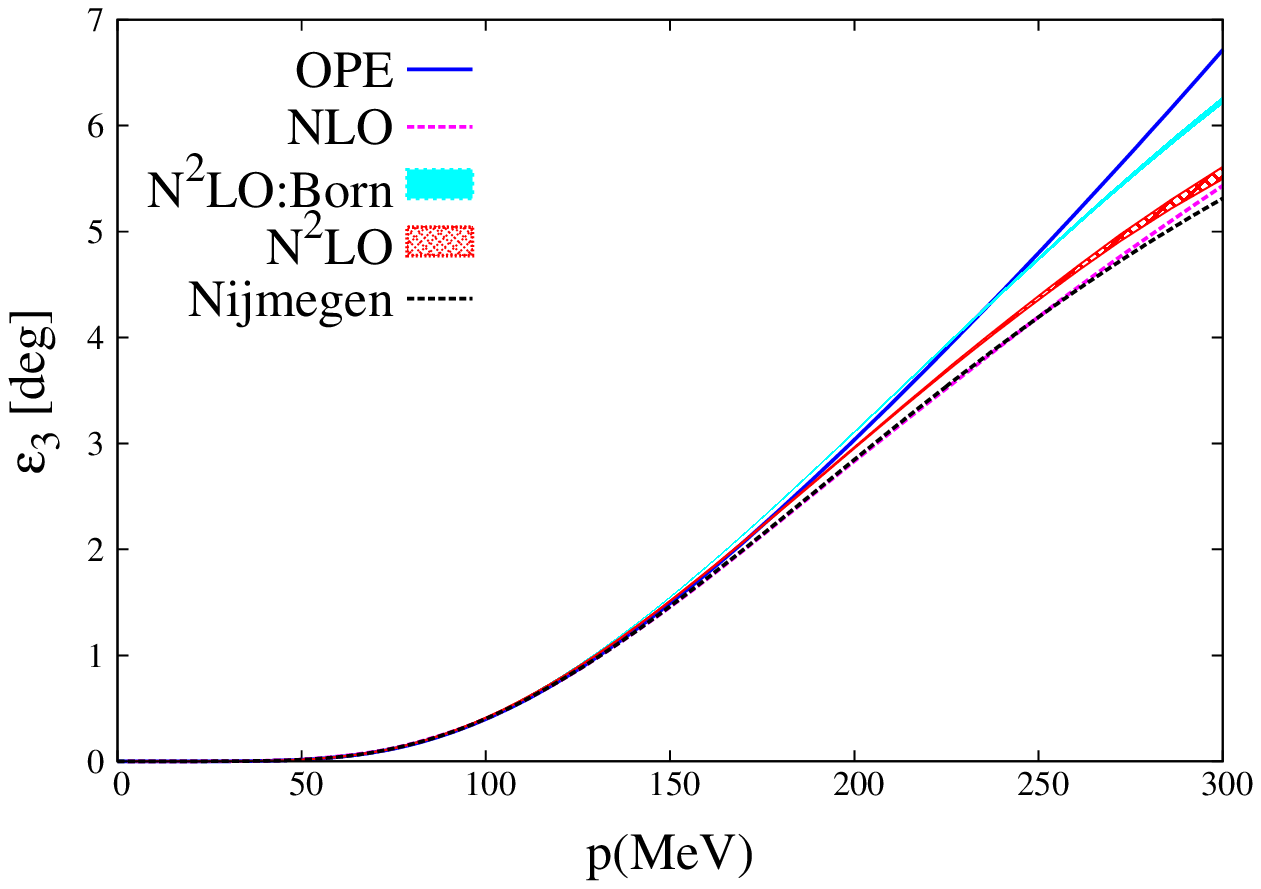}  
\end{tabular}
\caption[pilf]{\protect { (Color online.) From top to bottom and left to right: Phase shifts for $^3D_3$, $^3G_3$ and the mixing angle  $\epsilon_3$, in order.
The (red) hatched areas correspond to the NNLO results and the (cyan) filled bands are the leading order Born approximation.
 The NLO results are shown by the (magenta) dot-dashed lines and the LO ones are given by the 
 (blue) dotted lines. The Nijmegen PWA phase shifts correspond to the (black) dashed lines.}
\label{fig:3dg3} }
\end{center}
\end{figure}

For the study of the $^3D_3-{^3G_3}$ coupled waves we follow the formalism for coupled waves, Eqs.~\eqref{highdcc} and 
\eqref{highncc}, with $\ell_{11}=2$, $\ell_{12}=3$ and $\ell_{22}=4$, so that $\ell_{ij}$ subtractions are taken in the DRs for the  
coupled wave $ij$. 
 Regarding  the free parameters we follow the principle of maximal smoothness, although for the mixing wave 
the subtraction constants take their pure perturbative values. So that  we fit to data 
$D_{11}(-M_\pi^2)$ and $D_{22}^{(2)}(-M_\pi^2)$, with the resulting values:
\begin{align}
D_{11}(-M_\pi^2)&=0.90(5)~,\nn\\
D_{22}^{(2)}(-M_\pi^2)&=-0.09(1)~M_\pi^{-4}~,
\label{fit.3dg3}
\end{align}
 The interval of values in Eq.~\eqref{fit.3dg3} reflect the dependence on the set of  values considered for the $c_i$'s. 
The resulting phase shifts are shown by the (red) hatched areas in Fig.~\ref{fig:3dg3}.
 Importantly at NNLO the phase shifts for the $^3D_3$ wave   follow  closely the Nijmegen PWA phase shifts 
so that a remarkable improvement is obtained in comparison with both 
the NLO and Born results.
  Notice that this is accomplished without any need of dismissing the strength of
 TPE as directly obtained from the NLO $\pi N$ amplitudes. 
We have been able to improve the situation by taking into account the subtraction constant $ \delta_2$ or $D_{11}(-M_\pi^2)$, 
whose presence is required by the nonperturbative unitarity implementation\footnote{In more general terms, by generating the analytical properties 
associated with the RHC while respecting unitarity in the full amplitudes.} at NNLO, cf. Eq.~\eqref{summarypwc}.
  We also observe a good reproduction of the Nijmegen PWA results for the waves $^3G_3$ and $\epsilon_3$, 
which are already well reproduced at NLO \cite{gor2013} as shown  by the (magenta) dot-dashed lines.

\section{Coupled $^3F_4-{^3H_4}$ waves}
\label{gh4w}

\begin{figure}[h]
\begin{center}
\begin{tabular}{cc}
\includegraphics[width=.4\textwidth]{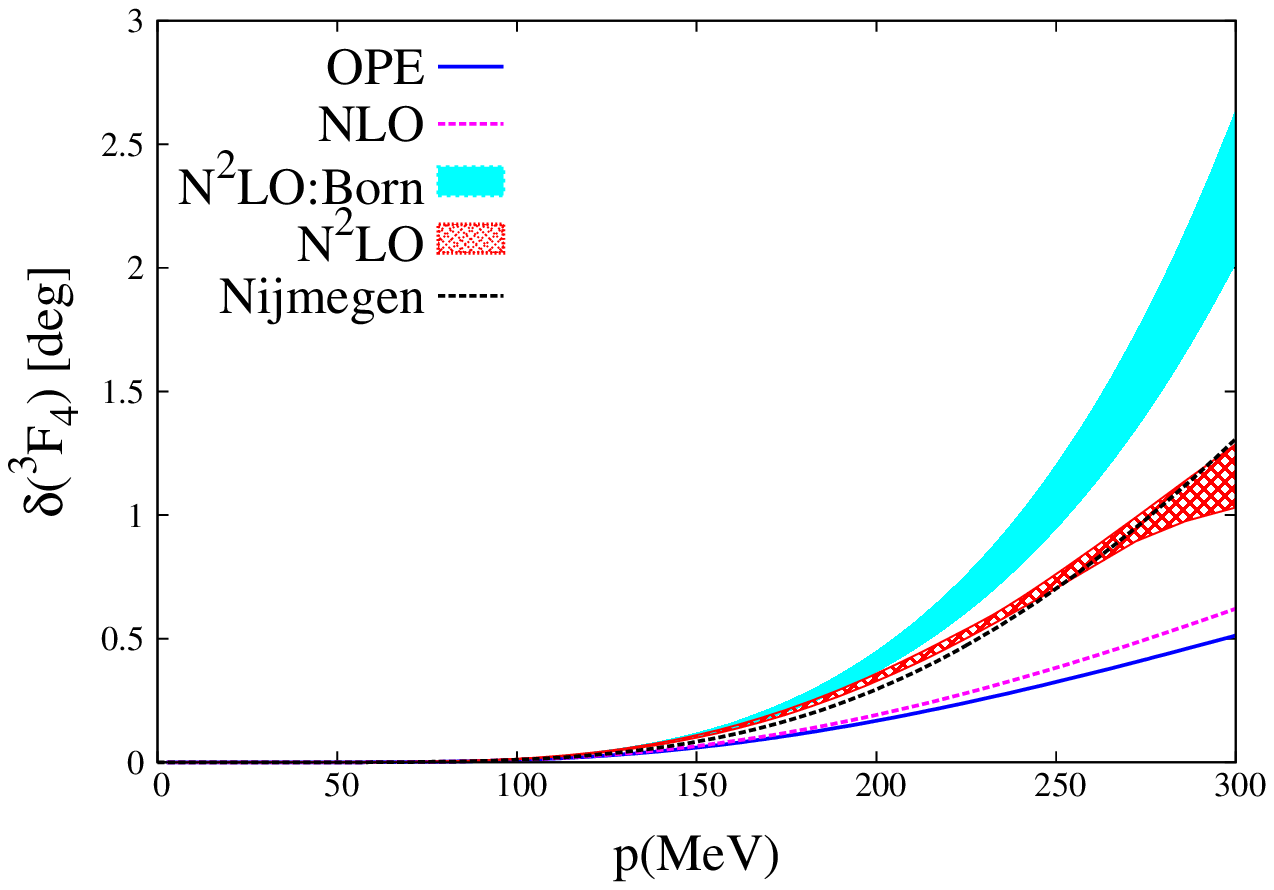} & 
\includegraphics[width=.4\textwidth]{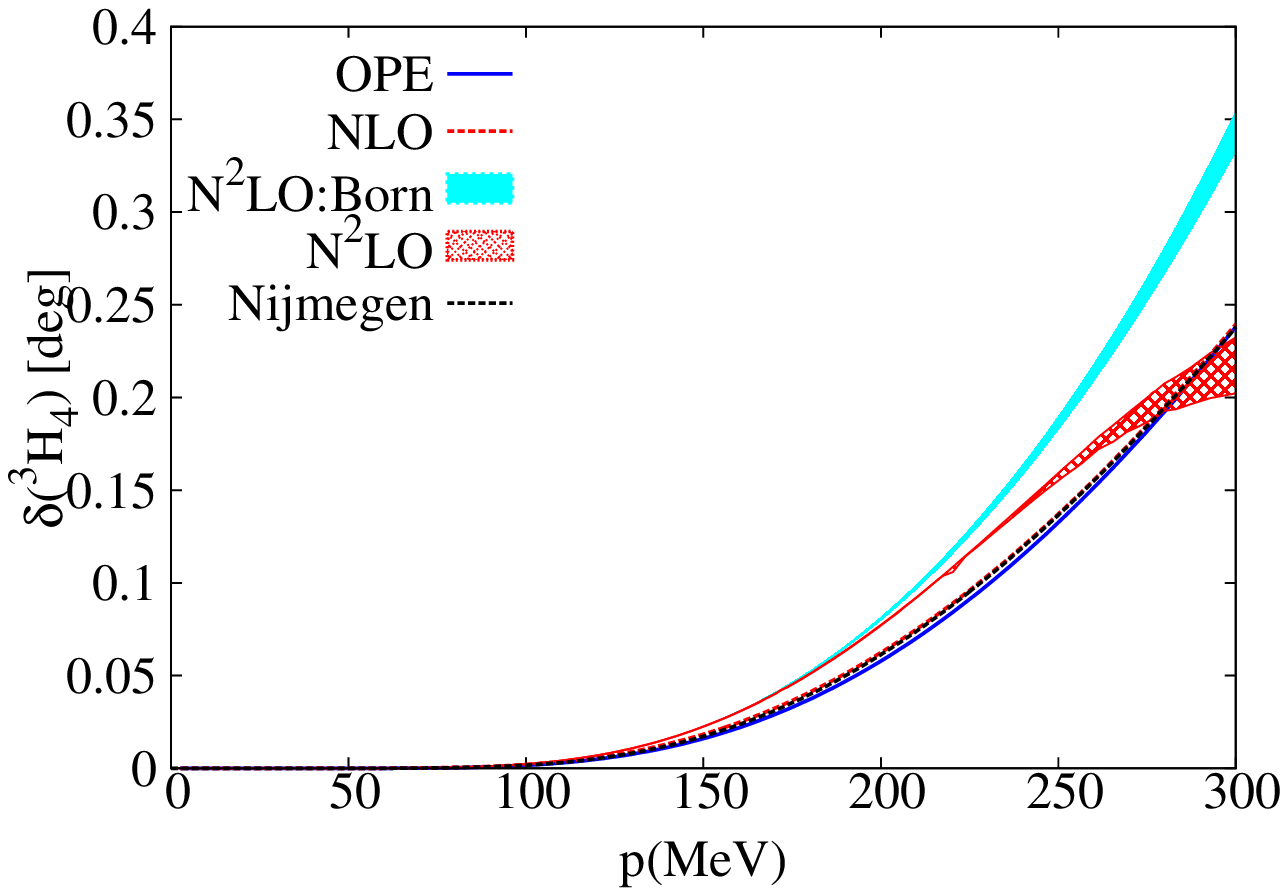}\\  
\includegraphics[width=.4\textwidth]{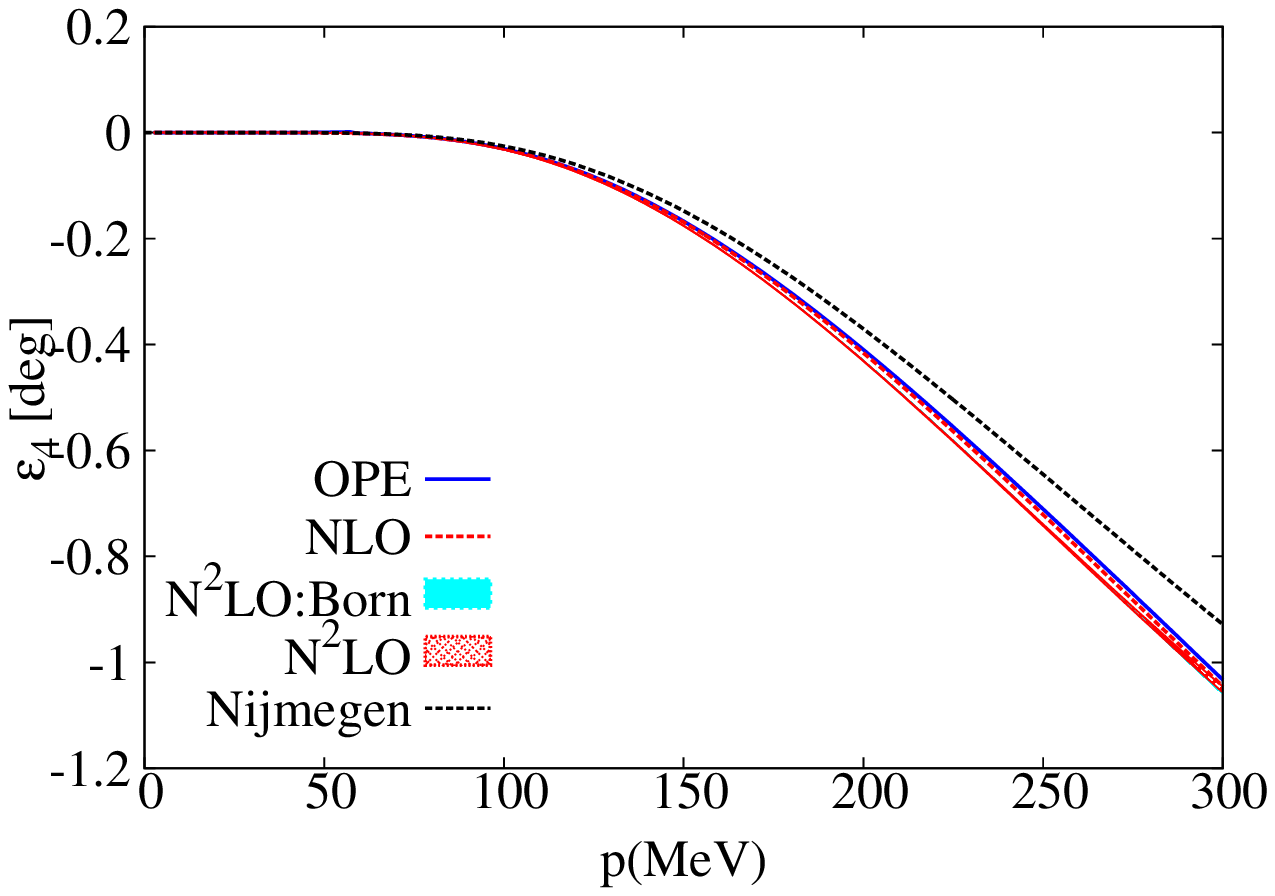}  
\end{tabular}
\caption[pilf]{\protect { (Color online.) From top to bottom and left to right: Phase shifts for $^3F_4$, $^3H_4$ and the mixing angle  $\epsilon_4$, in order.
The (red) hatched areas correspond to the NNLO results and the (cyan) filled ones to the leading
Born approximation.
 The NLO results are shown by the (magenta) solid line and the LO ones are given by the 
 (blue) dotted lines. 
The Nijmegen PWA phase shifts are given by the (black) dashed lines.}
\label{fig:3fh4} }
\end{center}
\end{figure}

The discussion of the $^3F_4-{^3H_4}$ coupled-wave system follows the  standard formalism for coupled waves, Eq.~\eqref{highdcc} and \eqref{highncc},
 with $\ell_{11}=3$, $\ell_{12}=4$ and $\ell_{22}=5$. The free parameters are then fitted to data according to the 
 principle of maximal smoothness. However, for $^3H_4$ and the mixing partial wave there is no improvement in the reproduction of data
 with respect to the situation in which the pure perturbative values are taken, so that 
at the end we only have to fit $D_{11}^{(1)}(-M_\pi^2)$  to the Nijmegen PWA results. 
The fitted value is
\begin{align}
D_{11}^{(1)}(-M_\pi^2)&= -0.009(3)~M_\pi^{-2}~.
\label{free.fh4}
\end{align}
 The resulting phase shifts and mixing angle are shown by the (red) hatched areas  in  Fig.\ref{fig:3fh4},
with the width of the band reflecting the dependence on values for the $\pi N$ NLO counterterms. 
One can observe  a clear improvement in the description of the $^3F_4$ phase shifts compared with the results from OPE (blue dotted lines), NLO 
(magenta dot-dashed lines) and leading Born approximation (cyan filled areas). 
 Similarly to the $^3D_3$ wave in the previous section, 
 this improvement is related with the effect of the subtraction constant $\delta_3^{(11)}$ which  
is not directly related with an improvement in  the 
calculation of $\Delta_{11}(A)$, and hence of the $NN$ potential. 
Let us recall that the subtraction constants $\delta_p^{(ij)}$ arise because of the rescattering process that the 
$N/D$  method allows to treat in a clear and well-defined way, 
overcoming the obscurities that still remain in the literature associated with the use of the cutoff regularized Lippmann-Schwinger  with 
a higher-order $NN$ potential. 
For the  mixing angle $\epsilon_4$ the quality in the reproduction of data is  similar to that obtained 
by the other approximations just quoted. 
However, for the  $^3 H_4$ phase shifts the outcome at NNLO is a bit worse than at 
NLO and OPE, though one should also notice the tiny values for the $^3H_4$ phase shifts so that this discrepancy 
is certainly small  in absolute value. 
We have also checked that it cannot be removed by releasing the other subtraction constants $\delta_p^{(22)}$, with 
$p=2$, 3 and 4. 
Likely, the origin of this difference in the $^3H_4$ phase shifts  between our full results and the Nijmegen PWA 
can be tracked back to the change in the leading Born approximation once the ${\cal O}(p^3)$ two-nucleon irreducible 
contributions are included in $\Delta_{22}(A)$.

\section{Coupled $^3G_5-{^3I_5}$ waves}
\label{gi5w}

\begin{figure}[h]
\begin{center}
\begin{tabular}{cc}
\includegraphics[width=.4\textwidth]{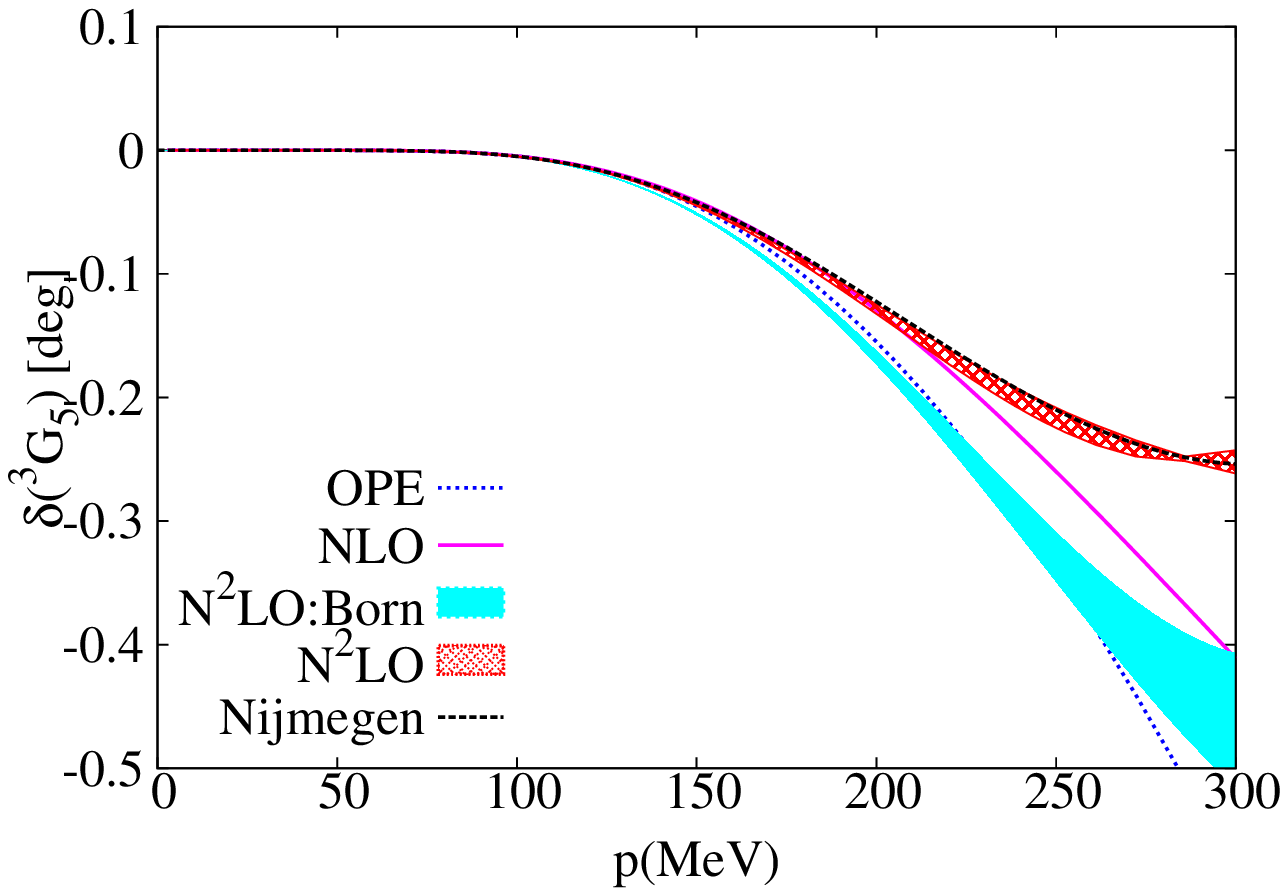} & 
\includegraphics[width=.4\textwidth]{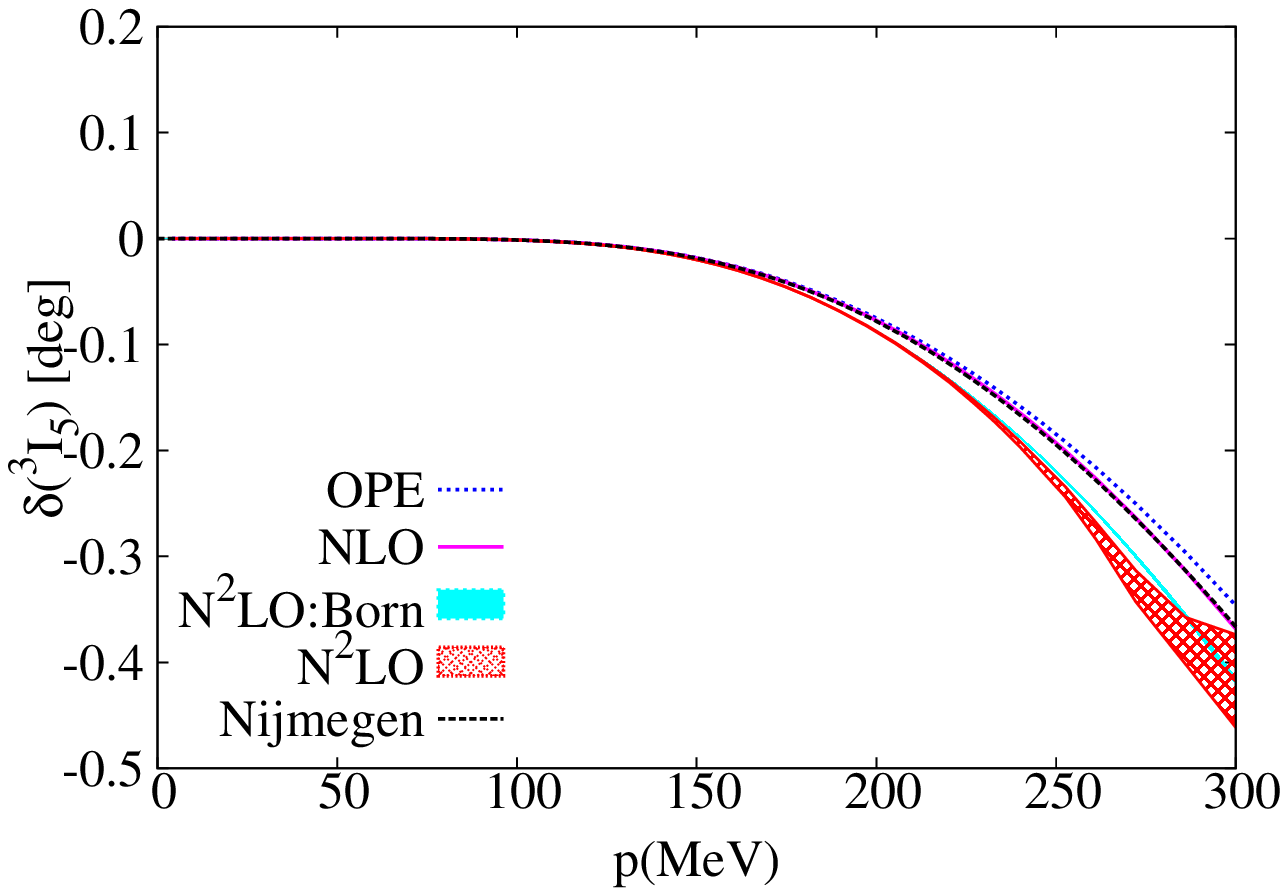}\\  
\includegraphics[width=.4\textwidth]{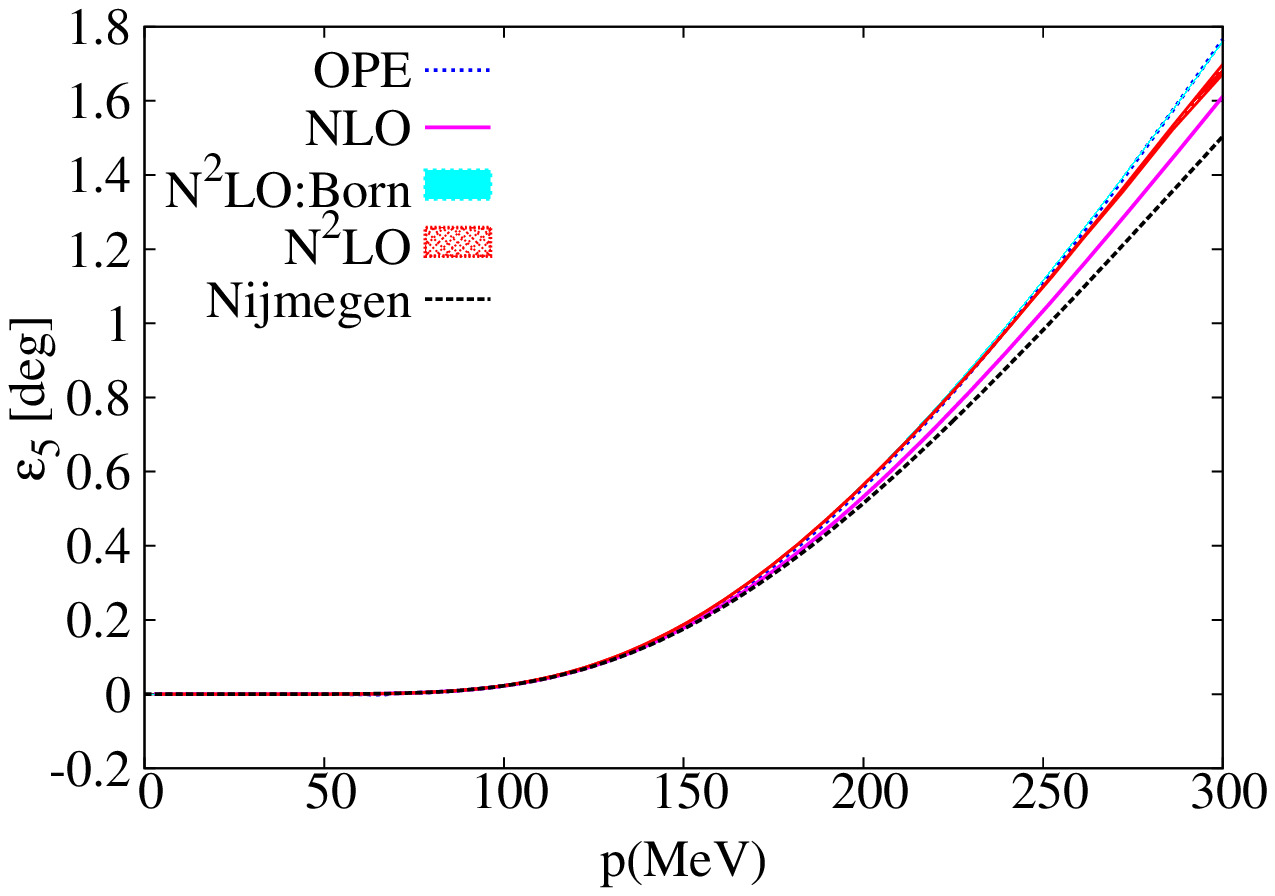}  
\end{tabular}
\caption[pilf]{\protect { (Color online.) From top to bottom and left to right: Phase shifts for $^3G_5$, $^3I_5$ and the mixing angle  $\epsilon_5$, in order.
The (red) hatched areas correspond to the NNLO results and the filled ones to the leading
Born approximation.
 The NLO results are shown by the (magenta) dot-dashed line and the LO ones are given by the 
 (blue) dotted lines.
 The Nijmegen PWA phase shifts are shown by the (black) dashed lines.}
\label{fig:3gi5} }
\end{center}
\end{figure}

The standard formalism for coupled waves with high angular momentum, Eqs.~\eqref{highdcc} and \eqref{highncc}, is followed here 
with  $\ell_{11}=4$, $\ell_{12}=5$ and $\ell_{22}=6$.
 The application of the  principle of maximal smoothness 
to fit the free parameters  provides a good reproduction of the Nijmegen PWA phase shifts \cite{Stoks:1994wp}.\footnote{For $5\leq J\leq 8$ the Nijmegen PWA phase shifts \cite{Stoks:1994wp} are those obtained from the $NN$ potential model of Ref.~\cite{obe}.}
The range of values obtained for the free parameters $D_{11}^{(2)}(-M_\pi^2)$ and $D_{22}^{4}(-M_\pi^2)$ is
\begin{align}
D_{11}^{(2)}(-M_\pi^2)&=-0.0025(5)~M_\pi^{-4}~,\nn\\
D_{22}^{(4)}(-M_\pi^2)&=-0.0125(5)~M_\pi^{-8}~,
\label{free.gi5}
\end{align}
while basically the same results are obtained for any  $D_{12}^{(3)}(-M_\pi^2)\leq 0~M_\pi^{-6}$. 
 The results are shown in Fig.\ref{fig:3gi5} by the (red) hatched areas whose widths take into account the 
uncertainty from the set of $c_i$'s   taken and some numerical noise from the iterative process. 
 A clear improvement results in the description of the $^3G_5$ phase shifts compared with the OPE (blue dotted lines), NLO (magenta dot-dashed lines)
 and leading Born approximation results (cyan filled areas). 
It is worth stressing that this partial wave cannot be well reproduced
even at NNNLO in the Weinberg potential scheme  neither by 
keeping a finite value for the three-momentum cutoff entering 
in the solution of the Lippmann-Schwinger equation \cite{epe042}, 
nor by  sending it to $\infty$ as in Ref.~\cite{zeoli}.
 A similar situation occurs too  
for the leading Born approximation results at NNLO, as shown by the (cyan) filled area in the first panel, 
 a result also obtained in Ref.~\cite{epe04}. 
Even more, the modification of the TPE mechanism proposed in this reference by making use of the 
so-called spectral-function regularization  is inoperative here to provide 
an improvement in the Born approximation results.
 A similar problem was also observed in the perturbative calculation at NNNLO in 
Ref.~\cite{entem}.
 From ours results this is not surprising because the improvement in the reproduction of the  
Nijmegen PWA phase shifts for the $^3G_5$ wave is accomplished through the subtraction constant $\delta_4^{(11)}$. 
This constant is directly related to the $NN$ rescattering (from which the final function $D_{11}(A)$ stems nonperturbatively) 
and not to the $NN$ potential or $\Delta_{22}(A)$.
 In the case of the mixing angle $\epsilon_5$ and the 
$^3I_5$ phase shifts  there is a slight worsening in the reproduction of Nijmegen 
PWA compared with the NLO ones, but still our results run very close to the Nijmegen PWA ones.

\begin{table}
\begin{center}
{\small
\begin{tabular}{|l|l|l|}
\hline
Wave & Type of DRs & Parameters \\
\hline
$^1S_0$ & 1DR & $\nu_1=30.69$ \\
       &  2DR & $\nu_1=30.69$~,~$\nu_2=-23(1)$, $\delta_2=-8.0(3)$ \\
\hline
$^3P_0$ & 3DR & $\nu_2=1.644$~,~$\delta_2=2.82(5)$~,~$\delta_3=0.18(6)$ \\
\hline
$^3P_1$ & 3DR & $\nu_2=-1.003$~,~$\delta_2=2.7(1)$~,~$\delta_3=0.47(3)$ \\
\hline
$^1 P_1$ & 2DR & $\nu_2=-1.723$~,~$\delta_2=0.4(1)$ \\
\hline
$^1D_2$ &  LTS & $D^{(1)}(0)=0.07(1)$  \\
\hline
$^3D_2$ & LTS &   $D^{(1)}(0)=-0.017(3)$  \\
\hline
$^1F_3$ & LTS & $D^{(2)}(0)=0.057(3)$  \\
\hline
$^3F_3$ & LTS & $D^{(2)}(0)= 0.035(5)$  \\
\hline
$^1G_4$ & LTS & $D^{(3)}(0)=-0.014(2)$  \\
\hline
$^3G_4$ & LTS & $D^{(3)}(0)=-0.055(5)$  \\
\hline
$^1H_5$ & LTS & $D^{(4)}(0)=0.156$  \\
\hline
$^3H_5$ & LTS & $D^{(4)}(0)=0.066$  \\
\hline
$^3S_1-{^3D_1}$ & $1$DR $^3S_1$, 2DR $^3D_1$, mixing & $E_d$ \\
               & 2DR all & $a_t$, $r_t$, $E_d$ \\
               & 2DR $^3S_1$, mixing, 3DR $^3D_1$ &  $a_t$, $r_t$, $E_d$, $\nu_3^{(22)}=-2.05(5)$ \\  
\hline
$^3P_2-{^3F_2}$ & 3DR for $^3P_2$ and LTS for the others & $\nu^{(11)}_2=0.178$~,~$D^{(1)}_{11}(-M_\pi^2)=0.025(5)$~,~
$\nu_3^{(11)}=0.155(5)$\\
               &  &  $D_{22}(-M_\pi^2)=0.011(4)$\\
\hline
$^3D_3-{^3G_3}$ & LTS & $D_{11}(-M_\pi^2)=0.90(5)$~,~$D^{(2)}_{22}(-M_\pi^2)=-0.09(1)$  \\
\hline
$^3F_4-{^3H_4}$ & LTS & $D_{11}^{(1)}(-M_\pi^2)=-0.009(3)$  \\
\hline
$^3G_5-{^3I_5}$ & LTS & $D_{11}^{(2)}(-M_\pi^2)=-0.0025(5)$~,~$D_{22}^{(4)}(-M_\pi^2)=-0.0125(5)$ \\
\hline
\end{tabular} }
\caption[pilf]{\protect {  We give in the columns from left to right, in order,  
the partial wave, the type of DRs employed to study it
 and the values for the free parameters involved.  }
\label{tab:allparam} }
\end{center}
\end{table}

Finally, we give in Table~\ref{tab:allparam} the values of the free parameters employed in the different partial waves according 
to the type of DRs employed, which is  indicated in the second column.
 This is done by following the notation, already introduced in Ref.~\cite{gor2013}, 
 $m$DR with $m=1,2,\ldots$, and it should be read as $m$-time subtracted DR.
 For the higher $NN$ partial waves we use the abbreviation  LTS to indicate that 
$\ell$ (or $J$ for the mixing partial waves) subtractions are taken  to satisfy the threshold behavior, following 
the standard formalism explained in Sec.~\ref{hpw}.
 According to the 
principle of maximal smoothness only the highest derivative $D^{(n)}(C)$ 
related to the subtraction constants in $D(A)$ is not fixed to its perturbative value 
(1 for $n=0$ and 0 for $n\neq 0$) and released, 
if appropriate.  The units correspond to appropriate powers of $M_\pi^2$, although they are not explicitly shown. 
There is a proliferation of free parameters for the $P$ waves because for them $\lambda>0$, Eqs.~\eqref{unlambdap} 
and \eqref{couplambdap}, so that, except for the $^1P_1$ wave,  three-time-subtracted DRs are needed.
 This could be a specific feature for the NNLO calculation of $\Delta(A)$ that has to be investigated for
 higher-orders.\footnote{If then $\lambda<0$ 
one would need to invoke  less free parameters for the $P$ waves than in Table~\ref{tab:allparam}.}

\section{Conclusions}
\label{conc}

We have discussed in this paper the application of the $N/D$ method when its dynamical input, namely, 
the imaginary part of the $NN$ partial waves along the LHC, is calculated in ChPT up to NNLO.  
It then comprises OPE, leading and subleading two-nucleon irreducible TPE and once-iterated OPE ~\cite{peripheral}. 
We have obtained a quite good reproduction of the Nijmegen PWA phase shifts and mixing angles, in better agreement 
than the one achieved in the previous lower order studies at LO \cite{paper1,paper2} and NLO \cite{gor2013}.   In particular, our NNLO results are able to reproduce 
 the phase shifts for the triplet waves with $\ell_{11}=J-1$, $^3P_2$, $^3D_3$, $^3F_4$ 
and $^3G_5$,  while at NLO they were not properly accounted for.
We do not need to modify the NNLO two-nucleon irreducible diagrams (or chiral $NN$ potential) in order to obtain such a good agreement 
with the Nijmegen PWA, contrary to common wisdom.  
The point that stems from our study is that one should perform in a well-defined way
 the iteration of diagrams along the RHC, which are responsible 
for unitarity and analyticity attached to this cut, rather than reshuffling  the $NN$ potential with contributions  
from higher orders. 
In this respect, the use of DRs allows one to perform the iteration of two-nucleon intermediate states independently of regulator. We have also compared our full results for the higher partial waves with the Born approximation. From this comparison, as well as from 
the direct study 
of the importance of the different contributions of $\Delta(A)$ to the dispersive integrals,  
it follows that  the $NN$  $D$ waves cannot be treated perturbatively.

It is also worth remarking that up to the order studied here we reproduce the long-range correlation between the 
effective ranges and the scattering lengths for the $NN$ $S$ waves when only once-subtracted DRs are applied. 
In this way one can predict values for the $S$-wave effective ranges in agreement with experiment up to around a $10 \%$. 
We have also elaborated a chiral power counting for the subtraction constants, so that twice-subtracted DRs 
are appropriate when $\Delta(A)$ is calculated at NLO and NNLO.  
From these considerations it turns out also that the chiral power expansion is made over a scale $\Lambda\sim 400$~MeV.  
One should consider further  the impact of higher orders in $\Delta(A)$, which are partially calculated already in 
the literature, as an interesting extension of the present work in order to settle the applicability of the $N/D$
method to $NN$ scattering in ChPT with a high degree of accurateness.

\section*{Acknowledgments}
 This work is partially funded by the grants MINECO (Spain) and ERDF (EU), grant FPA2010-17806 and the Fundaci\'on S\'eneca 11871/PI/09.
 We also thank the financial support from the EU-Research Infrastructure
Integrating Activity
 ``Study of Strongly Interacting Matter" (HadronPhysics2, grant n. 227431)
under the Seventh Framework Program of EU and   
the Consolider-Ingenio 2010 Programme CPAN (CSD2007-00042). 

\appendix{}

\section{Calculation of higher order shape parameters}
\label{appen:vs}
\def\theequation{\Alph{section}.\arabic{equation}}
\setcounter{equation}{0}

Let us explain first the method for the $^1S_0$ partial wave, which is then straightforwardly 
generalized to the $^3S_1$ case. 
Taking into account Eq.~\eqref{invTun} we have that  
\begin{align}
H(A)\equiv \frac{4\pi}{m T(A)}+i \sqrt{A}=\sqrt{A} \cot \delta
\label{def.ha}
\end{align}
is an analytical function of $A$ that has no (elastic) unitarity cut because it obeys the Schwarz 
reflection principle and it is real for $A>0$.  Then it admits a Taylor expansion around $A=0$ 
with a radius of convergence equal to $M_\pi^2/4$, since its first singularity is due 
to the onset of the LHC at $A=-M_\pi^2/4$. This expansion is the so-called ERE. 

We can calculate the function $H(A)$ for complex $A$ in a direct way from the DRs of Eqs.~\eqref{onceD}, 
\eqref{onceN},  for the once-subtracted case, and from Eqs.~\eqref{twiceD} and \eqref{twiceN} 
 in terms of twice-subtracted DRs. 
Nonetheless, care has to be taken when employing 
$g(A,k^2)$ from Eq.~\eqref{gdef} because one should guarantee that $\sqrt{A}$ is defined in the first Riemann sheet, 
that is, $\hbox{Im}\sqrt{A}>0$ must be enforced for all $A\in \mathbb{C}$.
 The same requirement should be also fulfilled by the $\sqrt{A}$ that appears explicitly in the definition of $H(A)$. 

The $n$th order derivative of $H(A)$ at $A=0$ can be calculated by making use of the Cauchy's integral formula
\begin{align}
H^{(n)}(0)=\frac{n !}{2\pi i}\oint_\gamma dz \frac{H(z)}{z^{n+1}}~,
\label{hn.deriv}
\end{align} 
where $\gamma$ is a close contour inside the ball of radius $M_\pi^2/4$ and taken counter-clockwise. 
In practical terms we take the contour $\gamma$ as a circle of radius $R<M_\pi^2/4$ with $z=R \exp i\phi$ and 
$\phi\in [0,2\pi]$. 
A good numerical check of the procedure is the stability of the derivative calculated from the previous 
equation independently of the value taken for $0<R<M_\pi^2/4$. Thus, we obtain
\begin{align}
a_s^{-1}&=-\frac{1}{2i \pi }\oint dz \frac{H(z)}{z}~,\nn\\
r_s&=\frac{1}{i \pi }\oint dz \frac{H(z)}{z^2}~,\nn\\
v_i&=\frac{1}{2i\pi}\oint dz \frac{H(z)}{z^{i+1}}~.
\label{vs1s0form}
\end{align}

We can proceed in the same way for the $^3S_1-{^3D_1}$ coupled wave system in terms of the eigenvalue $S_0$ given by 
\begin{align}
S_0&=\frac{1}{2}
\left[
S_{11} + S_{22}
+ (S_{11}-S_{22}) 
\sqrt{ 1 + \left( \frac{2 S_{12}}{S_{11}-S_{22}} \right)^2 }
\right]~.
\end{align}
Then, we define in terms of it the corresponding uncoupled partial wave 
\begin{align}
T_0(A)&=\frac{S_0-1}{2i\rho(A)}~,
\end{align}
where the definition of $\rho(A)$ in Eq.~\eqref{rhodef} should be taken 
in the first Riemann sheet.
 An analogous function to $H(A)$ in Eq.~\eqref{def.ha} 
is then constructed from $T_0(A)$ and  we can calculate the different parameters in the ERE 
of Eq.~\eqref{3s1.ere} as in Eq.~\eqref{vs1s0form}.


\end{document}